\documentclass[twocolumn]{aastex63}

\usepackage[utf8]{inputenc}

\usepackage{natbib}
\usepackage{graphicx}
\usepackage{amsmath}
\usepackage{booktabs}
\usepackage{longtable}
\usepackage{xspace}
\usepackage{hyperref}
\usepackage[T1]{fontenc}
\usepackage{lipsum}
\usepackage{comment}
\usepackage{xcolor}
\usepackage{multirow}
\usepackage{enumitem}
\usepackage{float}

\usepackage{IEEEtrantools}

\usepackage{makecell}
\setcellgapes[t]{3pt}
\setcellgapes[b]{1pt}

\providecommand{\mj}{\ensuremath{\,M_{\rm J}}}

\newcommand{\rearth}{\mbox{$R_{\ensuremath{\oplus}}$}}
\newcommand{\mearth}{\mbox{$M_{\ensuremath{\oplus}}$}}

\newcommand{\pdgcs}{$30^{+14}_{-12}\%$}
\newcommand{\pdg}{$16^{+2}_{-2}\%$}
\newcommand{\pdgMsiniCorrected}{$18^{+2}_{-2}\%$}

\newcommand{\pdgcsRosenthal}{$38^{+14}_{-13}\%$}
\newcommand{\pdgRosenthal}{$20^{+2}_{-2}\%$}

\newcommand{\pdgcsBryan}{$24^{+11}_{-10}\%$}
\newcommand{\pdgBryan}{$16^{+2}_{-2}\%$}

\newcommand{\rFb}{\ensuremath{5.24 \pm 0.09}}
\newcommand{\pFb}{\ensuremath{2.7}}
\newcommand{\mFb}{\ensuremath{42 \pm 3}}
\newcommand{\pFc}{\ensuremath{443^{+11}_{-13}}}
\newcommand{\mFc}{\ensuremath{0.26 \pm 0.02}}
\newcommand{\aFc}{\ensuremath{1.1 \pm 0.4}}

\newcommand{\rAb}{\ensuremath{3.39 \pm 0.07}}
\newcommand{\pAb}{\ensuremath{8.9}}
\newcommand{\mAb}{\ensuremath{10.4 \pm 0.9}}

\newcommand{\mAf}{\ensuremath{2.7 \pm 0.3}} 
\newcommand{\pAf}{\ensuremath{2744 \pm 146}} 
\newcommand{\aAf}{\ensuremath{3.6 \pm 0.2}} 

\newcommand{\rBb}{\ensuremath{1.3 \pm 0.6}}
\newcommand{\pBb}{\ensuremath{3.1}}
\newcommand{\mBb}{\ensuremath{4.5 \pm 0.5}}

\newcommand{\pBc}{\ensuremath{2101 \pm 3}}
\newcommand{\mBc}{\ensuremath{0.31 \pm 0.01}}
\newcommand{\aBc}{\ensuremath{3.06 \pm 0.04}}

\newcommand{\rCb}{\ensuremath{2.4 \pm 0.2}}
\newcommand{\pCb}{\ensuremath{2.7}}
\newcommand{\mCb}{\ensuremath{5 \pm 3}}

\newcommand{\pCc}{\ensuremath{502 \pm 16}}
\newcommand{\mCc}{\ensuremath{0.57 \pm 0.07}}
\newcommand{\aCc}{\ensuremath{1.13 \pm 0.03}}

\newcommand{\rDb}{\ensuremath{5.4 \pm 0.2}}
\newcommand{\pDb}{\ensuremath{3.8}}
\newcommand{\mDb}{\ensuremath{26 \pm 2}}

\newcommand{\pDc}{\ensuremath{389  \pm 4}}
\newcommand{\mDc}{\ensuremath{1.05 \pm 0.05}}
\newcommand{\aDc}{\ensuremath{0.98 \pm 0.02}}

\newcommand{\rEb}{\ensuremath{2.0 \pm 0.3}}
\newcommand{\pEb}{\ensuremath{0.7}}
\newcommand{\mEb}{\ensuremath{8.3 \pm 0.3}}

\newcommand{\pEc}{\ensuremath{5285 \pm 5}}
\newcommand{\mEc}{\ensuremath{3.84 \pm 0.08}}
\newcommand{\aEc}{\ensuremath{5.61 \pm 0.09}}

\newcommand{\dista}{95}
\newcommand{\radiusa}{2.3}
\newcommand{\pera}{9.0}
\newcommand{\trenda}{$-27.48 \pm 5.97$ m/s/yr}
\newcommand{\curva}{$14.48 \pm 2.89$ m/s/yr$^2$}

\newcommand{\distc}{111}
\newcommand{\radiusc}{3.0} 
\newcommand{\perc}{5.1}
\newcommand{\trendc}{$41.41 \pm 3.41$ m/s/yr}
\newcommand{\curvc}{$-13.65 \pm 1.40$ m/s/yr$^2$}

\newcommand{\distd}{73}
\newcommand{\radiusd}{3.9} 
\newcommand{\perd}{20.8}
\newcommand{\trendd}{$-22.09 \pm 2.11$ m/s/yr}
\newcommand{\curvd}{$-0.05 \pm 0.73$ m/s/yr$^2$}

\newcommand{\diste}{73}
\newcommand{\radiuse}{2.2}
\newcommand{\pere}{21.3}
\newcommand{\trende}{$30.20 \pm 1.87$ m/s/yr}
\newcommand{\curve}{$-7.22 \pm 0.72$ m/s/yr$^2$}

\newcommand{\distf}{83}
\newcommand{\radiusf}{3.2}
\newcommand{\perf}{3.6}
\newcommand{\trendf}{$-10.51 \pm 2.66$ m/s/yr}
\newcommand{\curvf}{$-7.50 \pm 3.55$ m/s/yr$^2$}

\newcommand{\distg}{72}
\newcommand{\radiusg}{8.1}
\newcommand{\perg}{38.8}
\newcommand{\trendg}{$-13.99 \pm 4.60$ m/s/yr}
\newcommand{\curvg}{$1.79 \pm 1.27$ m/s/yr$^2$}

\begin{document}

\title{The TESS-Keck Survey XXIV: Outer Giants may be More Prevalent in the Presence of Inner Small Planets}

\author[0000-0002-4290-6826]{Judah Van Zandt}
\altaffiliation{NASA FINESST Fellow}
\affiliation{Department of Physics \& Astronomy, University of California Los Angeles, Los Angeles, CA 90095, USA}

\author[0000-0003-0967-2893]{Erik A. Petigura}
\affiliation{Department of Physics \& Astronomy, University of California Los Angeles, Los Angeles, CA 90095, USA}

\author[0000-0001-8342-7736]{Jack Lubin}
\affiliation{Department of Physics \& Astronomy, University of California Los Angeles, Los Angeles, CA 90095, USA}

\author[0000-0002-3725-3058]{Lauren M. Weiss}
\affiliation{Department of Physics and Astronomy, University of Notre Dame, Notre Dame, IN 46556, USA}

\author[0000-0002-1845-2617]{Emma V. Turtelboom}
\affiliation{Department of Astronomy, 501 Campbell Hall, University of California, Berkeley, CA 94720, USA}

\author[0000-0002-3551-279X]{Tara Fetherolf}
\altaffiliation{NASA Postdoctoral Program Fellow}
\affiliation{Department of Earth and Planetary Sciences, University of California, Riverside, CA 92521, USA}

\author[0000-0001-8898-8284]{Joseph M. Akana Murphy}
\altaffiliation{NSF Graduate Research Fellow}
\affiliation{Department of Astronomy and Astrophysics, University of California, Santa Cruz, CA 95064, USA}

\author{Ian J. M. Crossfield}
\affiliation{Department of Physics \& Astronomy, University of Kansas, 1082 Malott, 1251 Wescoe Hall Dr., Lawrence, KS 66045, USA}

\author[0000-0003-0742-1660]{Greg Gilbert}
\affiliation{Department of Physics \& Astronomy, University of California Los Angeles, Los Angeles, CA 90095, USA}

\author[0000-0003-4603-556X]{Teo Mo\v{c}nik}
\affiliation{Gemini Observatory/NSF NOIRLab, 670 N. A'ohoku Place, Hilo, HI 96720, USA}

\author[0000-0002-7030-9519]{Natalie M. Batalha}
\affiliation{Department of Astronomy and Astrophysics, University of California, Santa Cruz, CA 95060, USA}

\author[0000-0001-8189-0233]{Courtney Dressing}
\affiliation{501 Campbell Hall, University of California at Berkeley, Berkeley, CA 94720, USA}

\author[0000-0003-3504-5316]{Benjamin Fulton}
\affiliation{NASA Exoplanet Science Institute/Caltech-IPAC, MC 314-6, 1200 E. California Blvd., Pasadena, CA 91125, USA}

\author[0000-0001-8638-0320]{Andrew W. Howard}
\affiliation{Department of Astronomy, California Institute of Technology, Pasadena, CA 91125, USA}

\author[0000-0001-8832-4488]{Daniel Huber}
\affiliation{Institute for Astronomy, University of Hawai`i, 2680 Woodlawn Drive, Honolulu, HI 96822, USA}
\affiliation{Sydney Institute for Astronomy (SIfA), School of Physics, University of Sydney, NSW 2006, Australia}

\author[0000-0002-0531-1073]{Howard Isaacson}
\affiliation{{Department of Astronomy,  University of California Berkeley, Berkeley CA 94720, USA}}
\affiliation{Centre for Astrophysics, University of Southern Queensland, Toowoomba, QLD, Australia}

\author[0000-0002-7084-0529]{Stephen R. Kane}
\affiliation{Department of Earth and Planetary Sciences, University of California, Riverside, CA 92521, USA}

\author[0000-0003-0149-9678]{Paul Robertson}
\affiliation{Department of Physics \& Astronomy, University of California Irvine, Irvine, CA 92697, USA}

\author[0000-0001-8127-5775]{Arpita Roy}
\affiliation{Space Telescope Science Institute, 3700 San Martin Drive, Baltimore, MD 21218, USA}
\affiliation{Department of Physics and Astronomy, Johns Hopkins University, 3400 N Charles St, Baltimore, MD 21218, USA}

\author[0000-0002-9751-2664]{Isabel Angelo}
\affiliation{Department of Physics \& Astronomy, University of California Los Angeles, Los Angeles, CA 90095, USA}

\author[0000-0003-0012-9093]{Aida Behmard}
\altaffiliation{Kalbfleisch Fellow}
\affiliation{American Museum of Natural History, 200 Central Park West, Manhattan, NY 10024, USA}

\author[0000-0001-7708-2364]{Corey Beard}
\altaffiliation{NASA FINESST Fellow}
\affiliation{Department of Physics \& Astronomy, The University of California, Irvine, Irvine, CA 92697, USA}

\author[0000-0003-1125-2564]{Ashley Chontos}
\altaffiliation{Henry Norris Russell Fellow}
\affiliation{Department of Astrophysical Sciences, Princeton University, 4 Ivy Lane, Princeton, NJ 08540, USA}
\affiliation{Institute for Astronomy, University of Hawai`i, 2680 Woodlawn Drive, Honolulu, HI 96822, USA}

\author[0000-0002-8958-0683]{Fei Dai} 
\affiliation{Institute for Astronomy, University of Hawai`i, 2680 Woodlawn Drive, Honolulu, HI 96822, USA}

\author[0000-0002-4297-5506]{Paul A.\ Dalba}
\affiliation{Department of Astronomy and Astrophysics, University of California, Santa Cruz, CA 95064, USA}

\author[0000-0002-8965-3969]{Steven Giacalone}
\affil{Department of Astronomy, California Institute of Technology, Pasadena, CA 91125, USA}

\author[0000-0002-0139-4756]{Michelle L. Hill}
\affiliation{Department of Earth and Planetary Sciences, University of California, Riverside, CA 92521, USA}

\author[0000-0001-8058-7443]{Lea A.\ Hirsch}
\affiliation{University of Toronto  Mississauga, 3359 Mississauga Road, Mississauga ON L5L 1C6 Canada}

\author[0000-0002-5034-9476]{Rae Holcomb}
\affiliation{Department of Physics \& Astronomy, University of California Irvine, Irvine, CA 92697, USA}

\author[0000-0002-2532-2853]{Steve~B.~Howell}
\affil{NASA Ames Research Center, Moffett Field, CA 94035, USA}

\author[0000-0002-7216-2135]{Andrew W. Mayo}
\affil{Department of Astronomy, University of California Berkeley, Berkeley, CA 94720, USA}

\author[0000-0003-2562-9043]{Mason G.\ MacDougall}
\affiliation{Department of Physics \& Astronomy, University of California Los Angeles, Los Angeles, CA 90095, USA}

\author[0000-0001-9771-7953]{Daria Pidhorodetska} 
\altaffiliation{NASA FINESST Fellow}
\affiliation{Department of Earth and Planetary Sciences, University of California, Riverside, CA 92521, USA}

\author[0000-0001-7047-8681]{Alex S. Polanski}
\affil{Department of Physics and Astronomy, University of Kansas, Lawrence, KS 66045, USA}

\author[0000-0001-7615-6798]{James Rogers}
\affiliation{Institute of Astronomy, University of Cambridge, Madingley Road, Cambridge CB3 0HA, United Kingdom}

\author{Lee J. Rosenthal}
\affiliation{Department of Astronomy, California Institute of Technology, Pasadena, CA 91125, USA}

\author[0000-0003-3856-3143]{Ryan A. Rubenzahl}
\affiliation{Center for Computational Astrophysics, Flatiron Institute, 162 Fifth Avenue, New York, NY 10010, USA}

\author[0000-0003-3623-7280]{Nicholas Scarsdale}
\affiliation{Department of Astronomy and Astrophysics, University of California, Santa Cruz, CA 95060, USA}

\author[0000-0003-0298-4667]{Dakotah Tyler}
\affiliation{Department of Physics \& Astronomy, University of California Los Angeles, Los Angeles, CA 90095, USA}

\author[0000-0001-7961-3907]{Samuel W. Yee}
\altaffiliation{51 Pegasi b Fellow}
\affiliation{Center for Astrophysics | Harvard \& Smithsonian, 60 Garden St, Cambridge, MA 02138, USA}

\author[0000-0003-1848-2063]{Jon Zink}
\affiliation{Department of Astronomy, California Institute of Technology, Pasadena, CA 91125, USA}

\begin{abstract}
We present the results of the Distant Giants Survey, a three-year radial velocity (RV) campaign to search for wide-separation giant planets orbiting Sun-like stars known to host an inner transiting planet. We defined a distant giant to have $a$ = 1--10 AU and $M_{p} \sin i = 70-4000$ \mearth~ = 0.2-12.5 \mj, and required transiting planets to have $a<1$ AU and $R_{p} = 1-4$ \rearth. We assembled our sample of 47 stars using a single selection function, and observed each star at monthly intervals to obtain $\approx$30 RV observations per target. The final catalog includes a total of twelve distant companions: four giant planets detected during our survey, two previously known giant planets, and six objects of uncertain disposition identified through RV/astrometric accelerations. Statistically, half of the uncertain objects are planets and the remainder are stars/brown dwarfs. We calculated target-by-target completeness maps to account for missed planets. We found evidence for a moderate enhancement of distant giants (DG) in the presence of close-in small planets (CS), P(DG|CS) = \pdgcs, over the field rate of P(DG) = \pdg. No enhancement is disfavored ($p \sim$  8\%). In contrast to a previous study, we found no evidence that stellar metallicity enhances P(DG|CS). We found evidence that distant giant companions are preferentially found in systems with multiple transiting planets and have lower eccentricities than randomly selected giant planets. This points toward dynamically cool formation pathways for the giants that do not disturb the inner systems.
\end{abstract}

\section{Introduction}
\label{sec:introduction}

Planets between the size of Earth and Neptune with orbital periods less than one year occur around the majority of Sun-like stars \citep{Petigura2018}. Meanwhile, giant planets with orbital periods longer than one year occur around 10--20\% of stars \citep{Rosenthal2022, Wittenmyer2020, Fischer2014, Cumming2008}. The spread in values arises from different stellar samples along with different definitions of what constitutes a ``distant giant planet.'' The close-in small planet (CS) population was compiled primarily using the transit method ({\em Kepler}/{\em K2}/{\em TESS}), while most distant giants (DG) were discovered with the radial velocity (RV) technique. Historically, transit and RV surveys have targeted nearly disjoint stellar populations \citep{Wright2012, WinnFabrycky2015}, resulting in few systems thoroughly searched for both planet types.

Different planet formation theories disagree on whether the occurrence rates of CS and DG planets should be positively or negatively correlated. \textit{In-situ} models predict solid-rich protoplanetary disks will facilitate the growth of planetary cores both interior and exterior to the ice line suggesting a positive correlation (e.g., \citealt{HansenMurray2012,Chiang2013}). By contrast, models that involve significant migration predict that distant giants could dynamically perturb the cores of nascent small planets, either barring them from inward migration \citep{Izidoro2015, Izidoro2018} or driving them into their host star \citep{BatyginLaughlin2015, Naoz2016}.

Multiple studies have sought to clarify this picture in recent years by measuring P(DG|CS), the conditional occurrence of DGs in systems known to host CSs. Using samples compiled from literature systems with archival RVs, \cite{ZhuWu2018} and \cite{Bryan2019} found enhancements of giants in CS-hosting systems over the field rate: P(DG|CS) $ \approx 30\%$ and P(DG|CS) = $39 \pm 7 \%$, respectively. \cite{Rosenthal2022} also found an enhancement of P(DG|CS) = $41 \pm 15 \%$ using a uniform sample of legacy RV targets from the California Legacy Survey (CLS; \citealt{Rosenthal2021}). In contrast, \cite{Bonomo2023} found no evidence for a correlation among a sample of 38 Kepler/K2 systems, P(DG|CS) = $9.3^{+7.7}_{-2.9}\%$. However, \cite{Zhu2024} noted that the average metallicity of the \cite{Bonomo2023} sample was sub-solar, and that correcting for this raised the conditional rate to $39^{+12}_{-11}\%$.

The Distant Giants Survey aims to measure P(DG|CS) in a homogeneously compiled sample of Sun-like stars hosting transiting CS planets detected by {\em TESS} \citep{Ricker2015}. We introduced the survey and presented the confirmed giant plants in our sample in \cite{VanZandt2023}. In this work, we present the completed Distant Giants Survey, including a uniform analysis of the partial orbits in our catalog, as well as our measurement of P(DG|CS). In Section \ref{sec:survey_review}, we review our survey's target selection function and observing strategy. We describe our planet detection algorithm in Section \ref{sec:detection_algorithm}. We summarize our catalog of full and partial orbit detections in Section \ref{sec:planet_catalog}, and describe the partial orbits in detail in Section \ref{sec:trend_analysis}. We characterize our survey sensitivity in Section \ref{sec:completeness}, and measure conditional occurrence framework in Section \ref{sec:cond_occ}. We present our results in Section \ref{sec:results} and discuss them in Section \ref{sec:discussion}.

\section{Survey Review}
\label{sec:survey_review}

\begin{figure*}
    \centering
    \includegraphics[width=0.62\textwidth]{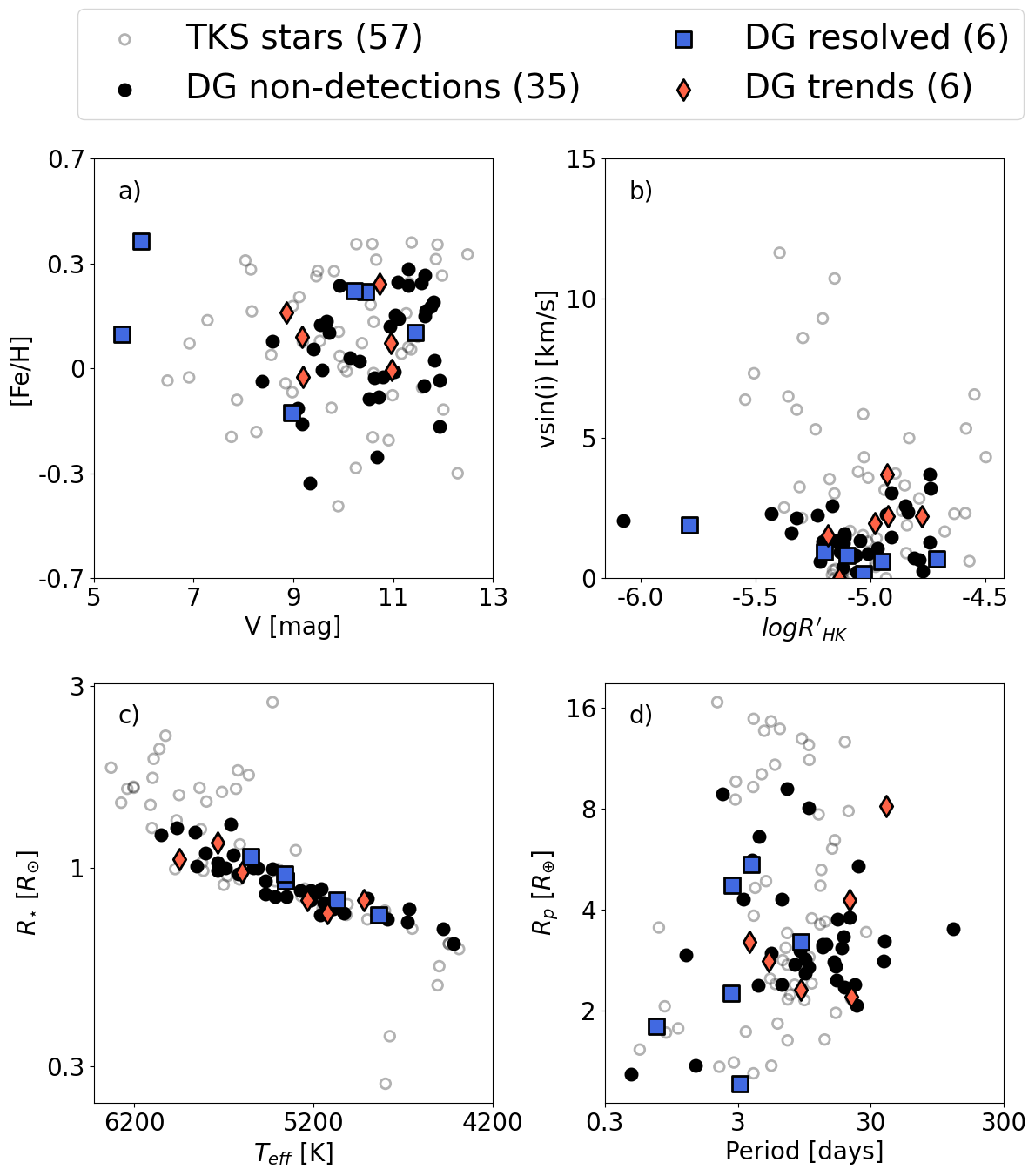}
    \caption{{\bf Stellar and transiting planet parameters of the Distant Giants Survey.} Panel (a): metallicity and $V$-band magnitude of stars with companions detected as resolved orbits (blue squares), companions detected as accelerations (red diamonds), and with no detected companions (black circles). Unfilled circles show other systems in the larger {\em TESS}-Keck Survey. Other panels are same as (a) but for for $V \sin i$ and  $\log R'_{HK}$ (panel b), stellar radius and temperature (panel c), and transiting planet radius and orbital period (panel d). For multi-transiting systems we show the first planet to pass our survey filter (lowest TOI number).
    }\label{fig:stellar_properties}
\end{figure*}

\begin{figure*}
    \centering
    \includegraphics[width=0.95\textwidth]{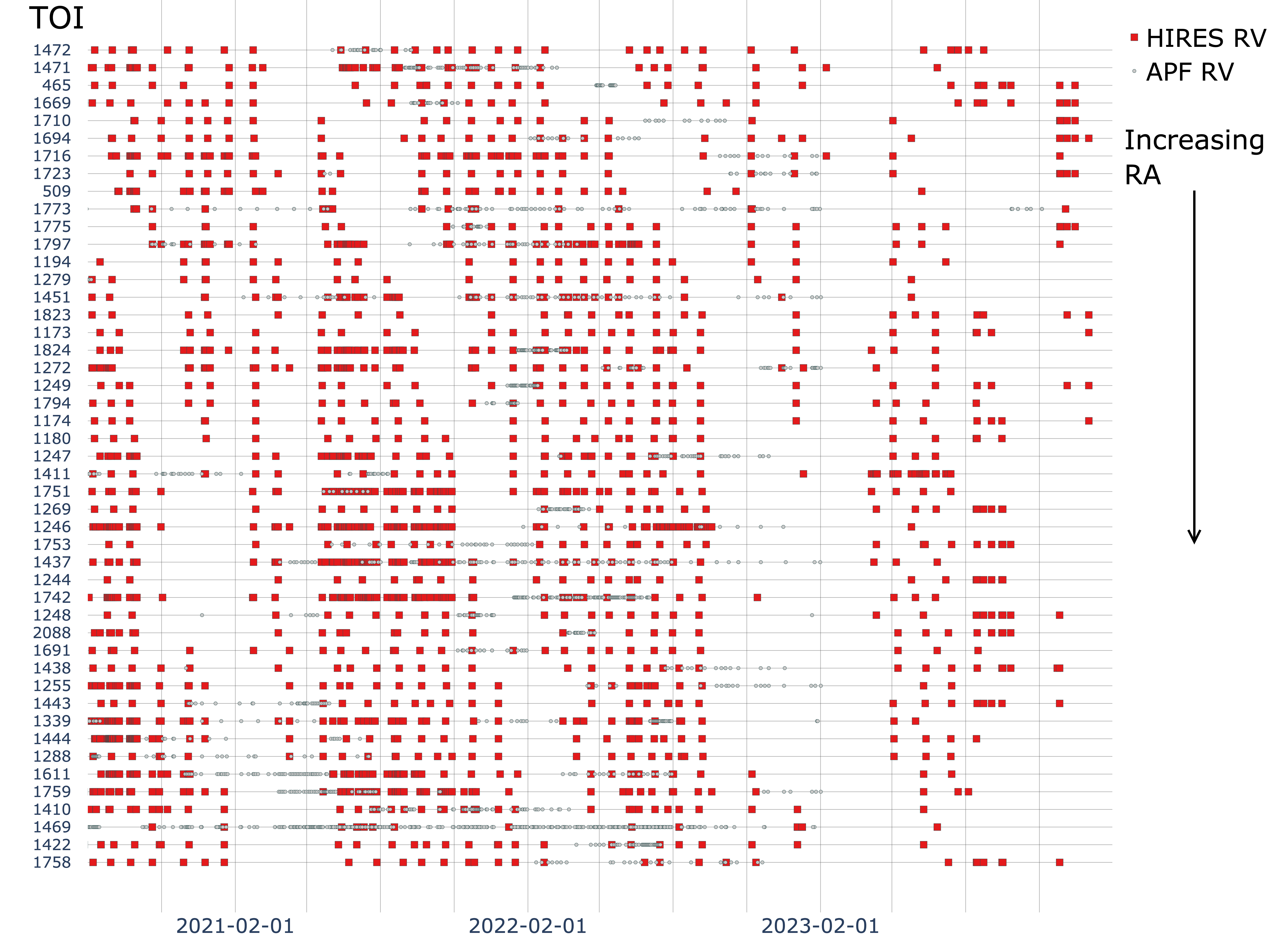}
    \caption{{\bf Radial velocity observations of the Distant Giants Survey}. In total we collected 1990 RVs from Keck/HIRES (red squares) and 2164 observations from APF Levy (gray circles) between 1 August 2020 and 31 January 2024. Note a few targets had prior RVs which are not shown. TOI identifiers are shown in the left margin, ordered by right ascension. Each year label on the $x$-axis marks February 1, the first day of the ``A'' observing semester. The typical target in our survey is not accessible from Keck for about three months per year, resulting in the diagonal cadence gaps. The eruption of Mauna Loa in November 2022, as well as unrelated damage to the lower shutter of the Keck I dome between November 2022 and May 2023, resulted in a substantial decrease in observation cadence for targets between RA $\sim16-22$ hr, which did not reach elevations >40$^{\circ}$ during this period.}\label{fig:abacus}
\end{figure*}

The Distant Giants Survey targeted 47 Sun-like ($M_{\star} = 0.5-1.5 \, M_{\odot}$, T$_{\text{eff}}<6250$ K) {\em TESS} targets, each hosting at least one transiting planet candidate \citep{Guerrero2021}, compiled to determine the conditional occurrence rate of long-period gas giants in the presence of inner small planets \citep{VanZandt2023}. We carried out our survey as part of the larger {\em TESS}-Keck Survey (TKS), a multi-institutional RV survey of over 100 {\em TESS} objects of interest \citep{Chontos2022}. We prioritized RV amenability in our sample, selecting stars with low activity ($\log R'_{\text{HK}}<-4.7$), low rotational velocity ($v \sin i<5.0$ km/s), and high declination ($\delta > 0^{\circ}$) to facilitate observations from Keck and Lick Observatories. We did not require our targets to have prior RV observations, nor did we exclude targets with extant RVs. We required that the transiting companion have $R_p<10 \, R_{\oplus}$ to include a few sub-Jovian size planets, but we apply further restrictions on inner planet radius in our occurrence calculations (see Section \ref{sec:cond_occ}). Our final sample exhibits a metallicity consistent with solar (median [Fe/H]=0.10, $\sigma_{\text{[Fe/H]}}$ = 0.17~dex). For stars with $T_{\text{eff}}>4800$ K, we report metallicity values calculated using \texttt{SpecMatch-Synthetic} \citep{PetiguraThesis}, while for stars with $T_{\text{eff}}\leq4800$ K, we report metallicities from \texttt{SpecMatch-Empirical} \citep{Yee2017}. We summarize the stellar properties of the targets in the TKS and our sample in Figure \ref{fig:stellar_properties}, and we provide stellar and transiting planet properties in Appendix \ref{appendix:transiting}.

We tailored our observing strategy to detect planets with long periods and large $K$-amplitudes: we observed each target once per month, primarily using the HIRES spectrograph coupled to the Keck I telescope \citep{Vogt1994}, with supplementary observations for bright targets ($V$<10) from the APF/Levy spectrograph at Lick Observatory \citep{Vogt2014}. We used the HIRES exposure meter to integrate to a minimum SNR of 110 per pixel. We set a goal of 30 total HIRES observations per target over the nominal three-year duration of the survey. We obtained median values of 37 RV observations, an 1109-day (3.0-year) observing baseline, and 1.7 m/s photon-limited RV uncertainty per target. We add HIRES's 2 m/s instrumental noise floor \citep{FultonThesis} to this last value in quadrature to obtain 2.6 m/s total RV uncertainty. We collected a total of 4154 RVs, 1990 of which were taken using Keck/HIRES. We reached at least 25 RVs and at least 1096-day (3.0-year) baselines for all of our 47 systems. We show our target cadence over the survey duration in Figure \ref{fig:abacus}.

\section{Planet Detection Algorithm}
\label{sec:detection_algorithm}

\begin{figure*}[]
\centering\includegraphics[width=0.6\textwidth]{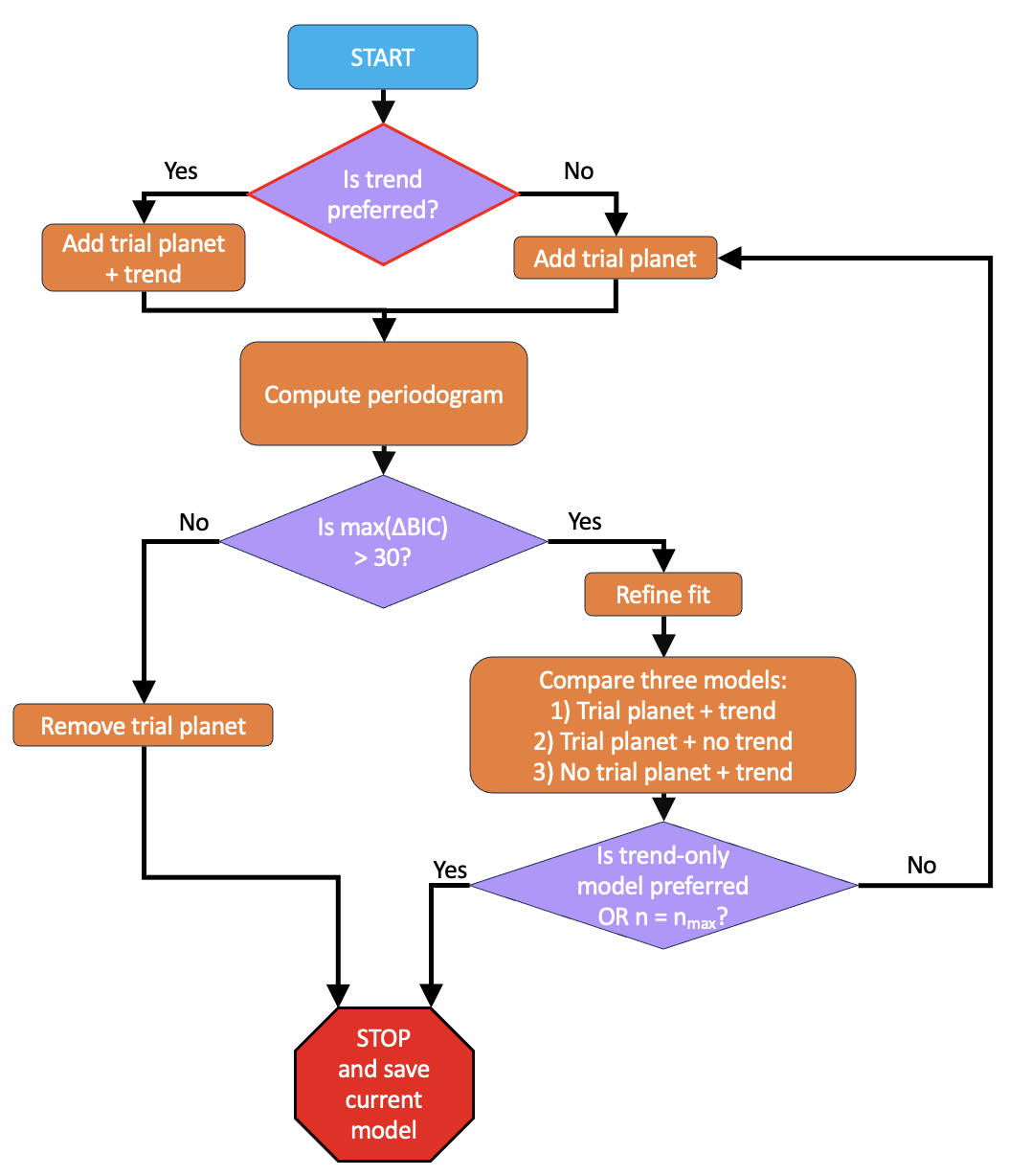}
\centering\caption{{\bf Flow diagram of our planet detection algorithm.} We used the same algorithm for detecting planets (Section~\ref{sec:planet_catalog}) and for computing completeness (Section~\ref{sec:completeness}). We began by initializing an orbital model with the RV data and the ephemerides of any known transiting planets. We then determined whether the data supported the inclusion of a trend, that is, if a trend model was favored over a flat model by $\Delta \text{BIC} > 5$, after which we added a new planet to the model. We computed ($\Delta$BIC) between the starting model and a model with a trial planet over a dense grid of trial periods. If the maximum $\Delta$BIC exceeded  30, we performed a final model comparison test to select a model with a planet, a trend, or both. We iteratively added planets to the model in this way until no more were found. For the injection/recovery experiments described in Section~\ref{sec:completeness}, we began with the final model from Section~\ref{sec:planet_catalog} to retain any planets found during the initial search. We then injected a synthetic planetary signal into the data and begin the recovery at the trend test step (outlined in red). When the search terminated, we checked whether the injected signal was recovered, either as a planet or as a trend, and recorded the result.}
  \label{fig:rvsearch_flowchart}
\end{figure*}

We detected planets using an automated algorithm that we applied uniformly to all RV timeseries. In broad strokes, our approach follows that of the \cite{Rosenthal2021} analysis of the California Legacy Survey (CLS). We used the \texttt{RVSearch} blind search algorithm to select an RV model by iteratively adding planet signals, then we fit the preferred model to the RV data using \texttt{radvel}. However, there are some key differences between our survey and CLS: in the CLS, the targets had more RVs ($N \sim 70$) recorded over longer observing baselines ($t \sim 20$~yr). We therefore tuned the search algorithm to the characteristics of our dataset. We summarize our procedure schematically in Figure~\ref{fig:rvsearch_flowchart} and provide further details below. 

We first established an initial model which included the transiting planet(s) along with their periods $P$ and times of conjunction $t_c$, which we retrieved from the {\em TESS} data validation reports. We optionally included a linear and/or quadratic term in our model to account for any long-term non-periodic variability. 

Next, we constructed a grid of trial periods. The spacing is such that there is at most a phase slip of 1 radian between the trial periods to ensure significant peaks are not missed. For each candidate period, we introduced an additional planet to the model, which we will refer to as the `trial planet.' With the trial planet's period fixed, we fit the remaining orbital parameters --- time of periastron $t_p$, eccentricity $e$, argument of periastron $\omega$, and RV semi-amplitude $K$. If a trend was included based on the prior step, we allowed its parameters to vary as well. During the fitting, we held parameters of all other planets fixed. We calculated the change in Bayesian Information Criterion ($\Delta$BIC; \citealt{Schwarz1978}) between each model and a model without the added planet. We repeated this step for all trial periods to produce a $\Delta$BIC periodogram. 

In principle, we may adopt any significance threshold to accept or reject periodic signals, provided that it is used in both the initial search and the completeness correction (described in Section~\ref{sec:completeness}). We identified $\Delta \text{BIC} > 30$ as a threshold that produced relatively few false positive and false negative detections across our sample. If the maximum $\Delta$BIC value did not exceed 30, we removed the trial planet from the model, designated the current model as preferred, and terminated the search. If the maximum $\Delta$BIC exceeded 30, we refined the fit using a finer period search, and performed a final comparison to select a trend or a planetary model. To do this, we generated three copies of the orbit model: (1) trial planet and no trend, (2) no trial planet and a trend, and (3) both trial planet and trend. From these, we selected the model with the highest $\Delta$BIC.

If our three-way model comparison favored a trend only, we designated the current model as preferred. Otherwise, we added another planet to our model and repeated the search until one of the termination conditions was met. As an additional termination condition, we set a maximum of eight planets on each system's model, though in practice never found evidence for more than two. After determining our final preferred model, we derived credible orbital solutions by sampling our posterior probability with \texttt{emcee} \citep{emcee2013}. 

\section{Planet Catalog}
\label{sec:planet_catalog}

\begin{figure*}[ht!]
\centering
\includegraphics[width=0.7\textwidth]{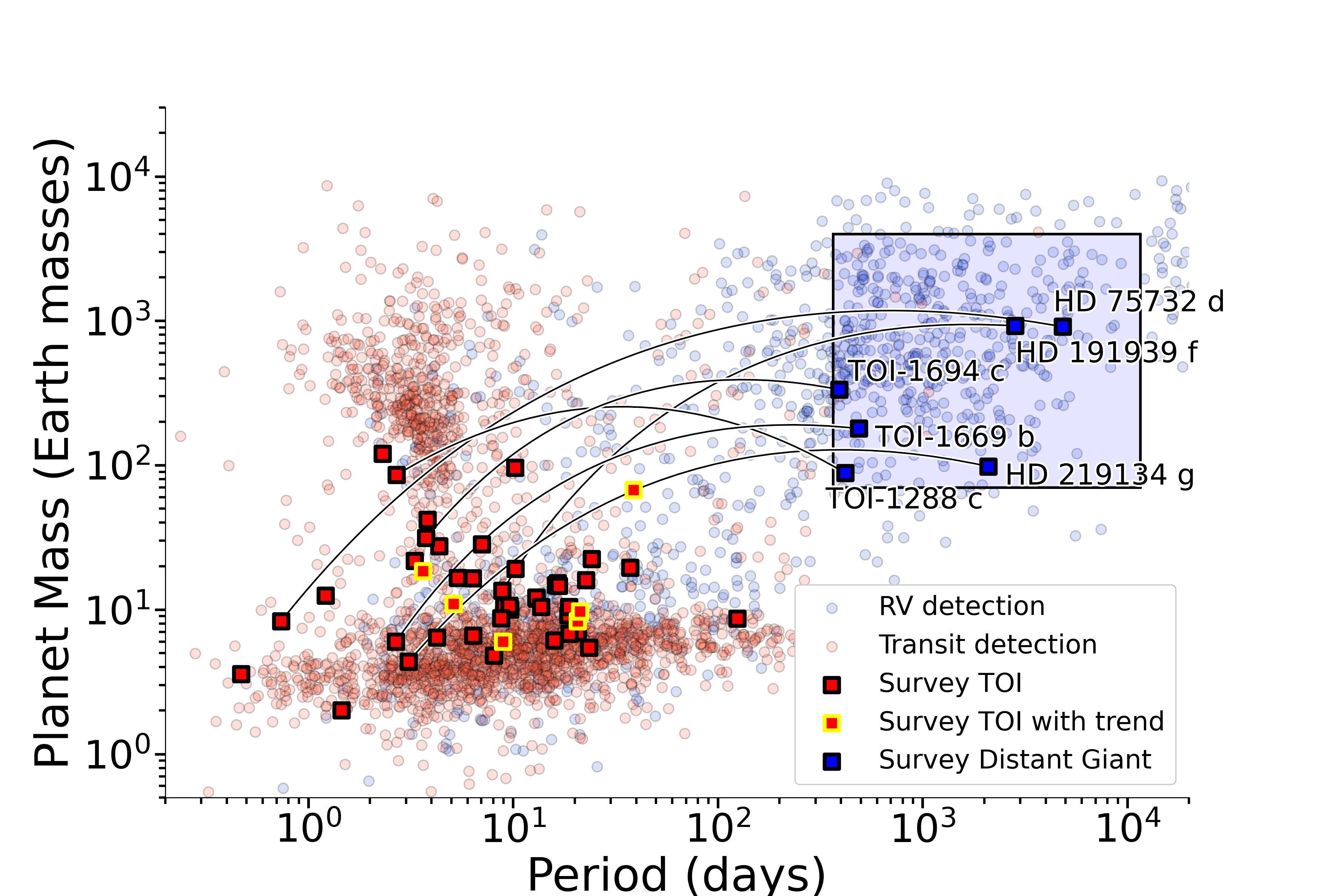}
\caption{{\bf Close-in small planets and their distant giant companions.}
Masses and periods of exoplanets from our survey (bold squares) and the NASA Exoplanet Archive (faded circles) for context . Red/blue points indicate planets discovered using the transit/RV method. Red squares show true masses of transiting planets in the Distant Giants Survey measured by RVs \citep{Polanski2024}, while blue squares show minimum mass measurements ($M\sin i$). For systems with multiple transiting planets, we show the parameters of the transiting planet with the lowest TOI designation that passed our filters. Red squares with yellow borders indicate systems in which we detected a linear/quadratic trend. Giant planets in our sample are connected to the inner planet in their system by a black line. The box corresponds to our nominal definition of a Distant Giant.}
    \label{fig:m_a_DG}
\end{figure*}

\begin{figure*}[ht!]
\centering
\includegraphics[width=0.9\textwidth]{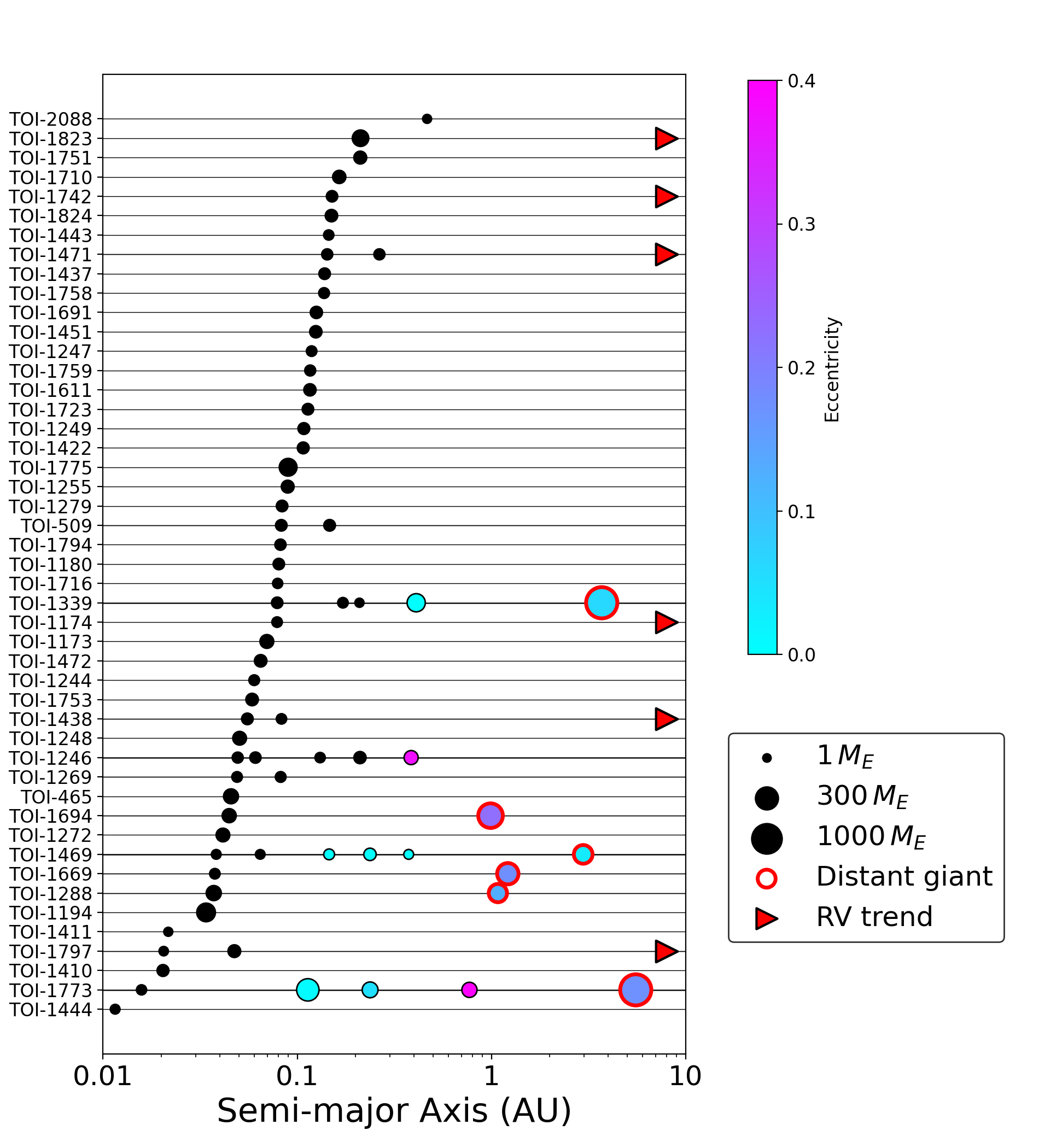}
\caption{{\bf Masses and orbital spacings of the planets in each system in our survey.} Systems are ordered according to the semi-major axis of the inner-most transiting planet. Transiting planets are shown with black markers and non-transiting planets are colored according to their eccentricity. Marker sizes are proportional to the square root of the planet true mass (for transiting planets) or minimum mass (for non-transiting planets). We use red borders to indicate planets which meet our definition of a distant giant. Systems with RV trends are marked with red triangles. Based on Figure 4 of \cite{Weiss2024}.}
    \label{fig:system_architectures}
\end{figure*}

\subsection{Six companions with resolved orbits}
\label{subsec:resolved}
We identified six giant planets with $P \lesssim t_{\text{base}}$ where we could fully resolve the orbits and measure planetary parameters with small fractional uncertainties. Two such planets, HD 219134~g \citep{Vogt2015} and HD 75732~d \citep{Fischer2008}, were known prior to the start of our survey. We announced the discovery of two more, TOI-1669~b and TOI-1694~c, in \cite{VanZandt2023}, and confirmed the mass of HD 191939~f in \cite{Lubin2024}. \cite{Knudstrup2023} independently resolved TOI-1288~c for a total of six giants. We display the masses and periods of these giants and their inner companions in Figure \ref{fig:m_a_DG} and Table \ref{tab:resolved_table}.

\begin{deluxetable*}{lrrrrrrrr}
\tabletypesize{\scriptsize}
\tablewidth{0pt}
\caption{Resolved Distant Giants planet properties}
\label{tab:resolved_table}
\tablehead{
            \colhead{} &
            \colhead{} &
            \multicolumn{3}{c}{Transiting Planet} \vspace{-0.3cm}&
            \multicolumn{3}{c}{Giant Planet} &
            \colhead{}\\
            \multicolumn{2}{c}{}\vspace{-0.32cm} &
            \multicolumn{3}{c}{\rule{5.4cm}{0.4pt}}\vspace{0.12cm} & 
            \multicolumn{3}{c}{\rule{6cm}{0.4pt}} \\
            \colhead{TOI} \vspace{-0.32cm} &
            \colhead{TKS Name} &
            \multicolumn{6}{c}{}\vspace{-0.32cm} &
            \colhead{Reference}\\
            \colhead{} &
            \colhead{} &
            \colhead{Period (days)} &
            \colhead{Radius (\rearth)} &
            \colhead{Mass (\mearth)} &
            \colhead{Period (days)} &
            \colhead{Separation (AU)} &
            \colhead{M$\sin i$ (\mj)} &
            \colhead{}}
\startdata
1288 & T001288 & $\pFb$ & $\rFb$ & $\mFb$ & $\pFc$ & $\aFc$ & $\mFc$ & 1\\
1339 & 191939 & $\pAb$ & $\rAb$ & $\mAb$ & $\pAf$ & $\aAf$ & $\mAf$ & 2\\
1469 & 219134 & $\pBb$ & $\rBb$ & $\mBb$ & $\pBc$ & $\aBc$ & $\mBc$ & 3, 4\\
1669 & T001669 & $\pCb$ & $\rCb$ & $\mCb$ & $\pCc$ & $\aCc$ & $\mCc$ & 5\\
1694 & T001694 & $\pDb$ & $\rDb$ & $\mDb$ & $\pDc$ & $\aDc$ & $\mDc$ & 5\\
1773 & 75732 & $\pEb$ & $\rEb$ & $$\mEb$$ &$\pEc$ & $\aEc$ & $\mEc$ & 4, 6\\
\enddata
\tablerefs{(1) \cite{Knudstrup2023}, (2) \cite{Lubin2024}, (3) \cite{Vogt2015}, (4) \cite{Rosenthal2021}, (5) \cite{VanZandt2023}, (6) \cite{DawsonFabrycky2010}}
\tablecomments{The right-most column gives the reference(s) for giant planet mass and period, as well as the transiting planet mass, if available. We list our own fitted giant planet separations. For TOI-1288, the reference is for all parameters of both planets. We cite transiting planet parameters for HD 191939 from \cite{Lubin2022}. We cite transiting planet parameters for the remaining systems from the {\em TESS} Data Validation Reports. Typical transiting planet period uncertainties are of order $10^{-4}-10^{-5}$ days. We include HD 191939, HD 219134, and HD 75732 in this table because we detected them in our full RV data sets. However, we treat these signals as trends in our homogeneous statistical analysis.}
\end{deluxetable*}

\subsection{Six companions with partial orbits}
\label{subsec:trends}
We detected six massive companions as long-term linear and/or quadratic RV trends and list them in Table~\ref{tab:trend_systems}. We adopted a threshold of $4\sigma$ to consider a fitted trend to be significant. The masses and orbital period of such objects have large uncertainties; often the RVs alone are insufficient to determine whether the object is a planet, brown dwarf, or star. We compute the relative probability of each scenario in Section \ref{sec:trend_analysis}, incorporating astrometric and imaging constraints where available. In Figure \ref{fig:system_architectures} we show the planet masses and separations for each system in our survey, and indicate the systems in which we detected a trend.

\begin{deluxetable*}{ccccccccc}
\tabletypesize{\scriptsize}
\tablecolumns{8}
\tablewidth{0pt}
\tablecaption{Distant Giants trend data}

\label{tab:trend_systems}

\tablehead{
    \colhead{TOI} &
    \colhead{TKS Name} &
    \colhead{$\dot{\gamma}$ (m/s/yr)} &
    \colhead{$\ddot{\gamma}$ (m/s/yr$^2$)} &
    \colhead{$\Delta \mu$ (mas/yr)} &
    \colhead{Direct Imaging?} &
    \colhead{P(planet)} &
    \colhead{P(BD)} &
    \colhead{P(star)}
    }
\startdata
1174 &      T001174 &  -27.0 $\pm$ 3.7 &  14.3 $\pm$ 2.1 &             --- &  True &     0.53 &     0.37 &       0.10 \\
1339 &       191939 &   26.9 $\pm$ 0.5 &  -9.9 $\pm$ 0.5 & 0.13 $\pm$ 0.03 &  True &     0.81 &     0.18 &       0.00 \\
1438 &      T001438 &   10.9 $\pm$ 1.4 & -13.5 $\pm$ 1.7 &             --- &  True &     0.33 &     0.52 &       0.15 \\
1469 &       219134 &   -4.4 $\pm$ 0.3 &   0.0 $\pm$ 0.0 & 0.15 $\pm$ 0.06 &  True &     1.00 &     0.00 &       0.00 \\
1471 &        12572 &  -22.0 $\pm$ 0.4 &  -0.5 $\pm$ 0.5 & 0.07 $\pm$ 0.05 &  True &     0.72 &     0.13 &       0.15 \\
1742 &       156141 &   13.1 $\pm$ 0.5 &  -6.4 $\pm$ 0.5 &             --- &  True &     0.35 &     0.57 &       0.08 \\
1773 &        75732 & -68.6 $\pm$ 12.5 &   6.8 $\pm$ 1.0 & 0.07 $\pm$ 0.06 & False &     1.00 &     0.00 &       0.00 \\
1797 &        93963 &   -9.4 $\pm$ 1.8 & -12.7 $\pm$ 1.8 &             --- &  True &     0.54 &     0.31 &       0.15 \\
1823 & TIC142381532 &   -8.6 $\pm$ 2.1 &   0.5 $\pm$ 0.9 &             --- &  True &     0.45 &     0.39 &       0.15 \\
\hline
Total &              &                  &                 &                 &       &     5.73 &     2.47 &       0.78 \\
\enddata
\tablecomments{RV, astrometric, and imaging information for the nine trend systems in our sample. We include HD 191939, HD 219134, and HD 75732 in this table despite knowing that their trends are planetary in origin because we treated their signals as trends in our statistical analysis. The three columns at the right give the probability of the measured signal in each system originating from a planetary, brown dwarf, or stellar companion between $3-64$ AU. We derived these probabilities by integrating the posterior distributions we calculated using \texttt{ethraid} over the appropriate mass interval. Summing the probabilities for each object type suggests that these nine systems host $5-6$ planets, $2-3$ brown dwarfs, and $\sim1$ stellar companion.}
\end{deluxetable*}

\subsection{Treatment of pre-survey data}
\label{subsec:pre_survey_data}

A handful of our targets' data sets significantly exceed our three-year, 30-observation criteria. For example, HD 219134 and HD 75732 each have >600 observations over $\sim$30 years. For such data sets, it is possible to detect many planets. We found that our detection pipeline, which was tuned for $N \sim 30$~observations and $t_{\text{base}} \sim 3$~yr, struggled to identify the correct orbital parameters of the smaller planets in these systems. We opted for a simple scheme by setting a maximum observing baseline of four years, which truncated the data sets of four systems: HD 207897, HD 191939, HD 219134, and HD 75732. The last three of these each host a distant giant with $P>5$ years, which presented as trends in the truncated data. We treated these signals as trends for our trend analysis (Section \ref{sec:trend_analysis}) and occurrence calculations (Section \ref{sec:cond_occ}), but recognized their planetary nature in our analysis of correlations between distant giants and inner small planet properties (Section \ref{sec:results}). HD 219134 and HD 75732 each have many four-year windows which we could have selected for our analysis. We conducted our occurrence calculation multiple times using different observing windows, and therefore sampling different phases of the giant planet in each system, to verify that our results were not sensitive to our choice of observing window.

\section{Trend Analysis}
\label{sec:trend_analysis}
\subsection{ethraid}
We characterized companions detected as trends using \texttt{ethraid} \citep{VanZandt2024a}. This code determines the masses and orbital periods that are consistent with a measured RV trend, imaging constraints, and/or astrometric accelerations through importance sampling. 

We assumed that the measured signal originated from a single object (as opposed to multiple bodies or stellar activity). We also assumed that its semi-major axis is between 3 and 64~AU; smaller orbits would be resolved as Keplerians and objects with larger orbits would have such high masses that they would be easily detected as stars. We considered companions between $0.1-1000 \mj$, covering planets, brown dwarfs, and low-mass stars. We adopted an informed $M$-$a$ prior, which we expand upon in Section~\ref{subsec:ethraid_prior}

We included astrometric constraints from the $Hipparcos-Gaia$ Catalog of Accelerations (HGCA, \citealt{Brandt2021}) and imaging constraints from \cite{Polanski2024}. The joint $M$-$a$ constraints for these objects are shown in Appendix \ref{appendix:individual_trend_systems} along with notes on each system. For each system, we collected the posterior samples output by \texttt{ethraid} and integrated the distribution over three mass intervals: $M < 13 \mj$, $13 \mj < M < 80 \mj$, and $80 \mj < M$. We report these fractions in Table \ref{tab:trend_systems} as the probability that the companion is a planet, brown dwarf, or star, respectively.

\subsection{Mass-Separation prior}
\label{subsec:ethraid_prior}
We implemented a prior on mass and separation to reflect the intrinsic prevalence of companions with different properties. We derived this prior distribution based on the occurrence rates of sub-stellar (CLS; \citealt{Rosenthal2021}) and stellar \citep{Raghavan2010} companions to Sun-like stars.

We chose to define this prior over the interval $0.03 - 64$ AU, $0.05 - 80 \mj$, excluding regions at very low mass or close separation, which are irrelevant to our trend analysis. We calculated survey completeness in this domain using the ensemble of injection/recovery experiments published by \cite{Rosenthal2021}.%
\footnote{Accessible at \url{https://github.com/leerosenthalj/CLSI/tree/master}}
Following~\cite{Petigura2018}, we divided this region into four intervals in mass and five intervals in semi-major axis, giving 20 cells. We employed the Poisson occurrence method of Section \ref{subsec:occ_model} to calculate the occurrence rate in each cell. 





\begin{figure*}
\centering
\includegraphics[width=0.9\textwidth]{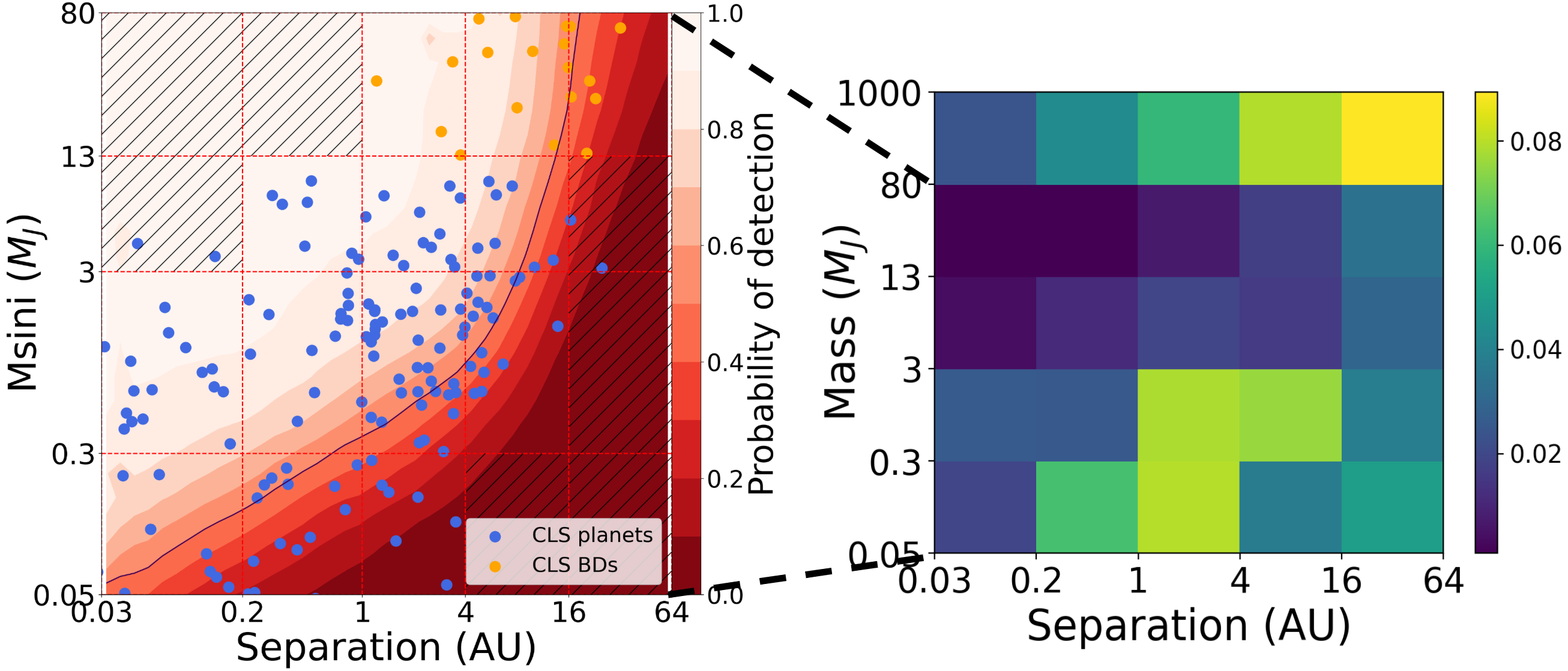}
\caption{{\bf Mass/separation prior informed by the \cite{Rosenthal2021} and \cite{Raghavan2010} surveys.}  \textbf{Left:} Minimum masses and orbital separations of the CLS \citep{Rosenthal2021} sample of planets (blue points) and brown dwarfs (orange points). We also show the survey-averaged completeness map (red contours) emphasizing the 50\% contour with a black line . We compute companion occurrence in the cells defined by the red dashed lines. Hatch marks show cells containing fewer than three objects, giving rise to highly uncertain occurrence rates. However, inspection of the RV constraints in Appendix \ref{appendix:individual_trend_systems} shows negligible overlap with these regions, so they do not affect our overall results. \textbf{Right:} The domain of $M$ and $a$ we explore in our companion search. Each $M$-$a$ sub-domain is colored with the number of objects per star and serves as our joint $M$-$a$ prior described in Section~\ref{subsec:ethraid_prior}. The highest mass bins are based on the \cite{Raghavan2010} rates.}
    \label{fig:mass_sep_prior}
\end{figure*}

For our stellar prior, we used the stellar period distribution of \cite{Raghavan2010}. They fit a normal distribution to a log-period histogram of 259 stellar companions detected among a sample of 454 Sun-like stars, finding that $\log P \sim \mathcal{N}(5.03, 2.28)$. We integrated this distribution to estimate the number of companions in each of our five semi-major axis intervals. We then applied the same occurrence model to these cells, approximating 100\% completeness. We illustrate our mass/separation prior in Figure \ref{fig:mass_sep_prior}.

\begin{figure*}
    \includegraphics[width=.50\textwidth]{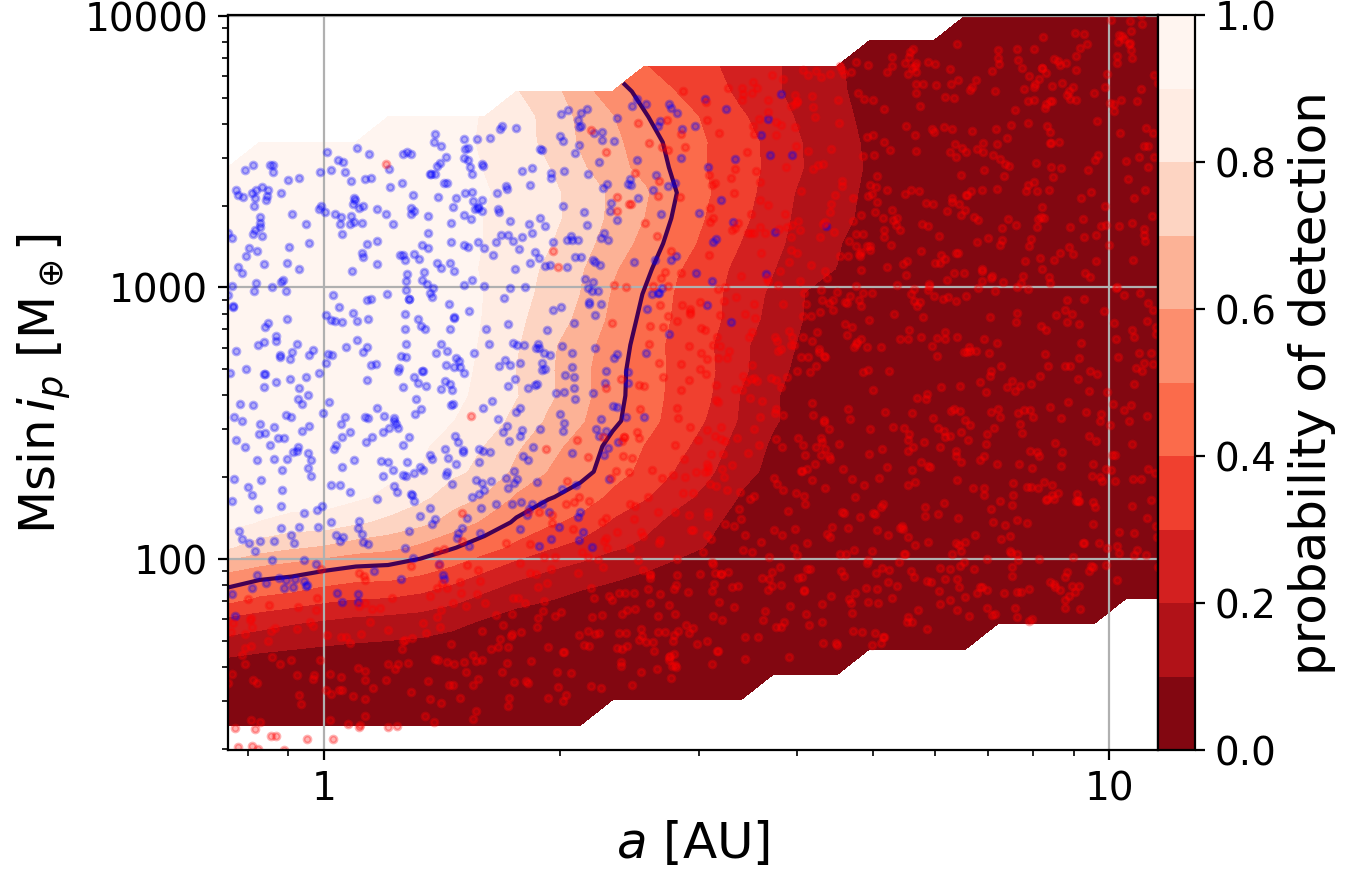}\hfill
    \includegraphics[width=.50\textwidth]{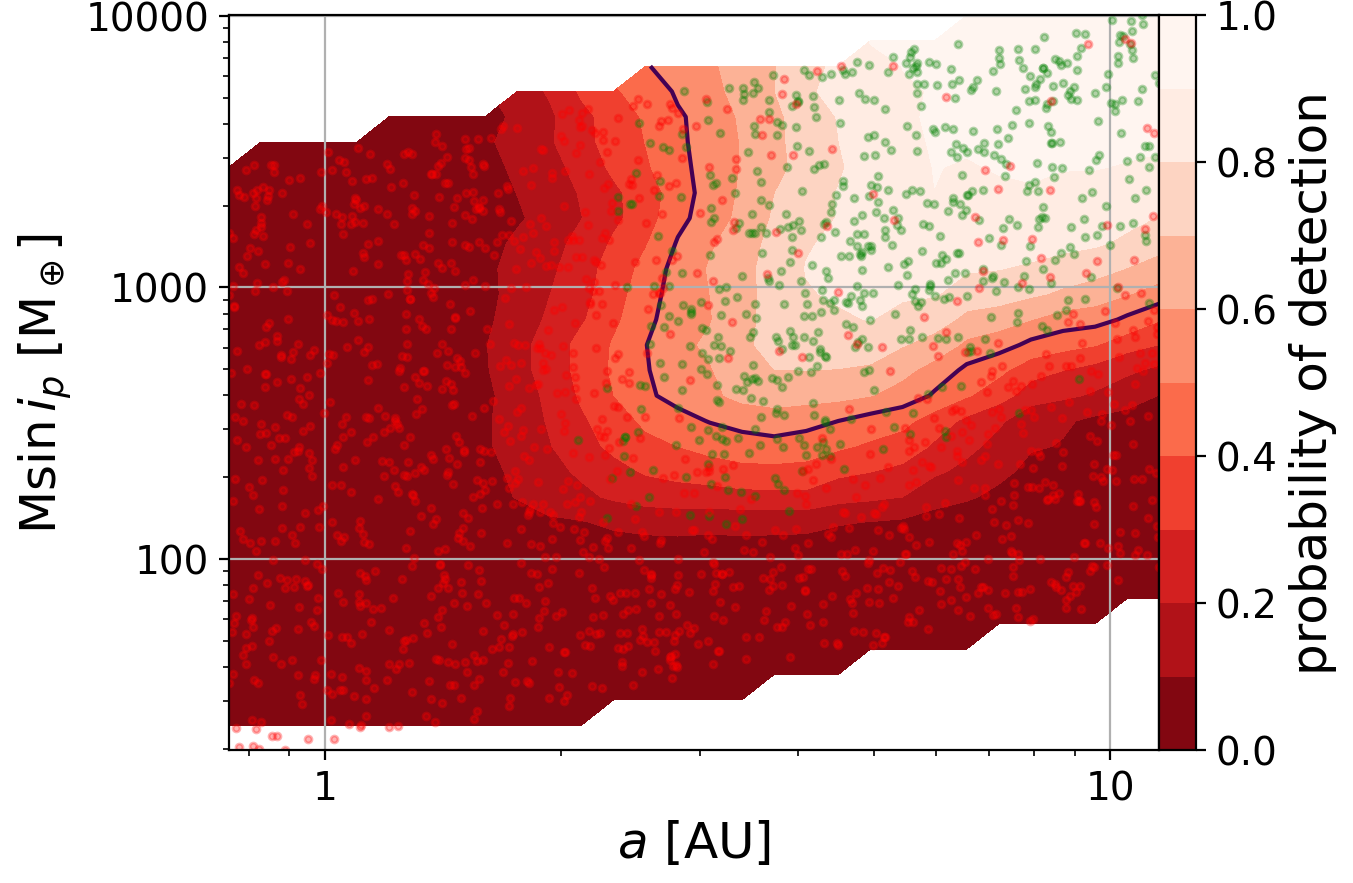}
    \\[\smallskipamount]
    \includegraphics[width=.50\textwidth]{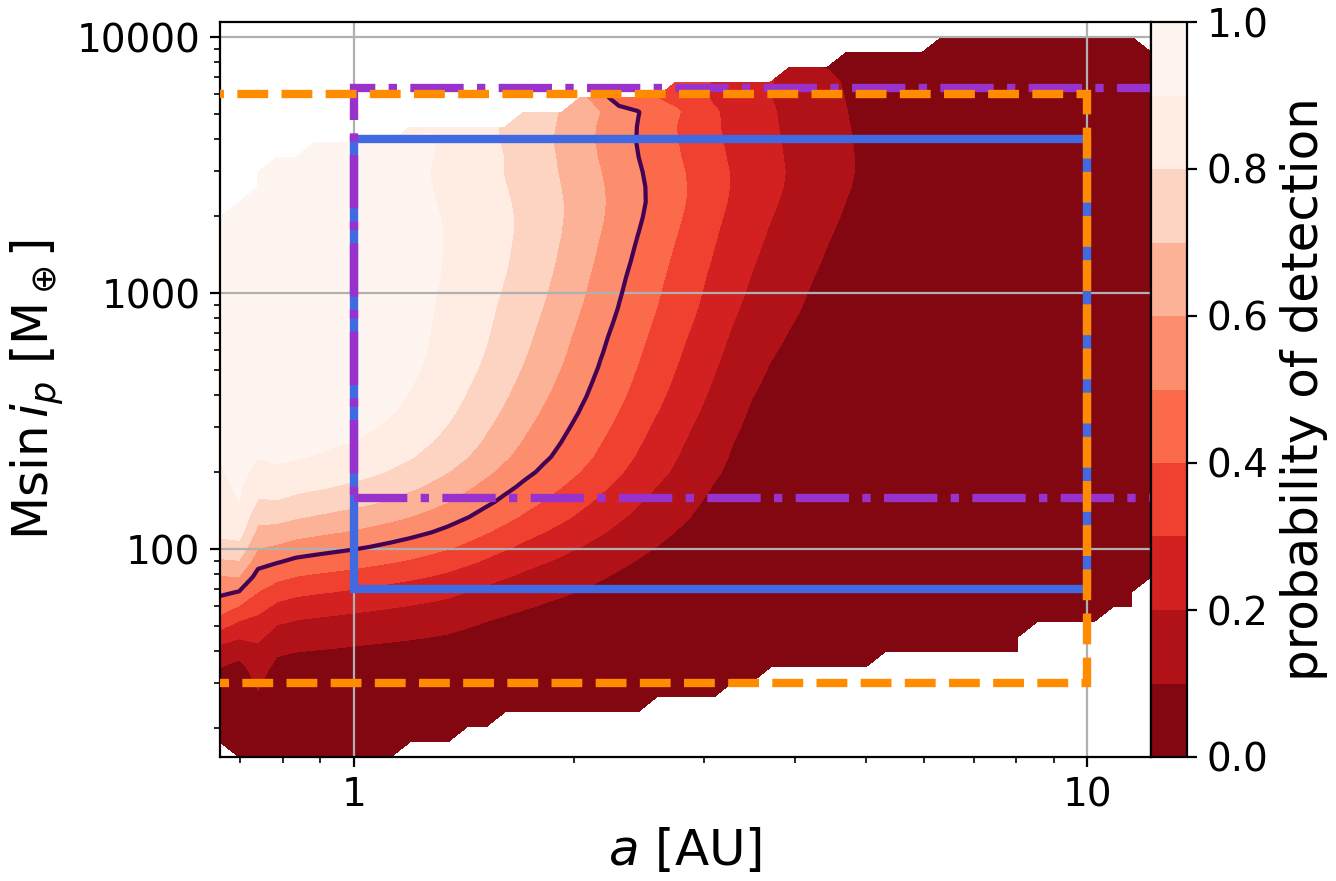}\hfill
    \includegraphics[width=.50\textwidth]{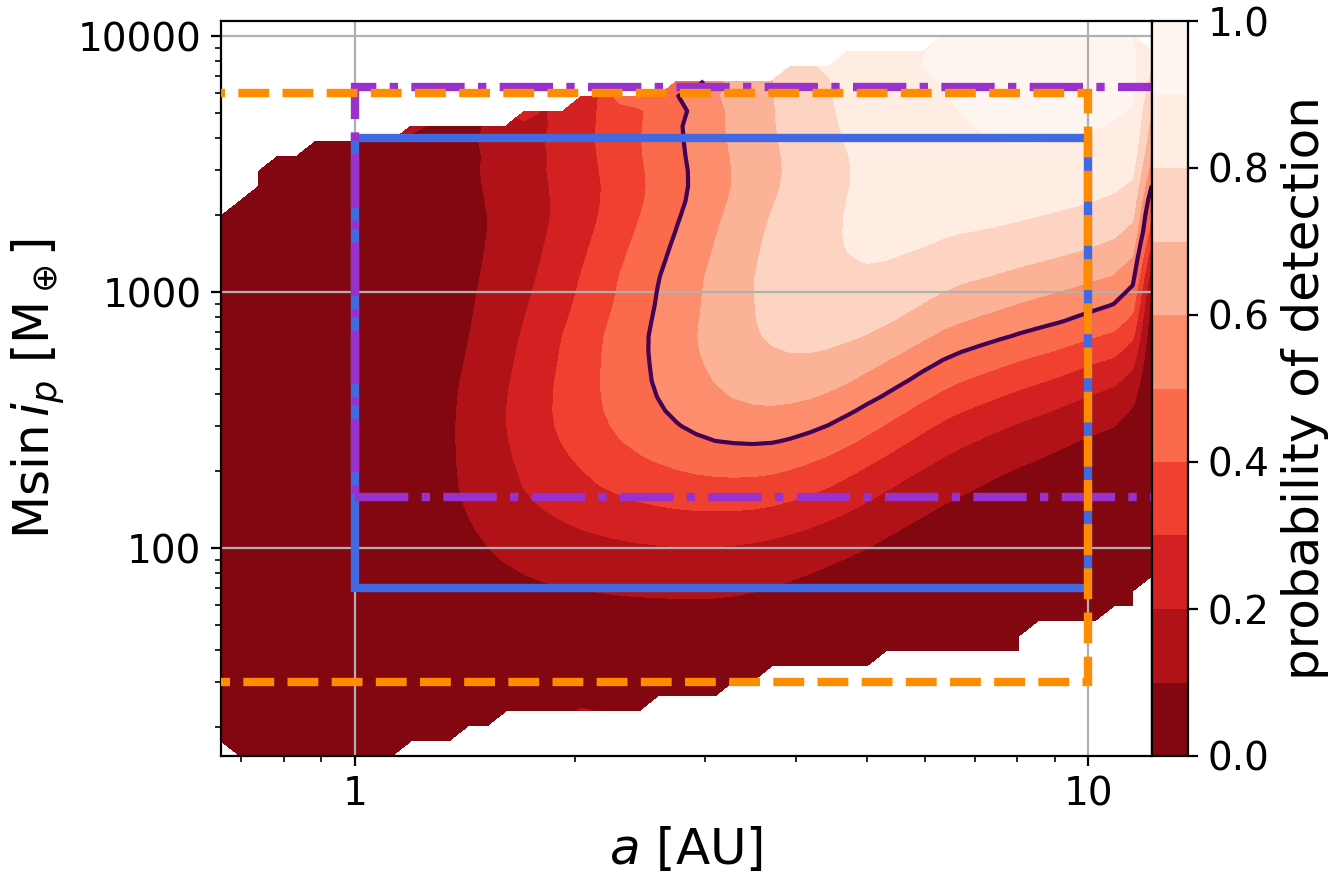}
    \\[\smallskipamount]
    \caption{Side-by-side comparison of injection/recovery results and calculated completeness to resolved orbits (left) and trends (right). We tracked whether our automated search algorithm recovered injected RV signals successfully (blue/green points for resolved orbits/trends), or unsuccessfully (red points). From this collection of recoveries, we calculated a sensitivity map, with contour lines marking completeness deciles, and the black contour denoting the 50\% completeness boundary. We injected 2000 signals for each system. \textbf{Top row:} Completeness for a typical system in our survey (TOI-1173: 3.9-year baseline, 27 observations). \textbf{Bottom row:} Average completeness of the 47 targets in our survey, with blue boxes indicating our nominal definition of a distant giant. Orange and purple dashed boxes indicate the giant planet definitions used by \cite{Rosenthal2022} and \cite{Bryan2019}, respectively. Note that we treat resolved orbits and trends as distinct detection classes, meaning that an orbit recovered as a trend is considered an unsuccessful recovery in the resolved recovery map, and vice versa.}\label{fig:injection_mosaic}
\end{figure*}

\section{Survey Sensitivity}
\label{sec:completeness}

\subsection{Distant Giants Survey}
While we designed the Distant Giants survey to yield a high uniformity in sensitivity to distant giants, each star has differences in RV noise properties, observational sampling, and other properties. We evaluated our sensitivity to both resolved and partial orbits on a star-by-star basis using an injection/recovery scheme. 

We began with the system's preferred orbital model (see Section \ref{sec:detection_algorithm}), subtracted any fitted trend/curvature from the data, and injected a synthetic planetary model. We generated these planets according to the following uniform distributions: $\log P \sim \mathcal{U}(\log P_{\text{min}}, \log P_{\text{max}})$, $\log K \sim \mathcal{U}(\log K_{\text{min}}, \log K_{\text{max}})$, $e \sim \mathcal{U}(0, 0.3)$, $t_p \sim \mathcal{U}(0, P)$, $\omega \sim \mathcal{U}(0, 2\pi$). Here, $P_{\text{min}}, P_{\text{max}}$ = (250 d, 15000 d) and $K_{\text{min}}, K_{\text{max}}$ = (2 m/s, 300 m/s). Figure~\ref{fig:injection_mosaic} shows the suite of experiments for TOI-1173 as an example. We permitted our algorithm to identify at most one additional planet via the same blind search described in Section \ref{sec:detection_algorithm} and in Figure \ref{fig:rvsearch_flowchart}, beginning by testing for a trend. 

When the search terminated, we recorded any signals recovered during the search. We considered a planet {\em successfully} recovered if $P$, $t_p$, and $K$ matched the injected values to 25\% or better. We considered a recovered trend significant if the fitted value corresponded to $\geq8$ m/s RV variation (i.e., three times the typical RV measurement error) over a three-year period. This threshold excluded low-significance trends in an analogous way to our $4\sigma$ trend threshold for our real catalog.

We computed completeness maps in $a$-$M\sin i$ space for each target by performing a moving average over the set of successful and unsuccessful detections (see Figure~\ref{fig:injection_mosaic} for an example). The survey sensitivity is the average of all individual maps (see bottom row of Figure~\ref{fig:injection_mosaic}). As a point of reference, our sensitivity to Jupiter-mass planets as resolved orbits is nearly 100\% at 1~AU and declines to 50\% by 2~AU (roughly the average baseline). Planets three times the mass of Jupiter are recovered as trends with 80\% completeness out to 10~AU.

\subsection{California Legacy Survey}
\label{subsec:cls_completeness}

\begin{figure}
    \includegraphics[width=.50\textwidth]{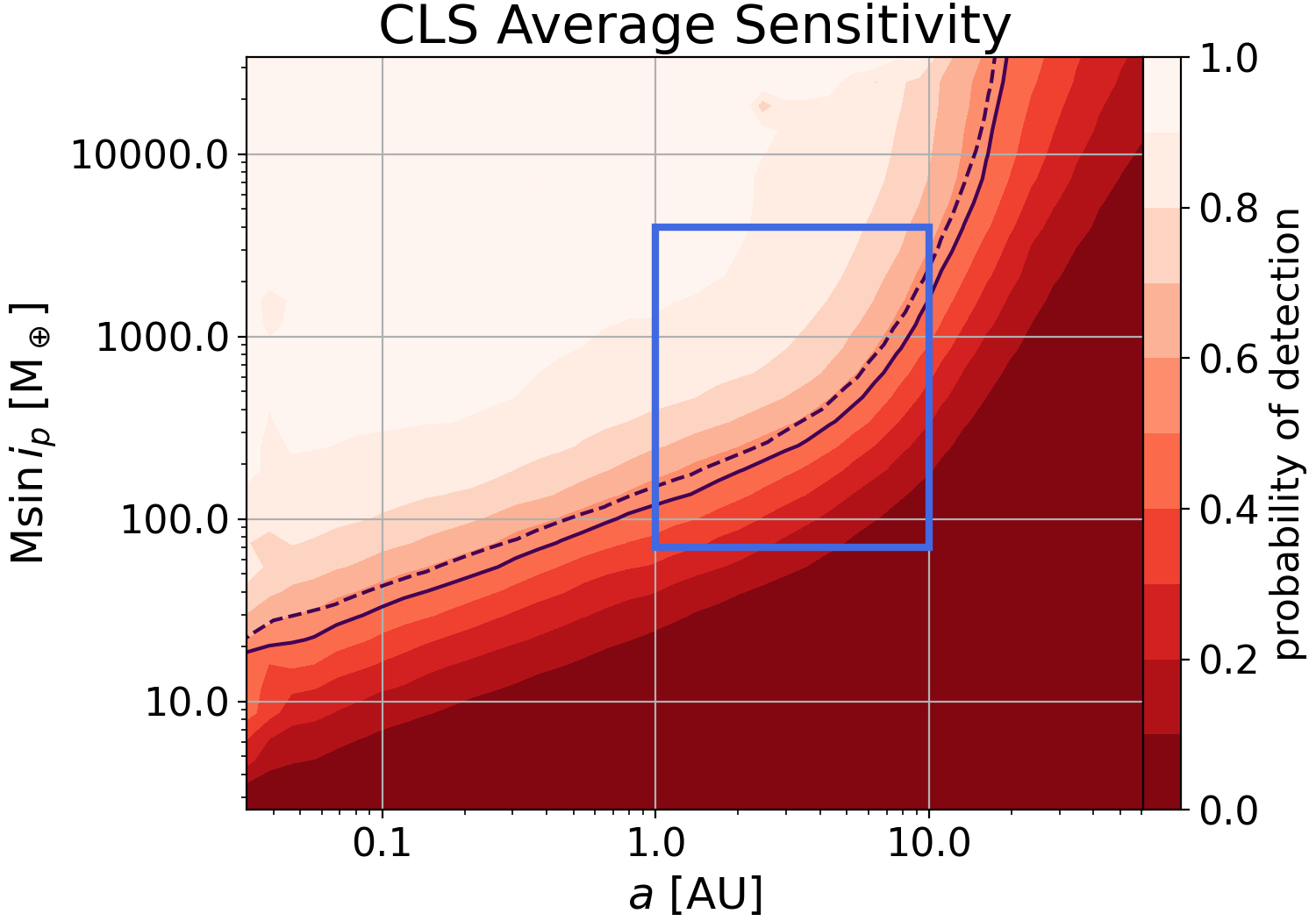}
    \caption{Average sensitivity to companions as a function of minimum mass and separation for the California Legacy Survey. Shaded regions show domains of constant detection probability, and the solid black contour shows the 50\% detection probability boundary. We recalculated this map after adjusting for inclination effects to determine the CLS sensitivity to true mass (see Section \ref{subsec:cls_completeness}). We show the 50\% boundary of this map as a dashed black line. We use a blue rectangle to show our nominal definition of a distant giant ($a = 1-10$, $M = 70-4000$ \mearth). The average CLS sensitivity to minimum/true masses within this domain is 59\%/52\%.}\label{fig:msini_correction}
\end{figure}

To calculate the field occurrence rate of distant giants, P(DG), we averaged together the 719 target-by-target completeness maps computed by \cite{Rosenthal2021} for the CLS sample. Their sensitivity is superior to ours due to their larger RV data sets and longer target baselines. For example, they maintain $\geq$50\% sensitivity to 1 \mj~ objects out to 6 AU (see Figure~\ref{fig:msini_correction}), whereas our sensitivity to Jupiter analogs drops precipitously beyond 2 AU. 

A direct comparison of Distant Giants conditional occurrence to CLS overall occurrence makes the implicit assumption that the two samples have the same inclination distribution. However, if planetary systems inside/outside 1~AU are highly aligned, our selection of systems with transiting planets will favor distant giant planets with edge-on orbits. All else being equal, we would expect increased sensitivity at a given mass in the Distant Giants sample since edge-on planets are easier to detect.

In the limiting case where all CS+DG systems are perfectly aligned (i.e., $\sin i = 1$), the completeness maps shown in Figure~\ref{fig:injection_mosaic} should be interpreted as $M$-$a$ (as opposed to $M \sin i$-$a$). On the other hand, if all systems have random orientations, then the maps are on equal footing with those of the CLS sample.

We modeled both extremes by recomputing the CLS completeness in the following manner: we duplicated each injection 10 times while assigning inclinations according to $\cos i \sim \mathcal{U}(0,1)$. We then converted  $M \sin i$ to $M$, maintaining each injection's status as a successful/unsuccessful recovery, and computed the survey-averaged completeness of the CLS in $M$-$a$ (as opposed to $M \sin i$-$a$). The difference between the corrected and uncorrected maps is minor: at 5 AU, the CLS sensitivity to $M \sin i$=1 \mj is 60\%, while the sensitivity to $M$=1 \mj is 50\%. We show both extremes in Figure~\ref{fig:msini_correction}.

The average completeness of the CLS survey to minimum masses within our nominal distant giant domain is 59\%, while the sensitivity to true masses in the same interval is 52\%. This change produces a $1\sigma$ (2\%) difference in our inferred field occurrence rate P(DG).

\section{Computing Planet Occurrence}
\label{sec:cond_occ}

\subsection{Definitions}
In this work, we define a `close-in' planet to have $a<1$~AU and a distant planet to have $a = 1-10$~AU. We define a `small' planet to have $R_{p} = 1-4 \, \rearth$ for CS planets and a `giant' planet to have $M = 70-4000$ \mearth~ ($0.22-12.6 \mj$). We also consider modified boundaries when making direct comparisons to previous studies. 

\subsection{Occurrence Model}
\label{subsec:occ_model}
Our goal is to measure {\em both} the conditional occurrence of giant planets in systems with small planets P(DG|CS) {\em and} the field occurrence rate of giant planets P(DG) and compare the two rates. For both rates we are considering the number of planets per star. 

Following the prescription in \cite{Rosenthal2022}, we modeled our observed planet catalog as a realization of a censored Poisson process. The process is censored because some planets are missed in regions of imperfect survey completeness ($Q (a, M) < 1$). Our task is to infer the parameters, $\boldsymbol{\theta}$, of the occurrence rate density function $\lambda (a, M | \boldsymbol{\theta})$, where the latter is defined as the number of planets per star per $\log a$-$\log M$ interval. We model $\lambda$ as log-uniform over the DG domain; $\boldsymbol{\theta}$ is thus a single number.


The appropriate likelihood has been described previously (e.g., \citealt{Rogers2021}) and is

\begin{gather}
    P(\{ \boldsymbol{\omega} \}, N_p|\boldsymbol{\theta}) = 
    \frac{e^{-\Lambda}\Lambda^{N_p}}{N_p!} \prod_{k=1}^{N_p}  \frac{Q(\boldsymbol{\omega}_k)\lambda (\boldsymbol{\omega}_k | \boldsymbol{\theta})}{\Lambda}.
    \label{eq:errorless_likelihood}
\end{gather}
Here, $N_p$ is the number of observed planets, $\boldsymbol{\omega}_k$ is the $(a, M)$ tuple for the $k$th planet, and $\Lambda$ --- the `intensity parameter' --- is 
\begin{gather}
    \Lambda = N_{\star} \int Q(\boldsymbol{\omega}_k)\lambda (\boldsymbol{\omega}_k | \boldsymbol{\theta}) \, d \log(a) d \log(M),
    \label{eq:intensity_parameter}
\end{gather}
\noindent where $N_{\star}$ is the number of host stars in our sample. Conceptually, the likelihood in Equation~\ref{eq:errorless_likelihood} can be understood as product of two terms: the first term (before the produce operator) is the probability of observing $N$ planets regardless of their parameters, the second is the probability of observing those planets with their specific $a$ and $M$ values.

Since companions are either detected as trends or resolved orbits, we construct separate likelihoods for each and multiply their results. 

\begin{gather}
    \begin{split}
        P(\{ \boldsymbol{\omega_{\text{pl}}}, \boldsymbol{\omega_{\text{tr}}} \}, N_{\text{pl}}, N_{\text{tr}}|\boldsymbol{\theta}) = 
        & P(\{ \boldsymbol{\omega_{\text{pl}}} \}, N_{\text{pl}}|\boldsymbol{\theta}) 
        \\
        \cdot &P(\{ \boldsymbol{\omega}_{\text{tr}} \}, N_{\text{tr}}|\boldsymbol{\theta})
    \end{split}
    \label{eq:trend_resolved_likelihood}
\end{gather}
Here, the subscripts ``pl'' and ``tr'' refer to the resolved and trend sub-samples, respectively.

We capture catalog uncertainties by sampling many catalog realizations from our full set of posteriors, where each realization comprises one sample from each of the 12 posteriors (three resolved orbits and nine trends). We discard any of the samples that fall outside of our occurrence domain, and derive the posterior surface using Equation \ref{eq:trend_resolved_likelihood}. We average together many such distributions to obtain a robust estimate of the occurrence rate density.

\subsection{Occurrence Computation}
\label{subsec:occurrence_computation}

We used this occurrence methodology to calculate P(DG|CS). We first collected the posterior distributions for all systems hosting a resolved distant giant or a trend. For the resolved planet posteriors, we used gaussian distributions defined by the planet parameters in Table \ref{tab:resolved_table}. For the trends, we used posterior distributions produced by \texttt{ethraid} (see Appendix \ref{appendix:individual_trend_systems}). We drew one sample from each posterior distribution, kept only the samples that satisfied our distant giant definition, and used Equation \ref{eq:trend_resolved_likelihood} to calculate planet occurrence with that realization of our catalog. We repeated this procedure 500 times and averaged the resulting planet occurrence estimates to account for our uncertainties.

\section{Results}
\label{sec:results}

\subsection{Distant giants may be enhanced in the presence of close-in small planets}
Using the procedure described in \ref{sec:cond_occ}, we found a conditional occurrence rate of P(DG|CS) = pdgcs. We then calculated P(DG)=\pdg$--$\pdgMsiniCorrected using the sample of \cite{Rosenthal2021}. The true field rate is likely intermediate between these two extremes. Our results suggest with $1\sigma$ confidence that P(DG|CS) is enhanced over P(DG) by a factor of $\lesssim2$.

\begin{figure*}
    \centering
    \includegraphics[width=0.7\textwidth]{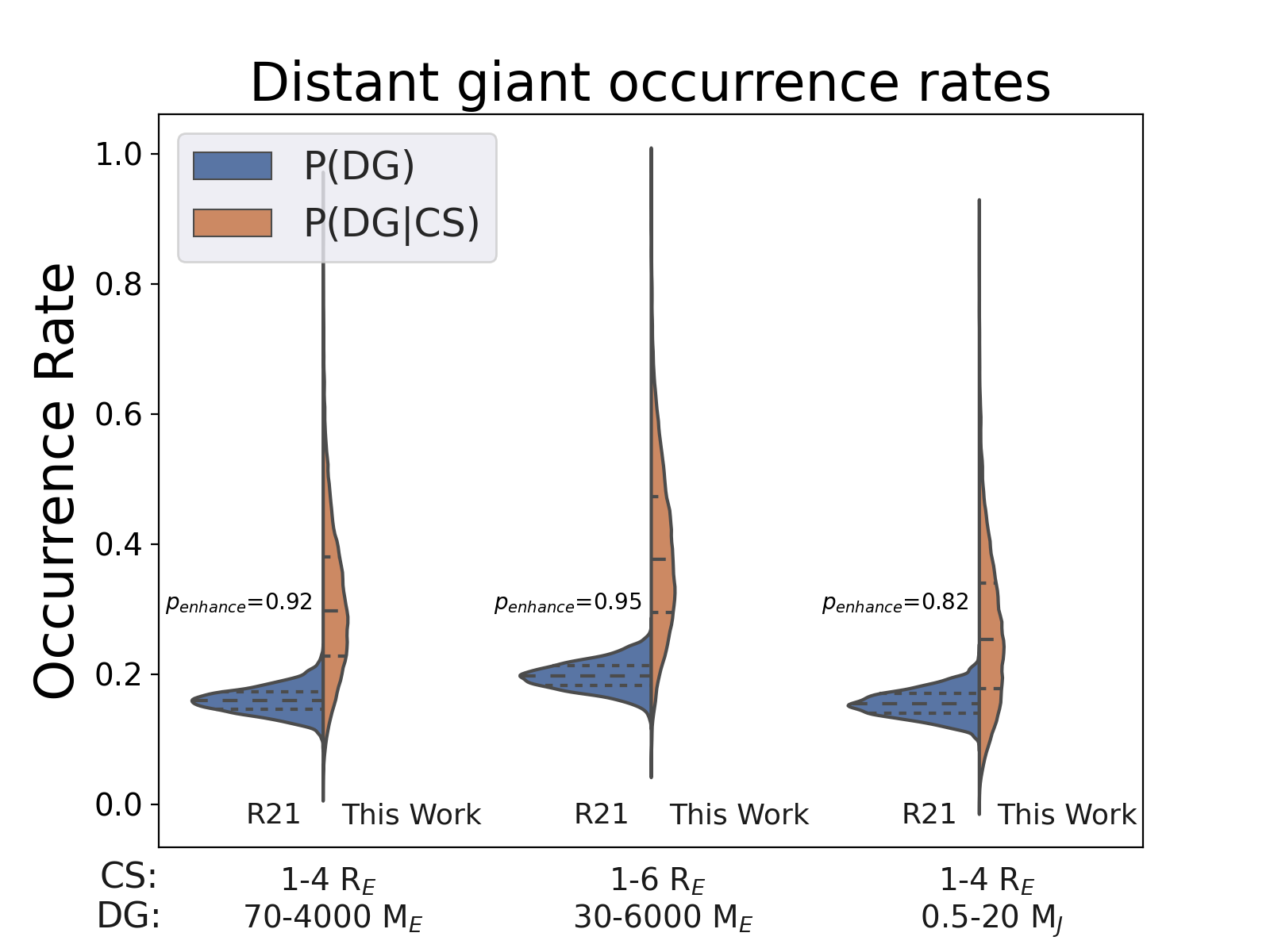}
    \caption{Measurements of the field (blue) and conditional (orange) occurrence of distant giants under different planet definitions. Black lines show distribution quartiles. The left set of distributions show occurrence rates for our nominal definitions: $a = 1-10$ AU, $M_{p} \sin i = 70-4000$ \mearth~ for DGs and $a > 1$ AU, $R_{p} = 1-4$ \rearth~ for CS planets. We annotate each distribution with the probability that the conditional rate is enhanced over the field rate.
    %
    }
    \label{fig:violin}
\end{figure*}

To quantify the significance of the enhancement, we randomly drew $10^4$ values from the P(DG|CS) and P(DG) posteriors and found  P(DG|CS) > P(DG) 92\% of the draws. An analogous experiment with the inclination-corrected P(DG) returned 90\%. We therefore conclude that P(DG|CS) is enhanced over P(DG) with $\geq90\%$ confidence, and that inclination disparities caused by our transit-hosting sample do not significantly affect this result. We summarize our results in Figure \ref{fig:violin}, and report our calculated occurrence rates under our nominal planet definitions in Table \ref{tab:dg_occurrence_rates}.

We repeated our analysis with planet definitions that more closely match those of \cite{Rosenthal2022} and \cite{Bryan2019}, hereafter R22 and B19, respectively (see bottom row of Figure \ref{fig:injection_mosaic}). R22 adopted the following definitions: DG --- $a = 0.23$--10~AU, $M_{p} \sin i = 30$--6000~\mearth; CS --- $a = 0.023-1$~AU, $M_{p} \sin i = 2-30$ \mearth. The 30~\mearth~boundary corresponds to $R_p\sim6 \, \rearth$ \citep{ChenKipping2017}. With this definition, we found P(DG|CS) = \pdgcsRosenthal and P(DG) = \pdgRosenthal, as well as a 95\% probability that P(DG|CS)>P(DG). Our conditional rate is also consistent with R22's finding of P(DG|CS) =$41^{+15}_{-13}\%$ using a different sample, though our field rate is $1\sigma$ higher than their value of P(DG) = $17.6^{+2.4}_{-1.9}\%$, likely owing to our exclusion of M dwarfs from the CLS sample. 

B19 required that $a = 1-20$ AU, $M_{p} \sin i = 0.5-20$ \mj~ for DGs and $R_{p} = 1-4$ \rearth~ for CS planets. Under this definition, we found P(DG|CS) = \pdgcsBryan, lower than their quoted rate of $39 \pm 7\%$, and marginally enhanced over the field rate of P(DG) = \pdgBryan. The probability of enhancement is only marginal at 82\%. We expect that this disagreement is due to differences both in our stellar samples and completeness correction procedures. Using the R22 and B19 definitions, we found evidence for an enhancement of P(DG|CS) at 95\% and 82\% confidence, respectively (see Figure \ref{fig:violin}). We list the results of these tests in Table \ref{tab:dg_occurrence_rates}. We did not perform $\sin i$ corrections when replicating the field rate calculations of \cite{Rosenthal2022} and \cite{Bryan2019} because these studies did not select purely for transiting inner planets in their stellar samples.

\begin{deluxetable*}{ccccccccc}
\tabletypesize{\footnotesize}
\tablecolumns{8}
\tablewidth{0pt}
\tablecaption{DG Occurrence Rates}

\label{tab:dg_occurrence_rates}

\tablehead{
    \colhead{DG mass limits (\mearth)} &
    \colhead{$N_{\star}$ (CLS)} &
    \colhead{$N_{\text{DG}}$ (CLS)} &
    \colhead{P(DG)} &
    \colhead{CS radius limits (\rearth)} &
    \colhead{$N_{\text{transiting}}$} &
    \colhead{$N_{\text{resolved}}$} &
    \colhead{$N_{\text{trend}}$} &
    \colhead{P(DG|CS)}
    }
\startdata
 70-4000 &         598 &       55 & $16^{+2}_{-2}\%$ &          1-4 &            35 &        1.0 &     5.2 & $30^{+14}_{-12}\%$ \\
         &             &          &                  &          1-6 &            42 &        2.2 &     5.9 & $33^{+12}_{-11}\%$ \\
 30-6000 &         598 &       74 & $20^{+2}_{-2}\%$ &          1-4 &            32 &        1.0 &     4.8 & $33^{+13}_{-12}\%$ \\
         &             &          &                  &          1-6 &            39 &        3.0 &     5.5 & $38^{+14}_{-13}\%$ \\
158-6356 &         598 &       54 & $16^{+2}_{-2}\%$ &          1-4 &            32 &        0.8 &     4.3 & $24^{+11}_{-10}\%$ \\
         &             &          &                  &          1-6 &            39 &        1.2 &     5.0 &  $23^{+11}_{-9}\%$ \\
\enddata
\tablecomments{Field and conditional distant giant occurrence rates under different planet definitions. The first DG mass limit is our nominal definition. The second and third match \cite{Rosenthal2022} and \cite{Bryan2019}, respectively. We require that CS and DG planets have $a<1$ AU and $a = 1-10$ AU, respectively. For the other two cases, we adopt the CS and/or DG separation limits given in \cite{Rosenthal2022} ($a = 0.023-1$ AU for CS and $0.23-10$ AU for DG) and \cite{Bryan2019} ($a = 1-20$ AU for DG). We calculated field occurrence rates using the CLS sample \citep{Rosenthal2021}, to which we applied a mass cut ($M_{\star} \geq 0.6 \, M_{\odot}$) to exclude M dwarfs.}
\end{deluxetable*}

\subsection{No strong evidence that high-metallicity systems exhibit a greater enhancement of distant giants}
\cite{Zhu2024} and \cite{BryanLee2024} reported an enhancement of DGs in the presence of CS planets specifically in metal-rich ([Fe/H]>0) systems, beyond the enhancement expected from the established occurrence-metallicity relation \citep{Fischer2005}. 
\cite{VanZandt2024b} tested this claim by repeating the analysis of \cite{BryanLee2024} but using a single sample to measure both the field and conditional occurrence in metal-rich systems. We did not find evidence that the enhancement was specific to metal-rich systems.

We tested the effect of metallicity on giant companion occurrence in the Distant Giants sample by repeating the analysis of Section \ref{sec:cond_occ} with only metal-rich systems. Of the 47 systems in our sample, 19 have super-solar metallicity and host a CS planet under $4 \, \rearth$. This subsample includes one of the three systems with a resolved giant with $P<5$ yr (TOI-1669), two of the three systems with a resolved giant with $P>5$ yr (HD 219134 and HD 75732), and three of the six trend systems (TOI-1438, HD 156141, and HD 93963). We calculated a metal-rich conditional occurrence rate of P(DG|CS, [Fe/H]>0) = $42^{+21}_{-18}\%$. We applied the same filters to the CLS sample and found a metal-rich field rate of P(DG|[Fe/H]>0) = $25^{+4}_{-3}\%$, yielding a probability of 88\% that P(DG|CS, [Fe/H]>0) is enhanced over P(DG|[Fe/H]>0), similar to our results using the non-metal rich sample. We conclude that there is not strong evidence that metal-rich systems exhibit a greater enhancement of the conditional distant giant occurrence rate over the field rate than field stars do.

We also calculated occurrence using the 17 metal-poor systems in our sample, 16 of which host a CS planet under $4 \, \rearth$. We found P(DG|CS, [Fe/H]<0) = $20^{+22}_{-13}\%$, against a field rate of P(DG|[Fe/H]<0) = $7 \pm 2\%$, giving an 86\% probability of enhancement. Our conditional rate is based on two systems, HD 191939 ([Fe/H]=-0.15) and TOI-1174 ([Fe/H]=-0.004), making it highly uncertain. Additionally, TOI-1174's metallicity is consistent with solar. Excluding this system from the calculation gives P(DG|[Fe/H]<0) = $15^{+14}_{-10}\%$ (82\% enhancement probability). Both of these cases show occurrence rates consistent with an enhancement, though we caution that they are derived using small samples. Despite large uncertainties, our results in analyzing the metal-rich, metal-poor, and full samples indicate that metallicity does not exert a strong influence on the relative enhancement of giants. Rather, both the conditional and field occurrence rates rise with metallicity, maintaining an approximately fixed ratio.

\subsection{Inner companions to resolved giants may be preferentially closer-in}
Inspection of Figures \ref{fig:stellar_properties}, \ref{fig:m_a_DG}, and \ref{fig:system_architectures} shows that close-in small planets with resolved distant giants have shorter periods on average than the parent sample. We conducted a two-sample Kolmogorov-Smirnov test \citep{KS1951} to determine whether the separations of close-in companions to resolved giants were drawn from the same distribution as the separations of the single companions. In systems with multiple transiting planets, we used the separation of the first TOI detected in the system. We found a $p$-value of 0.006, meaning that under the assumption that the two populations are drawn from the same distribution, we would expect discrepancies greater than or equal to those observed to occur with $0.6\%$ probability.e repeated this test using only the subset of our targets with transiting planet radii $<4 \, \rearth$, finding $p=0.015$. Our findings suggest that outer giants tend to have lower-separation inner companions, and that this trend may be slightly weaker for inner companions with smaller radii. We note that the transiting companion with the shortest period, HD 75732 e, is also accompanied by a 14-day warm Jupiter, which likely had a more significant dynamical impact on it than the distant giant in this system. We did not find a significant difference between the period distributions of inner planets in trend systems and of single inner planets (i.e., those in systems with no resolved giant and no trend). 

We note that using the first detected TOI in a system favors shorter separations in systems with multiple transiting planets. Thus, the pattern described above may be explained if outer giants are more likely to occur in systems with multiple inner planets (see Section \ref{subsec:giants_multis_correlation}).

\subsection{Outer companions may have preferentially low eccentricities}
To evaluate the eccentricity distribution of our detected giants compared with the broader giant planet population, we performed the following experiment. For each of the six resolved giants in our catalog, we drew one eccentricity value from a gaussian distribution centered on the planet's median eccentricity and with standard deviation equal to the derived eccentricity uncertainty. We then recorded the mean of these six eccentricities. We repeated this process 1000 times to account for the eccentricity uncertainty of each planet. We fit a gaussian distribution to the average eccentricity values, finding that $\langle e_{\text{sample}}\rangle = 0.11 \pm 0.03$.

\begin{figure}
    \includegraphics[width=0.5\textwidth]{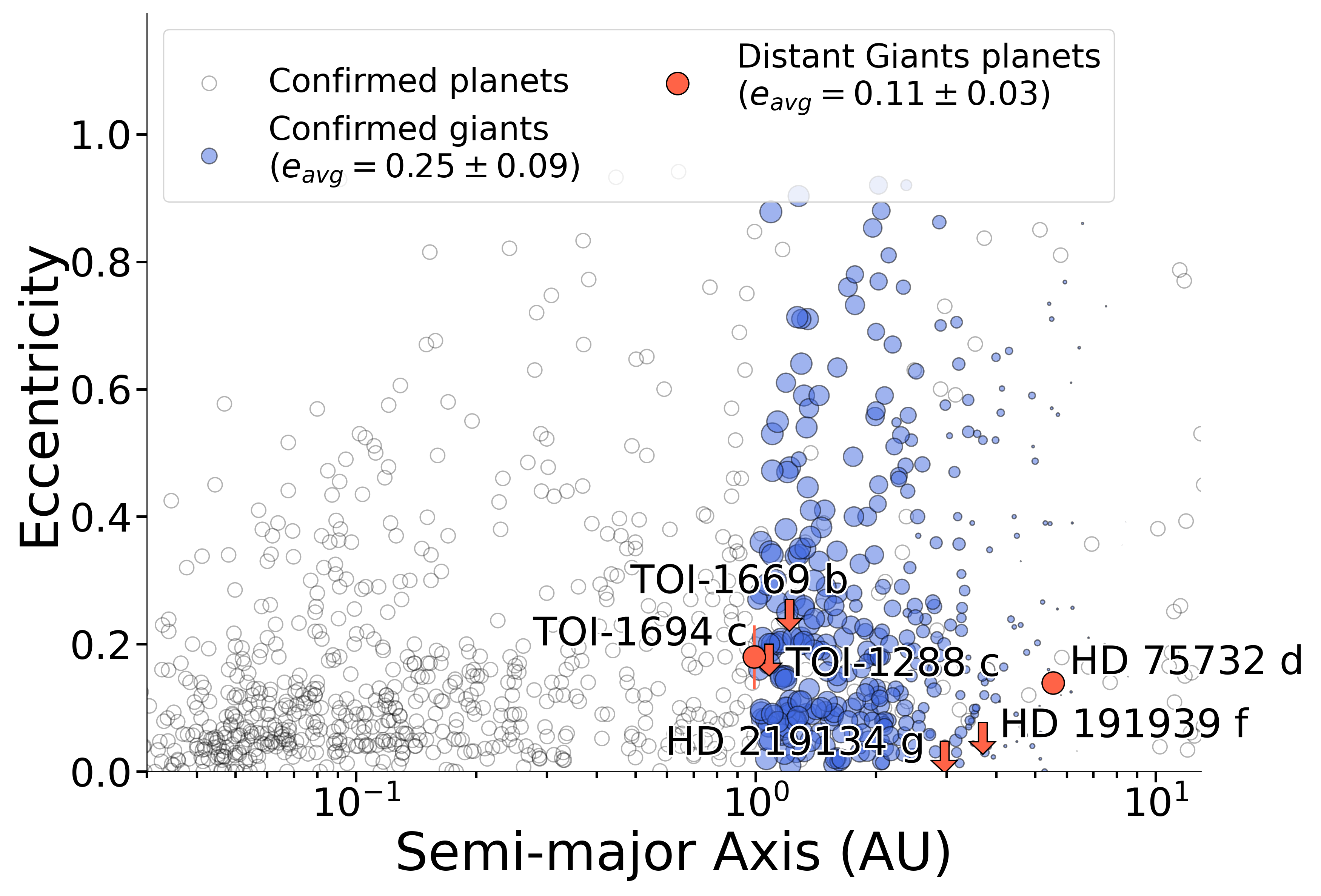}
    \caption{Distribution of eccentricity versus orbital separation for confirmed exoplanets in the NEA with $\sigma_e \leq 0.13$. Confirmed planets are shown as unfilled circles and confirmed distant giants (1-10 AU and 70-4000 $\mearth$) as blue circles. For confirmed giants, marker size is proportional to our survey-averaged completeness at the planet's mass and separation. We show giants in our survey in orange and indicate measurements with less than $3\sigma$ eccentricity precision with an arrow, placing the bottom of the arrow at the $1\sigma$ upper limit. The giant companions in our survey may have lower eccentricities than the average distant giant, but a larger sample is needed to draw a strong conclusion.}
    \label{fig:e_a}
\end{figure}

We used a similar process to quantify the eccentricities of distant giants around field stars. We began with all planets in the NASA Exoplanet Archive (NEA\footnote{https://exoplanetarchive.ipac.caltech.edu/}) that met our definition of a distant giant and had eccentricity uncertainty below 0.13, the maximum eccentricity uncertainty measured among the giants in our catalog. Note that we did not exclude giants with known inner planets. Because there are few systems hosting a confirmed giant and in which an inner planet can be ruled out at high significance, we chose to compute the field eccentricity distribution using all giants, irrespective of inner planet presence. To match the six resolved giants in our catalog, we drew six random planets from this pool, with probability proportional to our measured completeness to each planet (see Figure \ref{fig:injection_mosaic}). We then repeated the procedure we applied to our detected giants, obtaining a distribution of mean eccentricities for the six sampled giants. We again iterated this process by drawing 1000 such six-planet samples and calculating average eccentricity distributions for each of them. We found that the typical average eccentricity among a random sample of six planets from the NEA is $\langle e_{\text{field}}\rangle = 0.25 \pm 0.09$.
We repeated this analysis with only the four resolved giants whose close-in small planet had $R_p<4 \, \rearth$, and found a similar result: $\langle e_{\text{sample}}\rangle = 0.09 \pm 0.03$. Our findings show a discrepancy of $\sim1.5-2\sigma$, indicating that the giants in our sample may have lower eccentricities than field giants. We show the distributions of giant planet eccentricities in Figure \ref{fig:e_a}.

\subsection{Systems with multiple transiting planets may be more likely to host an outer companion}
\label{subsec:giants_multis_correlation}
Eight of the systems in our sample host more than one transiting planet. Four of these systems systems exhibit either a trend or a resolved orbit. We conducted a simple experiment to evaluate the statistical significance of the relationship of between inner planet multiplicity and outer companion occurrence. We randomly drew 12 of the systems from our survey, corresponding to the number of companions detected in our catalog either as resolved orbits or as trends, and counted how many of them belonged to the subset of systems with multiple transiting planets. In 10\% of our $10^4$ experiments, four or more of the sampled systems had multiple inner planets. This finding corresponds to a $p$-value of 0.1, providing a tentative indication that the systems exhibiting resolved orbits or trends have multiple transiting planets more often than average.

Because the nature of the trend systems is uncertain, we repeated the above experiment using only the resolved systems, two of which host multiple inner planets. We drew six systems from our target list, and calculated the fraction of times that two or more of them had multiple inner planets. In this case, there was no evidence of a correlation between inner planet multiplicity and outer giant presence ($p$=0.27). We found similar results when considering only systems hosting inner planets with $R_p<4 \, \rearth$.

\subsection{Outer companion occurrence does not correlate with stellar parameters}
We conducted KS tests for a variety of stellar parameters to see they correlated with distant giant occurrence. We found no significant correlations between the resolved giants in our sample and $T_{\text{eff}}$ ($p$=0.57), $logR'_{\text{HK}}$ ($p$=0.95), or radius ($p$=0.94). We found tentative evidence that stars hosting resolved giants have lower-than-average $v\sin i$ ($p$=0.06) and higher-than-average stellar metallicity ($p$=0.09), in agreement with the the established occurrence-metallicity relation \citep{Fischer2005}. We found similar results when restricting our analysis to the systems with transiting planets smaller than $4 \, \rearth$.

\section{Discussion}
\label{sec:discussion}

\subsection{Distant giant occurrence}
Our finding of a possible positive correlation between CS and DG planets is consistent with most previous studies of this relationship \citep{Rosenthal2022, Bryan2019, ZhuWu2018}. Each of these studies used different --- though overlapping --- target samples compiled according to distinct criteria and analyzed by different methods, and although none was large enough to conclusively measure P(DG|CS), the overarching agreement between them points to a linked formation history between these classes.
We derived a lower enhancement factor of P(DG|CS) over P(DG) than prior works, which may be due to differences between our stellar samples. We constructed the Distant Giants sample using a uniform selection function, and we built up the RV baselines for most of our targets from scratch. Further, our requirement of transiting inner planets necessitated a more involved completeness correction to account for potential inclination biases. We expect that these choices resulted in higher accuracy in our inferred occurrence rates, but they also reduced statistical power by restricting our sample size.

The recently-completed Keck Giant Planet Search (KGPS; \citealt{Weiss2024}) is the largest survey (63 stars) yet used to address conditional giant planet occurrence, and also targeted transiting planet hosts with a uniform selection function and consistent observing strategy. Due to its similarity to the Distant Giants Survey, the KGPS will serve as a useful point of comparison for the results presented here.

We calculated our conditional and field occurrence rates using two different stellar samples, resulting in potential offsets stemming from different stellar parameter distributions (e.g. mass, metallicity, temperature). The sample sizes of current long-baseline RV surveys ($\lessapprox$1000 stars) limits the possibility of measuring both P(DG|CS) and P(DG) in a single sample. Large future surveys of statistically identical stellar samples will alleviate this problem, in addition to providing more accurate and precise occurrence measurements.

\subsection{Distant giants and metallicity}
The occurrence-metallicity relation has been known for two decades: gas giant planets are more prevalent around metal-rich stars \citep{Fischer2005}. This pattern implies that the high densities of solid material in the protoplanetary disks of metal-rich stars facilitate the formation of giant planets. Our study and others before it suggest that giants are also more prevalent in systems hosting an inner small planet.

A natural question is what the interplay between these two effects is. For example, do systems with a metal-rich host star {\em and} a close-in companion show the same relative enhancement over metal-rich field stars as non-metal-rich stars with close-in planets show over non-metal-rich field stars? Recently, \cite{BryanLee2024} reported an increased relative enhancement in metal-rich systems, suggesting that high-metallicity environments are especially well-suited to producing DG-CS systems. In contrast, our findings suggest that metallicity does not strongly influence the relative enhancement; rather, systems of all metallicities exhibit a similar enhancement of distant giants in the presence of close-in small planets.

\subsection{Distant giants and inner planet properties}
A positive correlation between DG and CS planets could indicate that DGs help inner planets form, or that both planets develop independently in similar environments. Whether the DG-CS relation is causative may be encoded in the dynamical characteristics of the systems that host them.

In Section \ref{sec:results}, we found preliminary evidence that, in systems with both an inner transiting planet and a distant giant, the inner planet(s) have shorter periods than average, and the giants have lower eccentricities than average. If real, these patterns could shed light on the formation history of this class of systems. For example, these giants may have excited the eccentricities of their inner companions, initiating high-eccentricity migration to shorter periods through the eccentric Kozai-Lidov mechanism \citep{Li2014, Naoz2016}. On the other hand, this picture requires a high mutual inclination between the inner and outer planets, in tension with the possible overrepresentation of giants in multi-transiting systems. Another explanation is that the giants underwent early Type II disk migration \citep{Lin1986}, entraining gas and planetesimals in the inner disk and driving them to shorter separations \citep{BatyginLaughlin2015}.

We also found that outer giants may be more common in systems with multiple inner transiting planets. This is somewhat unexpected, given that a misaligned outer giant could dynamically perturb the multi-transiting geometry \citep{Naoz2016}. The fact that the system configurations endured suggests that their giants have low mutual inclinations. This feature, coupled with the observed tendency for giant companions to have lower eccentricities, points to a preference for CS/DG systems to either maintain or settle into dynamically cool final configurations, much like the Solar System.

Obliquity offers another indication of dynamical evolution. Of the six systems hosting resolved outer giants, four have measurements of the sky-projected spin-orbit angle between the host star and the inner transiting planet: HD 191939 ($\lambda=3.7\pm5$ deg; \citealt{Lubin2024}), HD 219134 ($\lambda=0-20$ deg; \citealt{Folsom2018}), TOI-1694 ($\lambda=9^{+22}_{-18}$ deg; \citealt{Handley2024}), and HD 75732 ($\lambda=10^{+17}_{-20}$ deg; \citealt{Zhao2023}). The high degree of alignment in these systems comports with a picture involving low mutual inclinations and gentle planetary migration mechanisms.

Many of the findings presented in this work are suggestive at the $2-3\sigma$ level, but not statistically unassailable. To confirm or refute them, similar studies must be performed using larger stellar samples. For example, the number of {\em TESS} candidate hosts recently surpassed 7000, enabling the construction of a quadrupled (200-star) sample under our target selection criteria. Meanwhile, next-generation RV spectrographs such as the Keck Planet Finder \citep{Gibson2016}, NEID \citep{Schwab2016}, and the Habitable Zone Planet Finder \citep{Mahadevan2012} in the north, as well as ESPRESSO \citep{Pepe2021} in the south, offer vastly improved throughput over their predecessors, and more than enough precision to detect long-period giants. Their enhanced efficiency would permit a survey of an additional 150 systems using the same amount of telescope time needed to observe our original 47-star sample. Such a survey would open the door to conditional occurrence measurements at the $3-5\sigma$ level, providing dispositive evidence for or against a correlation.

Additionally, the fourth data release of the $Gaia$ mission \citep{Gaia2016} is expected to yield tens of thousands of giant companion detections (e.g., \citealt{Feng2024, Wallace2024}). These detections will enable precise estimates of the field rate of super-Jupiter planets and, combined with small planet detections from RV and/or transit missions, may help constrain their conditional occurrence as well.

\subsection{Brown dwarf occurrence}
The mass/separation prior we derived in Section \ref{subsec:ethraid_prior} sheds light on the occurrence rate of brown dwarfs as a function of orbital separation. Our work takes advantage of the CLS's sensitivity to brown dwarfs at wide separations ($\lessapprox64$ AU), which extends into the discovery space of high-contrast imaging surveys at $\sim5-1000$ AU (e.g. \citealt{Bowler2020, BowlerNielsen2018, Chauvin2018}). \cite{Nielsen2019} measured the prevalence of brown dwarfs from $10-100$ AU, finding that $0.8^{+0.8}_{-0.5}\%$ of stars host such a companion. They also found that brown dwarfs and giant planets ($5-13 \, \mj$) exhibit different semi-major axis distributions, with planets peaking in occurrence between $1-10$ AU and brown dwarfs favoring wider separations.

We integrated our mass-separation prior between $13-80$ \mj~ and $10-30$ AU, finding that brown dwarfs occur with a frequency of 1.6\% in this interval. Assuming the occurrence rate is log-uniform out to 100 AU, we estimate that 3.2\% of stars host a brown dwarf between $10-100$ AU, significantly greater than the finding of \cite{Nielsen2019}. On the other hand, we observed a distinction between giant planet and brown dwarf occurrence, in agreement with \cite{Nielsen2019}. We found that giant planets, which we define as having $M = 3-13 \, \mj$, peak in occurrence in the interval $1-4$ AU, and decline at greater separations. By contrast, brown dwarf occurrence may increase at greater separations, reaching its maximum in the interval $16-64$ AU. Like the distinct eccentricity distributions between brown dwarfs and giant planets fit by \cite{Bowler2020}, our finding of disparate separation distributions supports the idea that these objects follow different formation pathways.

It is important to note that a number of effects may have influenced our calculated occurrence rates. First, stellar companions on inclined orbits may masquerade as lower-mass objects in RV surveys. We simulated this effect by applying random orbital orientations to a set of stars following the distribution of \cite{Raghavan2010}. We found that on average, one of the nine brown dwarfs we used in our calculation are likely to be stars, insufficient to explain the disagreement with \cite{Nielsen2019}. Nevertheless, the small number of brown dwarfs means that significant contamination remains a possibility. Second, despite its multi-decade baseline, the CLS has limited sensitivity to companions at tens of AU. Many of the detections are partial orbits with large mass and separation uncertainties, and may therefore not fall within the bin in which we counted them. For example, the mass of HD 28185 c was recently revised from $40^{+43}_{-28}$ \mj~ to $6\pm0.6$ \mj~ through the incorporation of HGCA astrometry \citep{Venner2024}. We reserve a more detailed analysis and a firmer conclusion for future work.

\section{Conclusion}
\label{sec:conclusion}

We presented the results of a three-year RV survey to search for outer giant planets around 47 Sun-like stars with known inner planets. Our final catalog includes six RV trends and six well-characterized giants. We incorporated all of these detections into a Poisson likelihood model to calculate the conditional occurrence of distant giants in the presence of close-in small planets, P(DG|CS). We corrected for missed planets by characterizing our detection sensitivity in each system. We found that \pdgcs~ of stars that host a CS planet ($a<1$ AU, $R_{p} \leq 4$ \rearth) also host a DG ($a = 1-10$ AU, $M_{p} \sin i = 70-4000$ \mearth). Meanwhile, using the larger CLS sample of \cite{Rosenthal2021}, we determined that between \pdg~ and \pdgMsiniCorrected~ of stars host a DG planet irrespective of the presence of CS planets. Our findings give tentative evidence for a $1.5-2$x enhancement of giants in CS-hosting systems, suggesting that outer giants and inner small planets may be positively correlated, with giants even promoting the formation of inner planets.

Sample size and homogeneity are vital components of a precise and accurate measurement of conditional giant occurrence. Studies of this topic to date, including this one, have had to prioritize one of these components at the expense of the other. However, advancements over the last few years have made possible a dramatic increase in sample size without compromising sample purity. Making use of this progress will bring the nuances of planetary formation into sharper focus.

\section{Acknowledgments}
J.V.Z. acknowledges support from NASA FINESST Fellowship 80NSSC22K1606. J.V.Z. and E.A.P. acknowledge support from NASA XRP award 80NSSC21K0598.

J.M.A.M. is supported by the National Science Foundation (NSF) Graduate Research Fellowship Program (GRFP) under Grant No. DGE-1842400. J.M.A.M. and N.M.B. acknowledge support from NASA’S Interdisciplinary Consortia for Astrobiology Research (NNH19ZDA001N-ICAR) under award number 19-ICAR19\_2-0041.

E. V. T. acknowledges support from a David \& Lucile Packard Foundation grant.

T.F. acknowledges support from an appointment through the NASA Postdoctoral Program at the NASA Astrobiology Center, administered by Oak Ridge Associated Universities under contract with NASA.

M.L.H. would like to acknowledge NASA support via the FINESST Planetary Science Division, NASA award number 80NSSC21K1536.

This work was supported by NASA Keck PI Data Awards, and NASA Key Project Mission Support proposals administered by the NASA Exoplanet Science Institute. Data presented herein was obtained at the W. M. Keck Observatory from telescope time allocated to the National Aeronautics and Space Administration through the agency's scientific partnership with the California Institute of Technology and the University of California. The Observatory was made possible by the generous financial support of the W. M. Keck Foundation. 

The authors wish to recognize and acknowledge the very significant cultural role and reverence that the summit of Maunakea has always had within the indigenous Hawaiian community. We are most fortunate to have the opportunity to conduct observations from this mountain.

\bibliography{bib.bib}

\appendix

\section{Companions Detected as Trends}
\label{appendix:individual_trend_systems}

\subsection{TOI-1174}
TOI-1174 is a K2 dwarf at a distance of $\dista$ pc hosting a transiting $\radiusa \, \rearth$ sub-Neptune with a $\pera$-day period. We measured RV trend and curvature of \trenda~ and \curva~ in this system, indicating the presence of an outer companion. Although we were unable to precisely constrain $a$ and $M_{p}$ in this system due to the lack of astrometry data, 832 nm speckle imaging observations from the 'Alopeke imager coupled to the 8-m Gemini North telescope \citep{Scott2021} and reduced according to \cite{Howell2011} ruled out luminous companions beyond $\sim$40 AU and more massive than $\sim$200 $\mj$. We depict the direct imaging constraints by converting the measured contrast curves to mass/separation space, assuming circular face-on orbits for simplicity as explained in \cite{VanZandt2024a}. We found that the source of the measured RV variability is most likely planetary: P(planet) = 53\%.

\begin{figure}[H]
    \includegraphics[width=.49\textwidth]{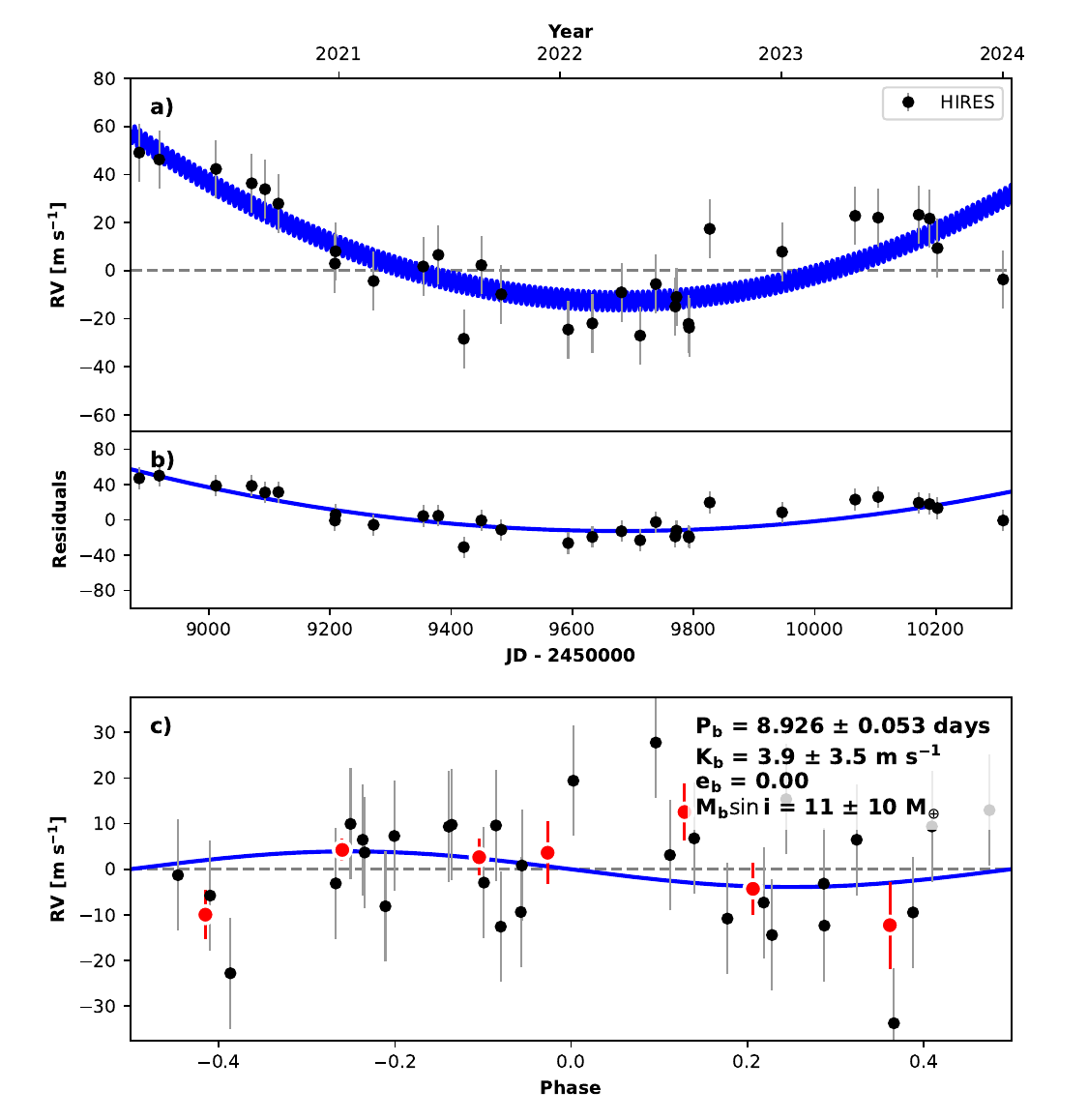}\hfill
    \includegraphics[width=.49\textwidth]{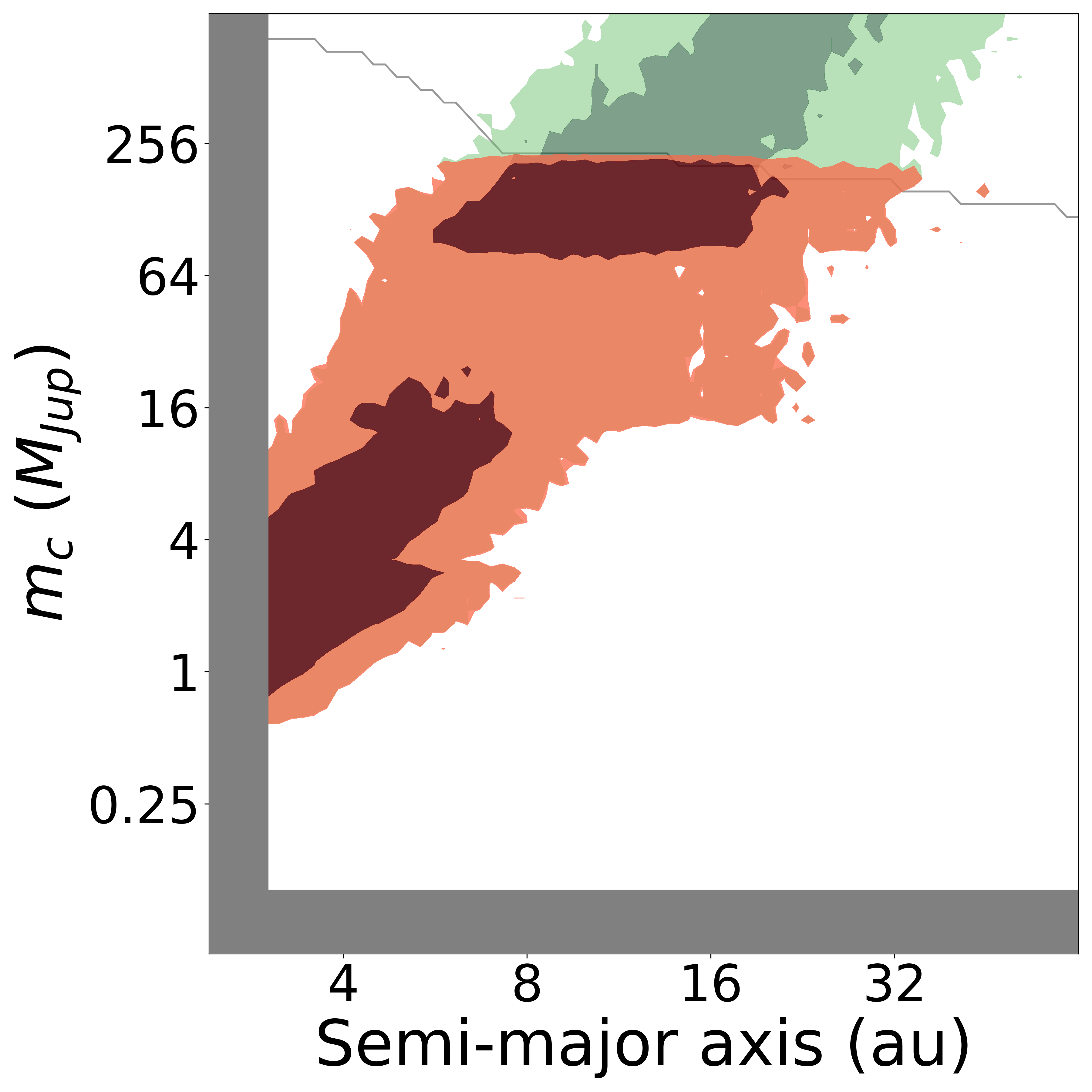}
    \caption{\textbf{Left:} Our orbital fit to the TOI-1174 system using \texttt{radvel}. Panel a) shows the full RV time series and errors with black circles, with our preferred model as a blue line. Panel b) shows the residuals to the planetary model, isolating the fitted trend/curvature. Panel c) shows the RV time series phase-folded to the period of the inner planet in this system. Red points give binned RV values. We did not recover the inner transiting planet signal in the RVs. \textbf{Right:} Our \texttt{ethraid} posterior surface derived using the measured trend. Green regions show models consistent with the RV trend, and red regions show models consistent with both the RVs and the direct imaging for this system, which revealed no luminous companions. Dark (light) regions indicate 68\% (95\%) confidence intervals. The gray line approximates the contrast limits imposed by imaging for a circular, face-on companion. Gray panels at low mass and short separations show companion parameters incompatible with the observed trend due to our observing baseline.}
    \label{fig:T001174_radvel_ethraid}
\end{figure}

\subsection{HD 191939}
\begin{figure}[H]
    \includegraphics[width=.46\textwidth]{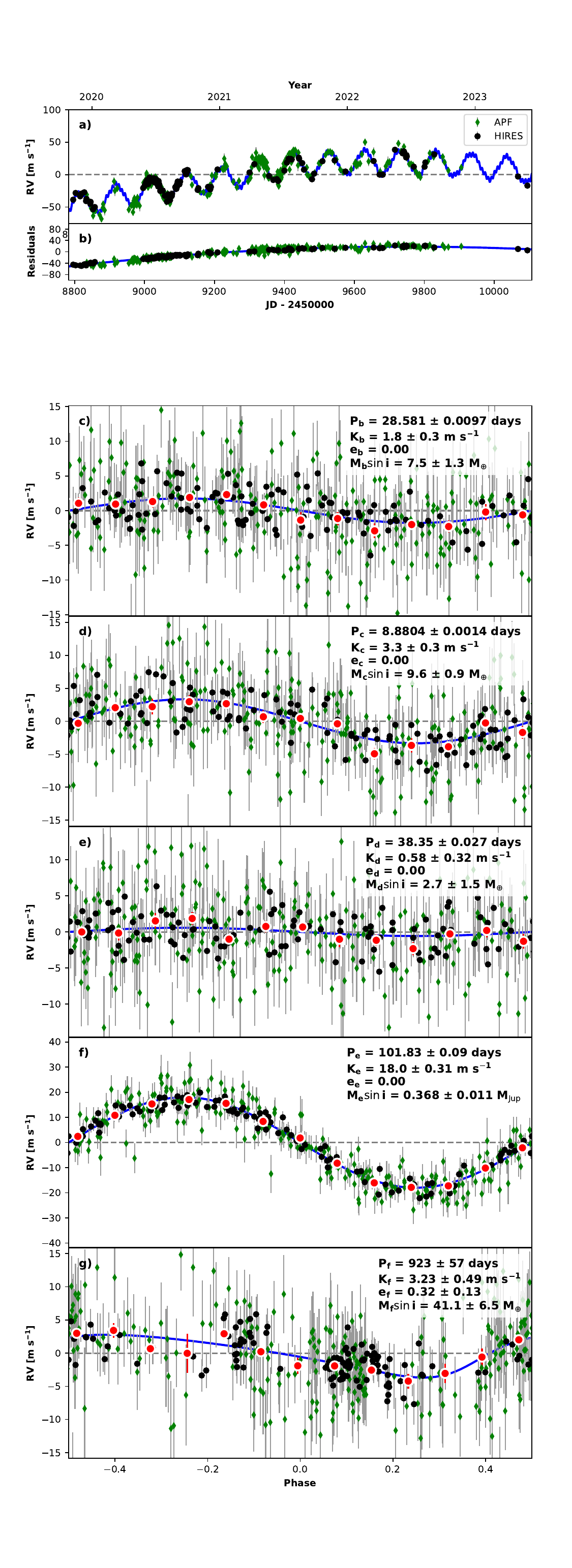}\hfill
    \includegraphics[width=.46\textwidth]{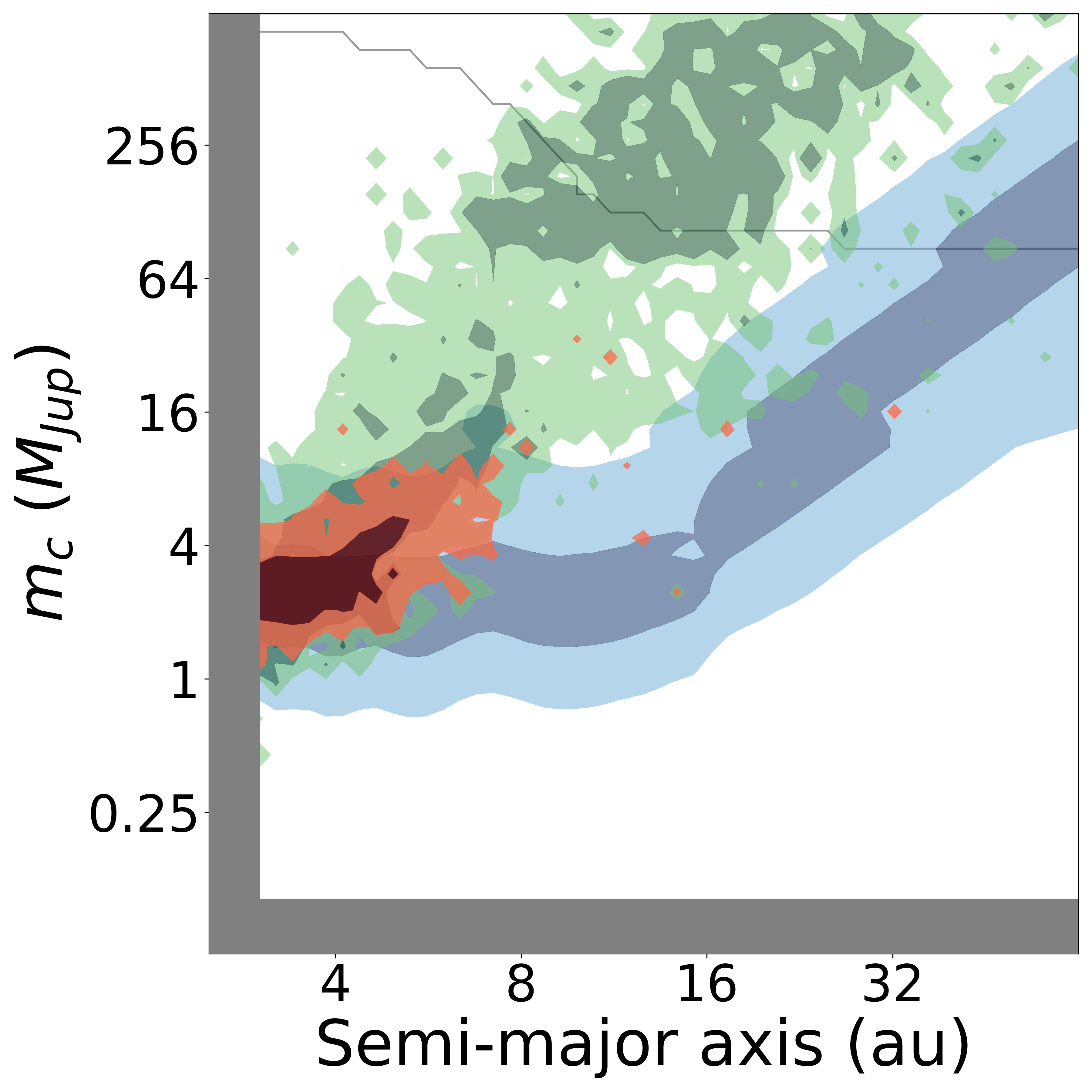}
    \caption{Same as Figure \ref{fig:T001174_radvel_ethraid} for HD 191939. We included the three transiting planets in our model, and our blind search algorithm detected the known 100-day super-Saturn as well as a spurious long-period planet. The blue posterior surface shows constraints imposed by the astrometric acceleration measured in this system. The trend and curvature in our truncated RV time series along with the astrometric acceleration, yield a high planetary odds ratio.}
    \label{fig:191939_radvel_ethraid}
\end{figure}

HD 191939 is a G0 dwarf 54 pc away hosting three transiting sub-Neptunes with periods of 8.9, 28.6, and 38.4 days. Multiple studies have probed this system with RVs \citep{BadenasAgusti2020, Lubin2022, Lubin2024}, resulting in mass measurements of the transiting planets, a determination that the system is aligned, and the discovery of a 100-day super-Saturn (e) and an eight-year super-Jupiter (f). We truncated this system's time series to four years, causing HD 191939 f to present as a trend. We combined this trend with an HGCA astrometric acceleration of $0.13 \pm 0.03$ mas/yr, which resulted in an 81\% probability that the trend's origin is planetary. We included adaptive optics imaging from Gemini/NIRI in our analysis, though it does not rule out any companion models. Figure \ref{fig:191939_radvel_ethraid} shows our orbital fit and trend analysis for this system. Our automated search algorithm recovered planet e, as well as a spurious 900-day planet. This planet demonstrates a shortcoming of our automated algorithm, which is not designed to be sensitive to multiple signals. Nevertheless, with an RV semi-amplitude of 3.2 m/s, it did not detract significantly from the signal of planet f, which has $K=47$ m/s.

\subsection{T001438}
T001438 is a K1 dwarf at $\distc$ pc hosting two transiting sub-Neptunes, the inner of which has a radius of $\radiusc \, \rearth$ and a period of $\perc$ days (Persson et al. in prep.). T001438 showed the largest RV trend in our sample: $\dot{\gamma} =$ \trendc, $\ddot{\gamma} =$ \curvc. Our trend analysis, along with 832 nm speckle imaging from 'Alopeke, indicated that these signals may originate from a planetary (33\%), brown dwarf (52\%) or stellar companion (15\%).

\begin{figure}[H]
    \includegraphics[width=.49\textwidth]{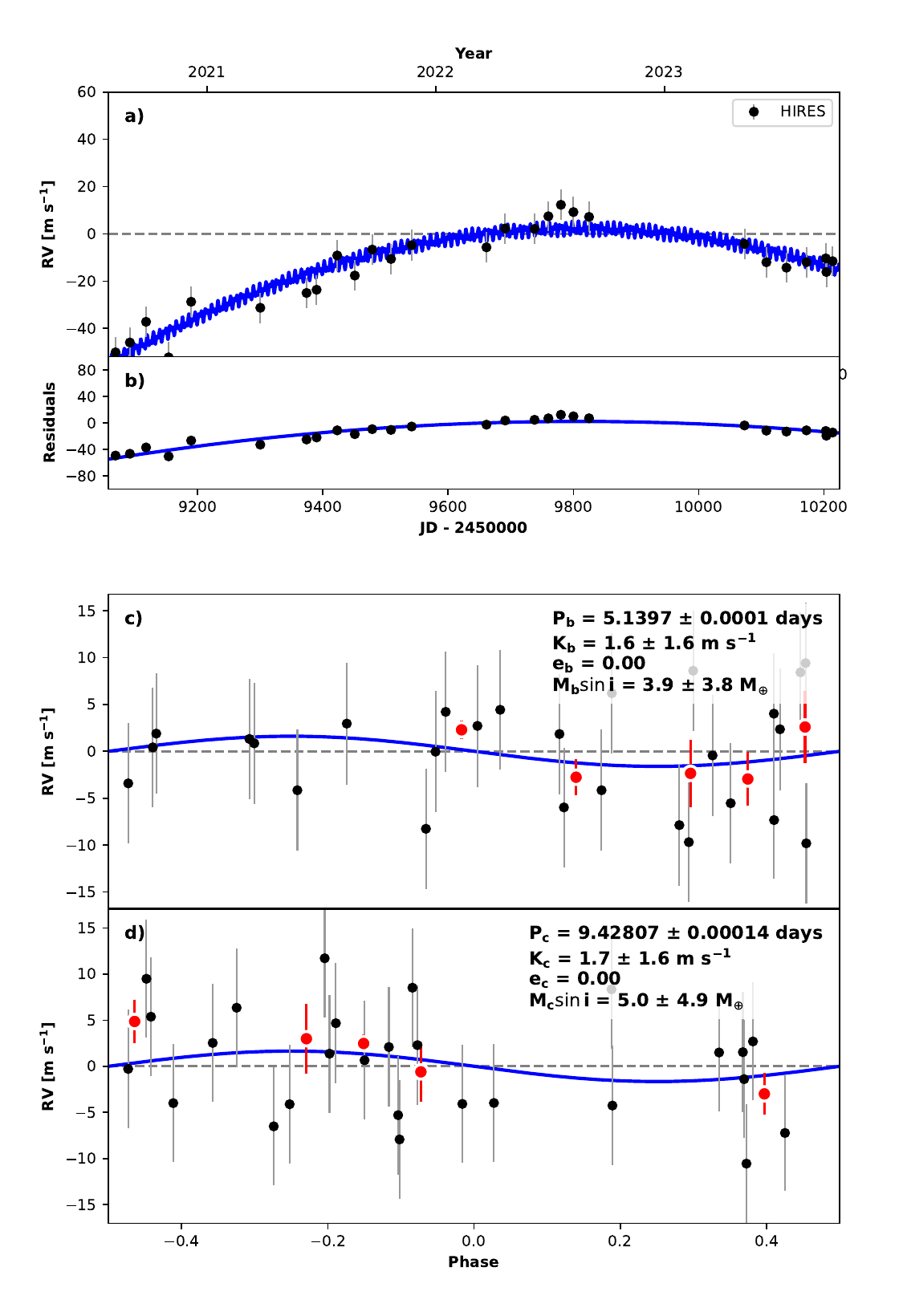}\hfill
    \includegraphics[width=.49\textwidth]{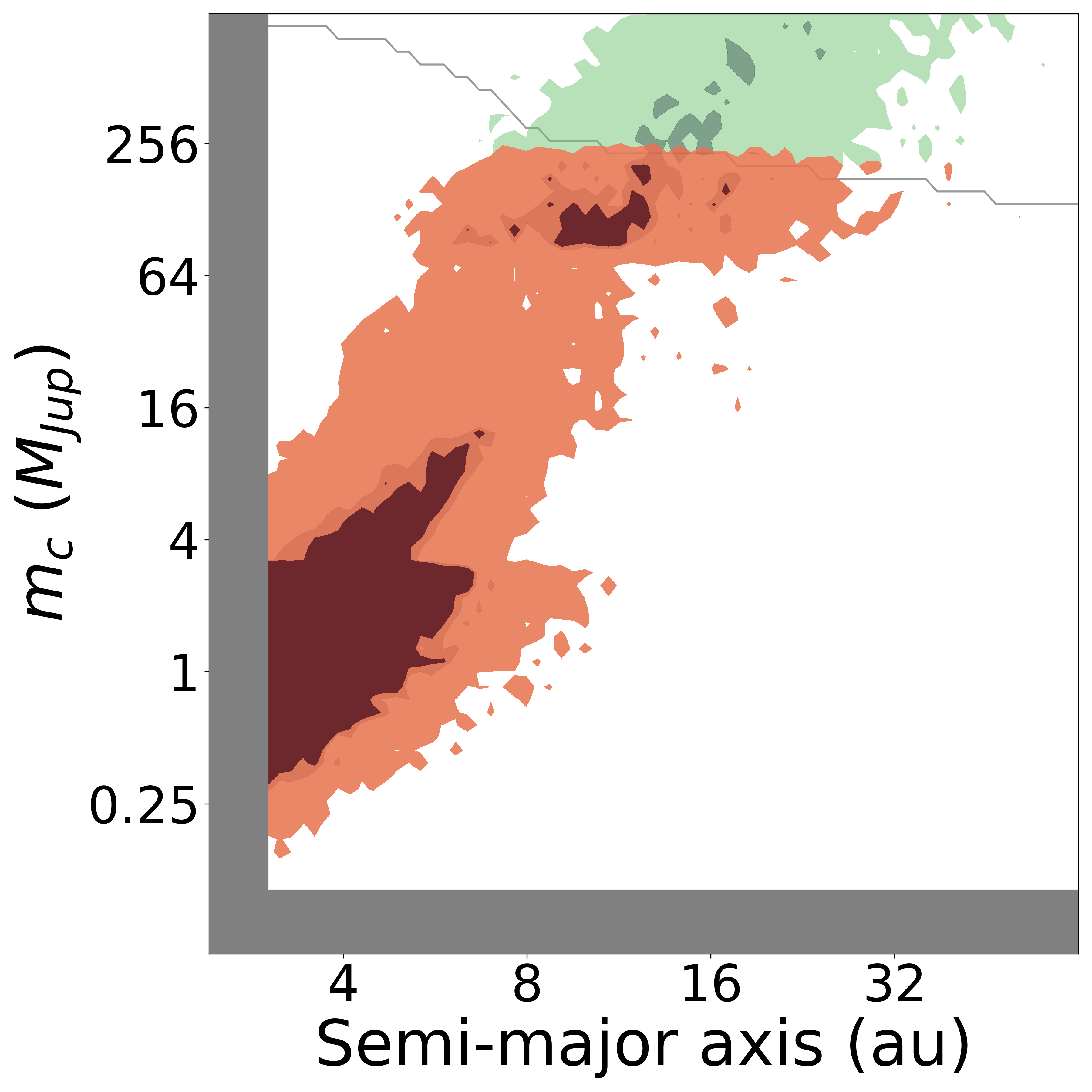}
    \caption{Same as Figure \ref{fig:T001174_radvel_ethraid} for TOI-1438. We were unable to detect the transiting planets in this system, but we measured strong trend and curvature, consistent with planetary, brown dwarf, or stellar companion models.}
    \label{fig:T001438_radvel_ethraid}
\end{figure}

\subsection{HD 219134}
\begin{figure}[H]
    \includegraphics[width=.49\textwidth]{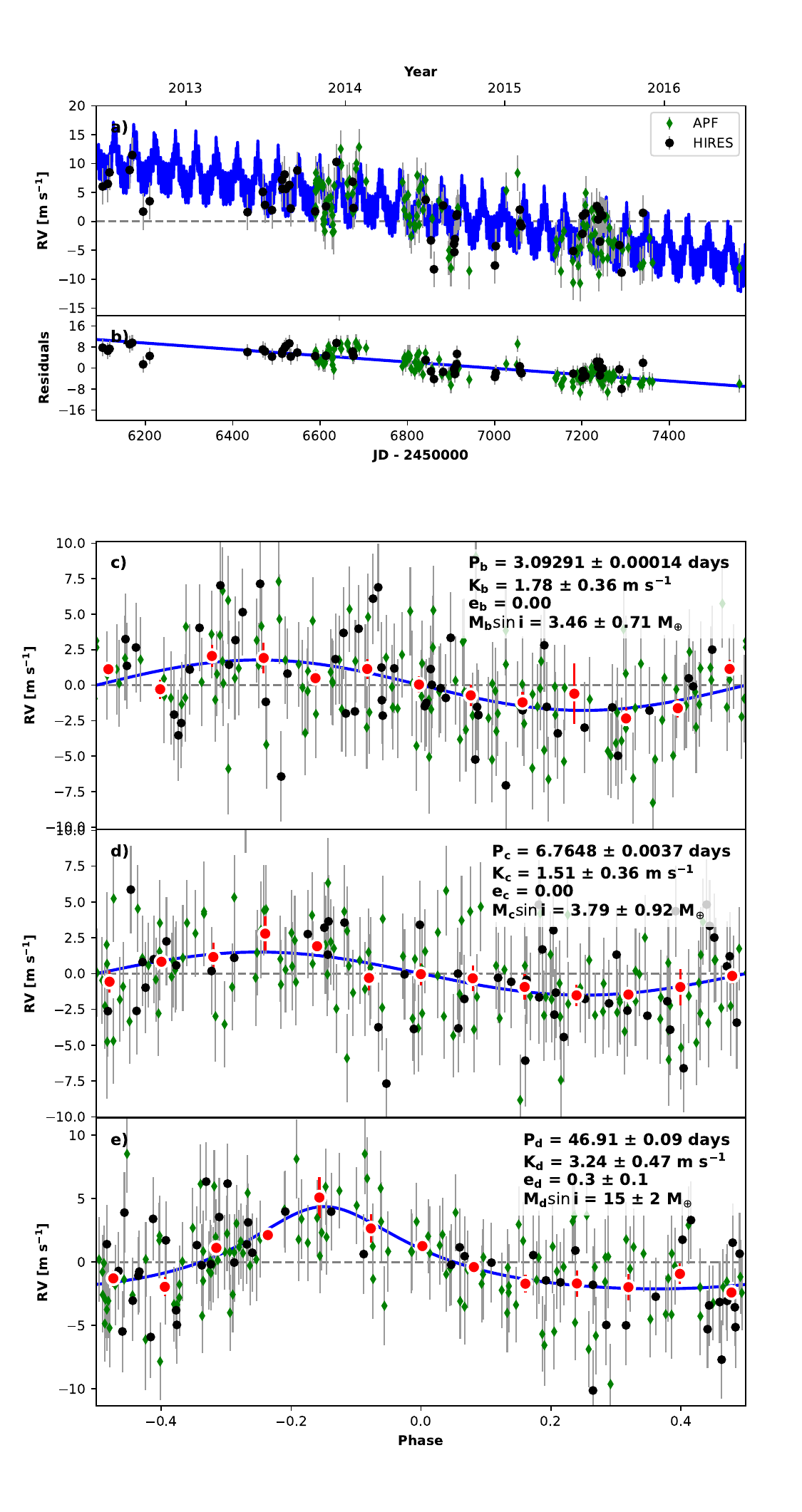}\hfill
    \includegraphics[width=.49\textwidth]{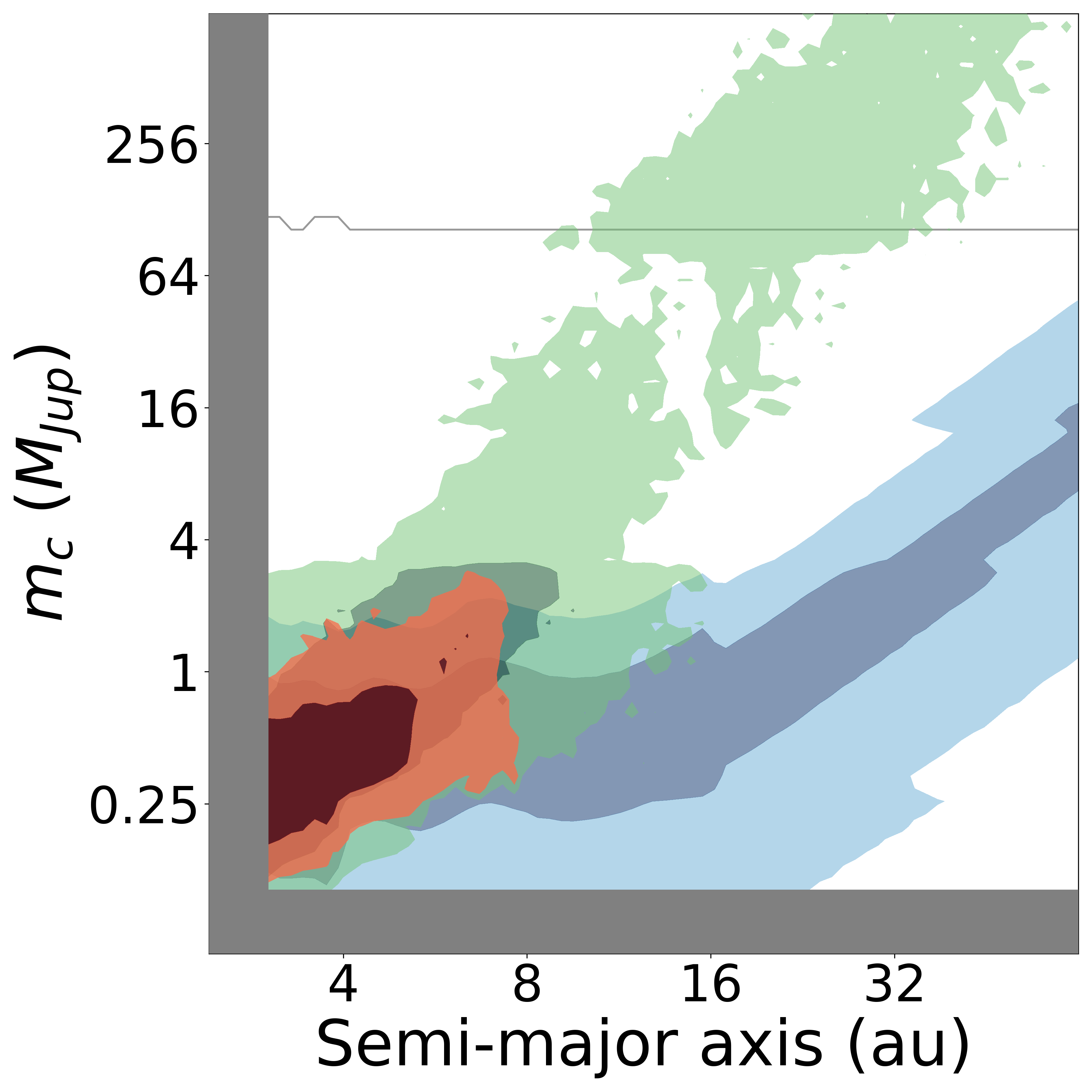}
    \caption{Same as Figure \ref{fig:T001174_radvel_ethraid} for HD 219134. We chose RV measurements from an arbitrary four-year span of this system's full data set. We recovered one of the non-transiting planets using a blind search, but missed two others. We analyzed the measured trend and HGCA astrometry to calculate a probability near 1 of this signal originating from a planet. We also tested other four-year spans and verified that our choice did not strongly influence our final odds ratio.}
    \label{fig:219134_radvel_ethraid}
\end{figure}

HD 219134 is a nearby (6.5 pc) K3 dwarf hosting two transiting super-Earths with periods of 3.1 and 6.8 days. This system has been observed for multiple decades, prividing detections of four additional planets, including a six-year super-Saturn meeting our distant giant definition \citep{Vogt2015}. As with HD 191939, we truncated this system's baseline to four years and performed our blind search. We tested the dependence of our results on our choice of truncation window, and found that it had a negligible effect. Our automated algorithm detected the 47-day Neptune analog (e), but missed a known super-Earth with a 23-day period (d) and a 94-day super-Earth (f). It also found a strong trend due to HD 219134 g, which we analyzed together with an HGCA acceleration to find a high planetary odds ratio ($p_{pl}\sim1$). Figure \ref{fig:219134_radvel_ethraid} shows the results of our full and partial orbit fits.

\subsection{HD 12572}
HD 12572 is a G9 dwarf at a distance of $\distd$ pc hosting two transiting sub-Neptunes. The inner planet, HD 12572 b, has a radius of $\radiusd \, \rearth$ and a $\perd$-day period \citep{Osborn2023}. This star's high brightness ($V=9.2$) allowed us to obtain contemporaneous APF RVs alongside our HIRES observations. We measured RV trend and curvature of \trendd~ and \curvd~ in this system, as well as a marginally significant astrometric acceleration of $\Delta \mu = 0.07 \pm 0.05$ mas/yr. Coupled with Br $\gamma$ direct imaging from NIRC2, we calculated a $72\%$ probability that the outer companion in this system is a planet. Our results are consistent with \cite{Osborn2023}, who concluded that the outer companion is a brown dwarf between 15-50 AU.

\begin{figure}[H]
    \includegraphics[width=.49\textwidth]{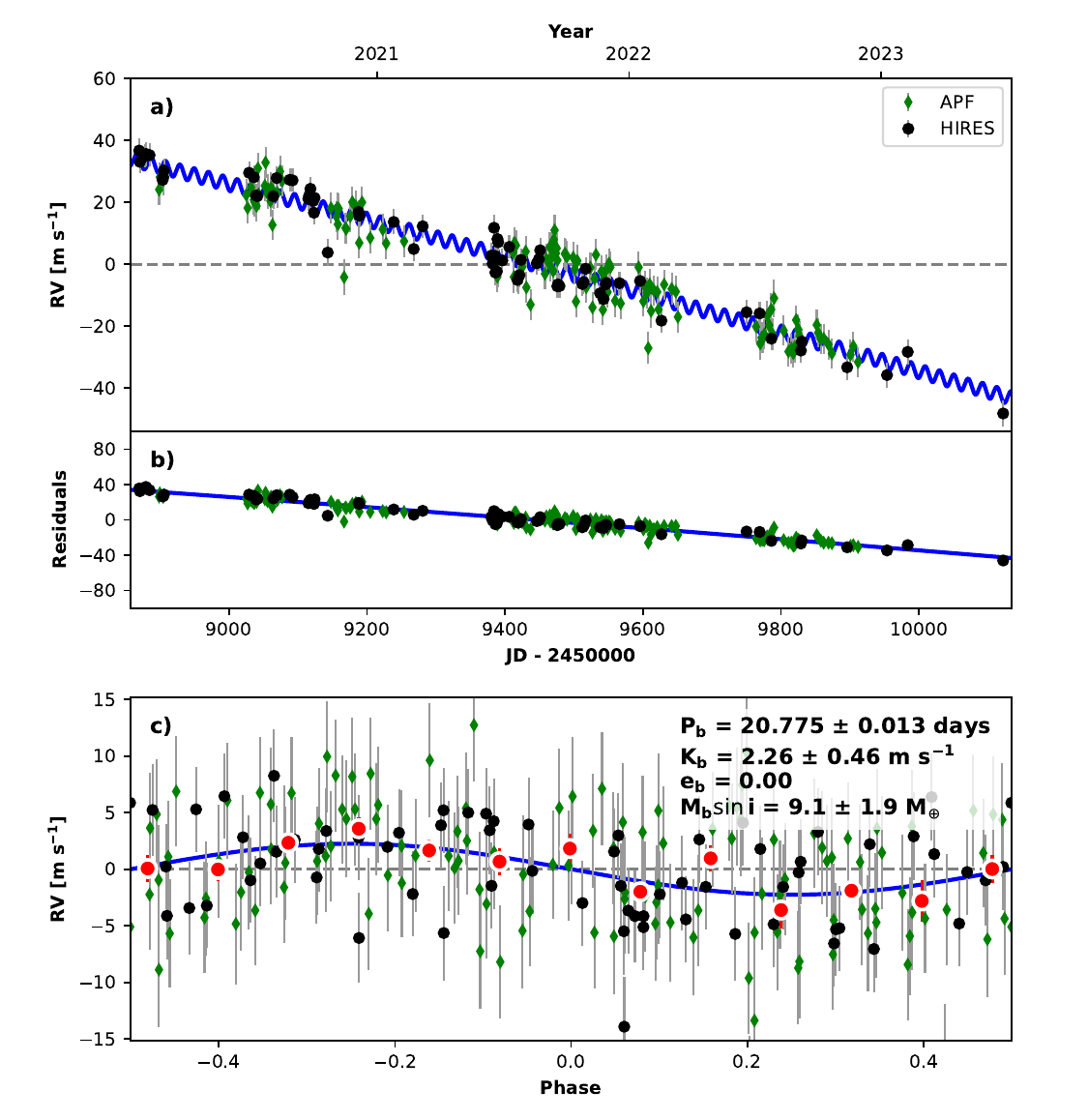}\hfill
    \includegraphics[width=.49\textwidth]{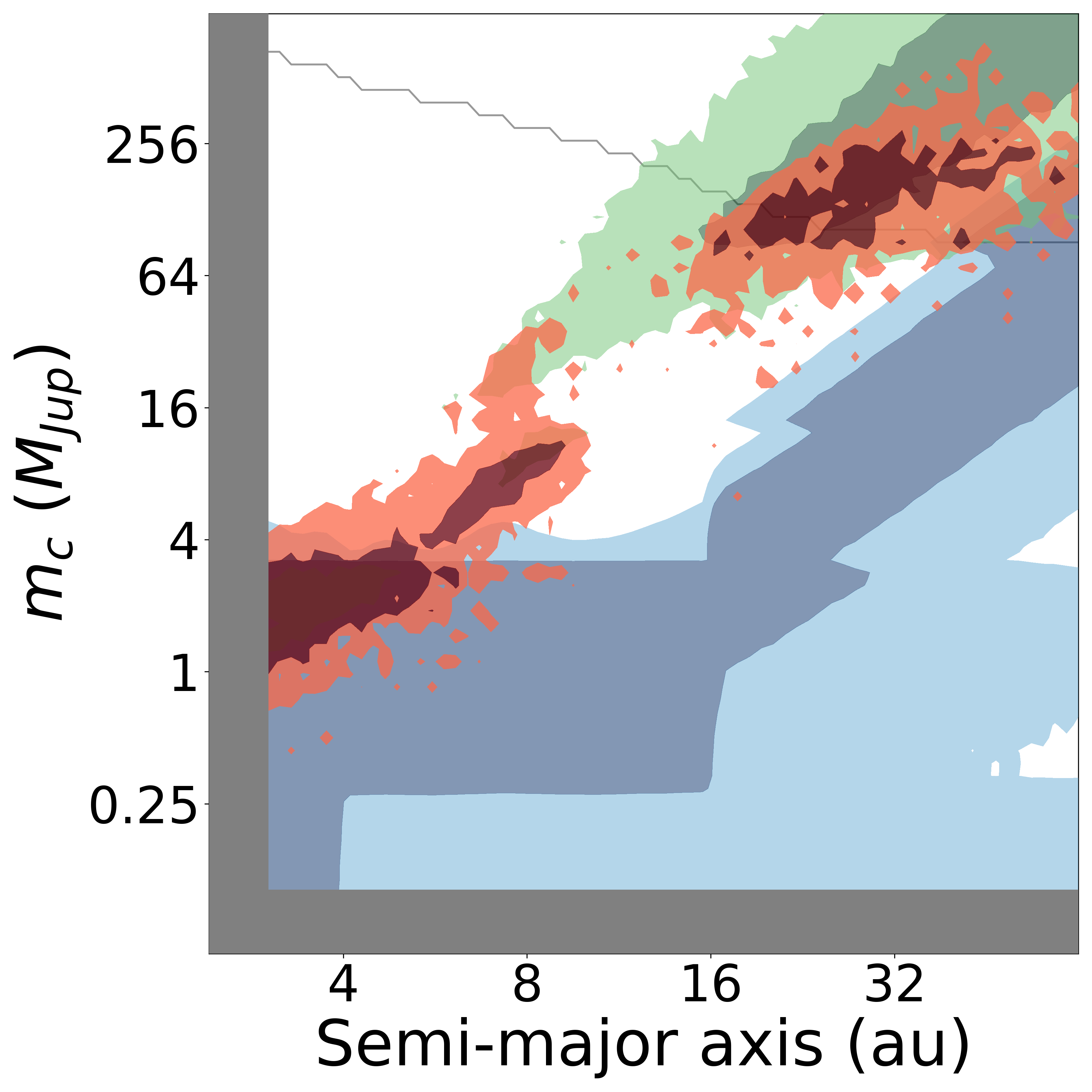}
    \caption{Same as Figure \ref{fig:T001174_radvel_ethraid} for HD 12572. We measured the mass of the 20-day sub-Neptune in this system and also found a strong linear trend with no significant curvature. We also measured a marginal astrometric acceleration, which imposed added constraints (blue contours). Note that the gray line marks orbital denotes orbital models which are ruled out by direct imaging under the assumption of a circular, face-on orbit. Companions with non-zero inclinations and eccentricities may lie beyond the line without being ruled out.}
    \label{fig:12572_radvel_ethraid}
\end{figure}

\subsection{HD 156141}
HD 156141 is a solar analog (G2) at a distance of $\diste$ pc hosting a transiting $\radiuse \, \rearth$ sub-Neptune with a $\pere$-day period. We measured RV trend and curvature of \trende~ and \curve, and ruled out high-mass stellar models using NIRC2 Br $\gamma$ imaging. We found that the outer companion in this system has a 35\% probability of being a planet.

\begin{figure}[H]
    \includegraphics[width=.49\textwidth]{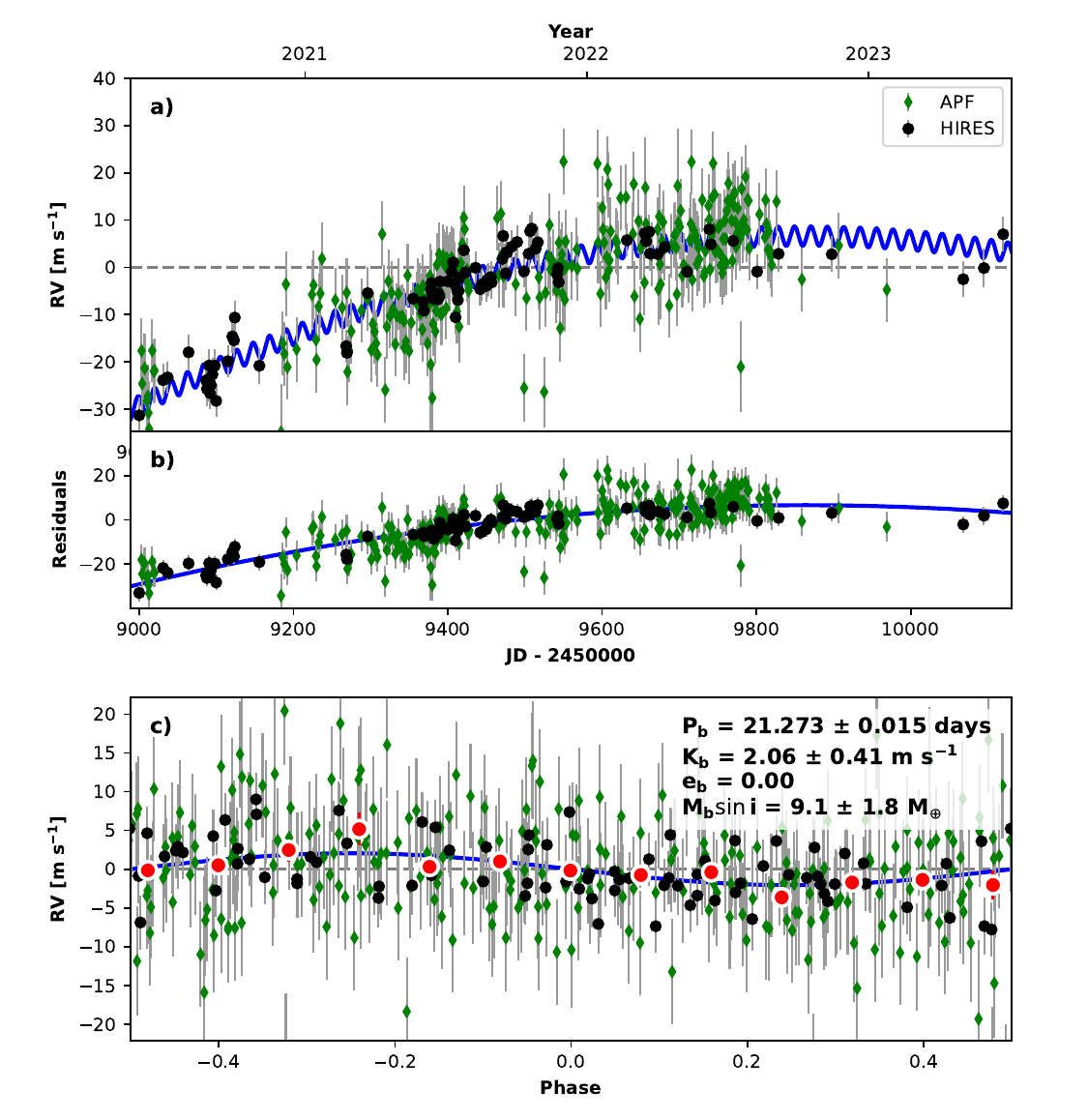}\hfill
    \includegraphics[width=.49\textwidth]{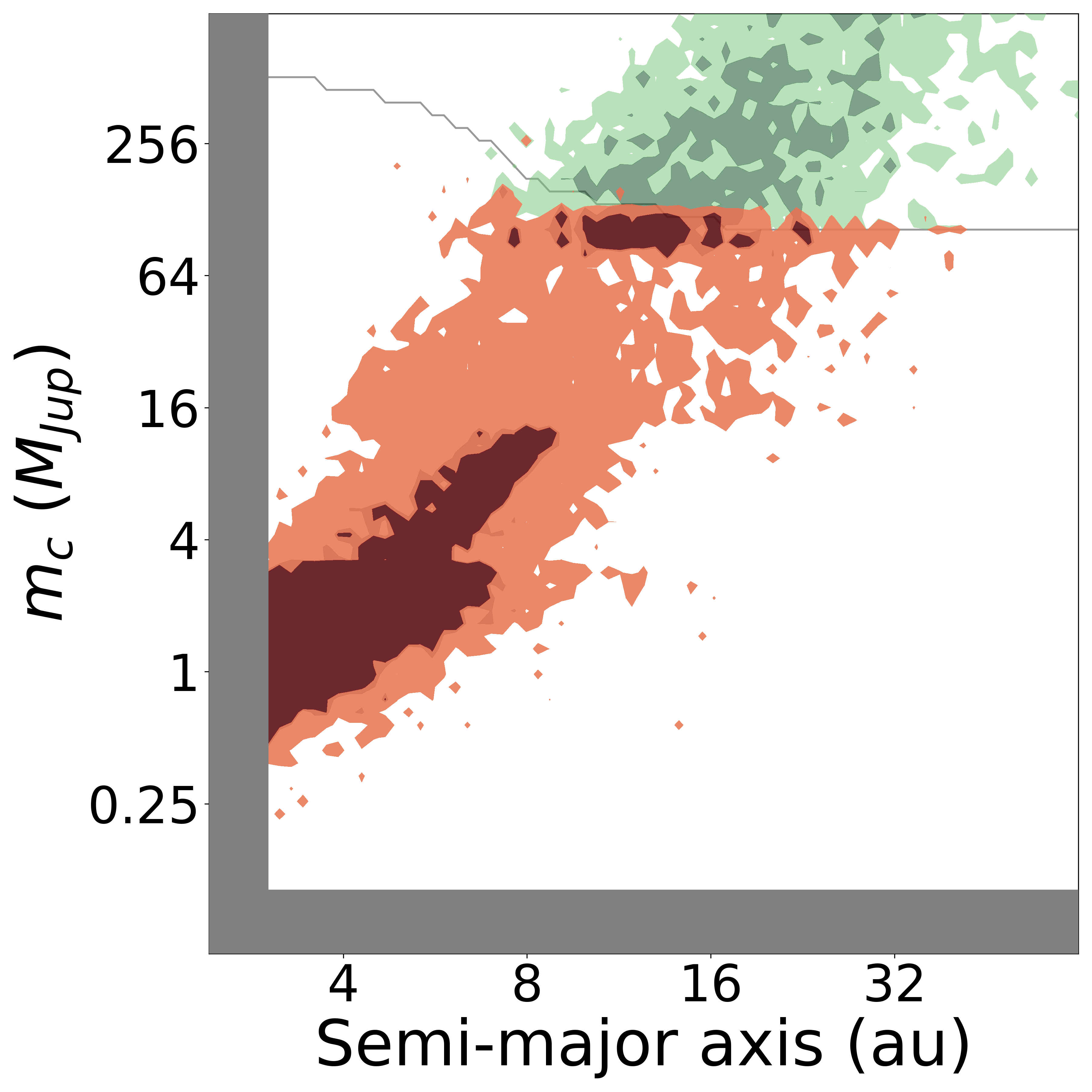}
    \caption{Same as Figure \ref{fig:T001174_radvel_ethraid} for HD 156141. We attained a marginally significant measurement of the inner transiting planet's mass, and high-significance trend and curvature measurements. The long-term signals in this system are consistent with planets and brown dwarfs, whereas stellar models are nearly ruled out with the aid of direct imaging.}
    \label{fig:156141_radvel_ethraid}
\end{figure}

\subsection{HD 75732}
\begin{figure}[H]
    \includegraphics[width=.49\textwidth]{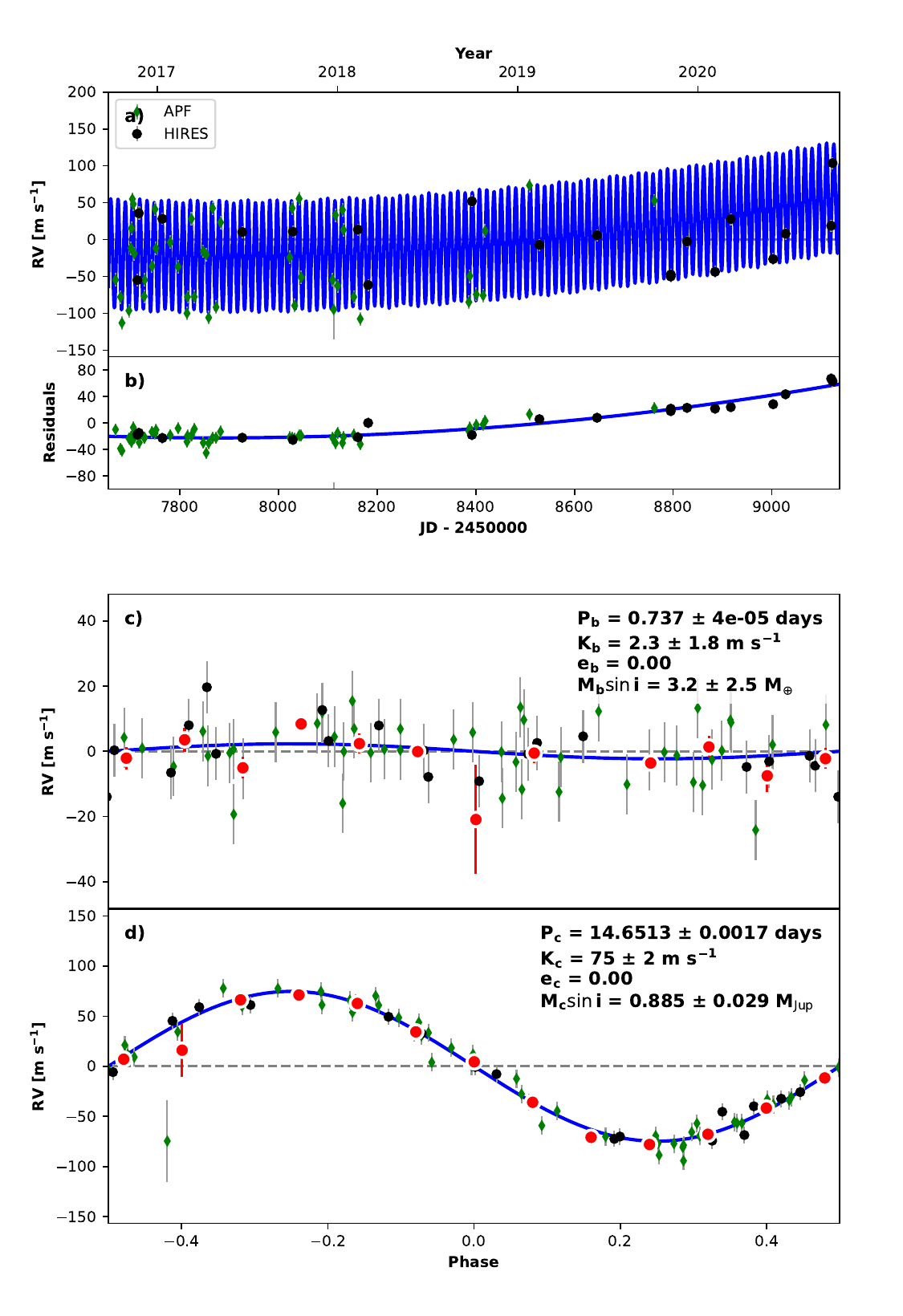}\hfill
    \includegraphics[width=.49\textwidth]{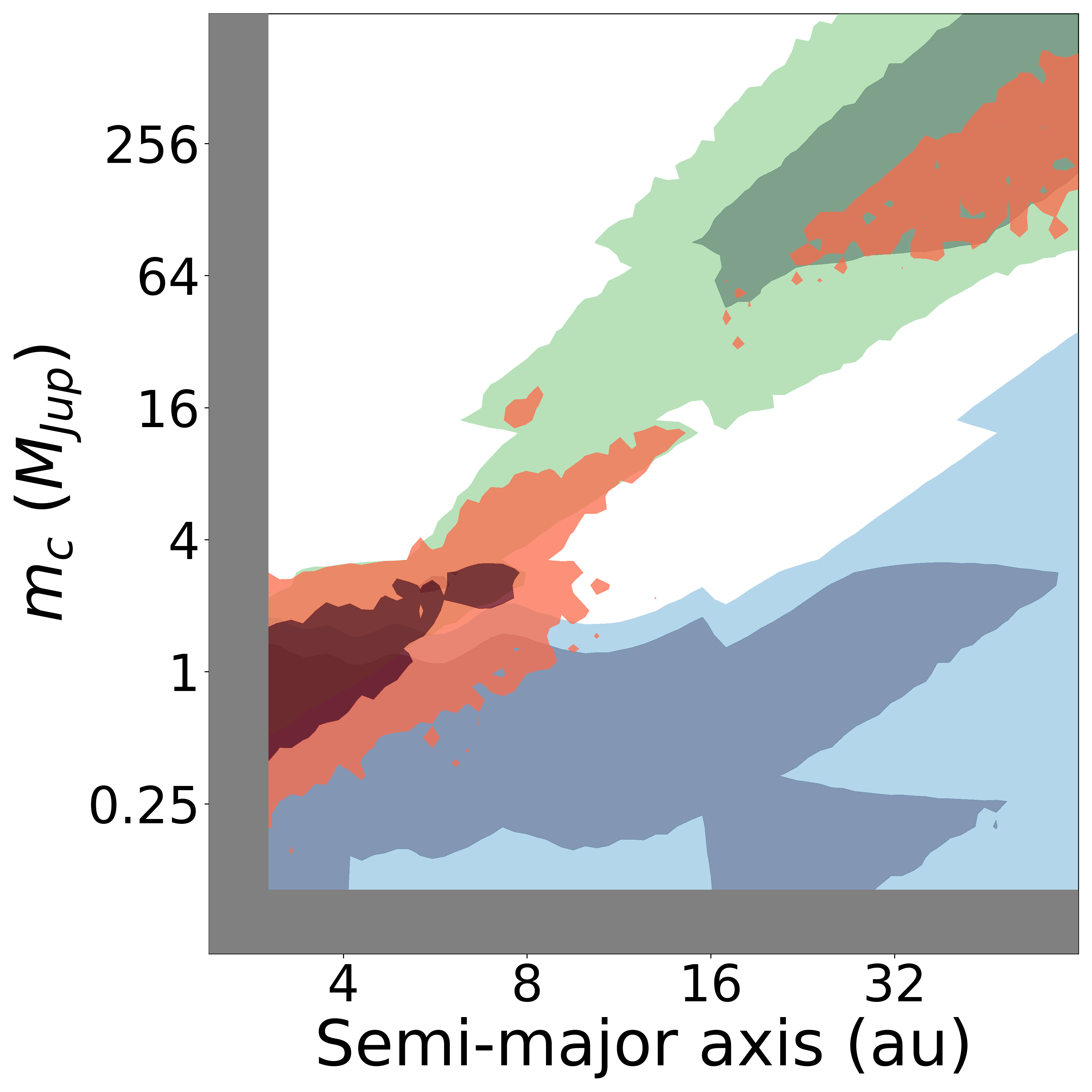}
    \caption{Same as Figure \ref{fig:T001174_radvel_ethraid} for HD 75732. We detected only one of the four non-transiting planets in this system using our blind search algorithm. The residual trend, together with a low-significance astrometric acceleration, constrains the mass-separation posterior primarily to the planetary regime.}
    \label{fig:75732_radvel_ethraid}
\end{figure}

HD 75732 is a nearby (12.5 pc) K0 dwarf hosting a transiting ultra-short period (0.74-day) super-Earth. Like HD 219134, this system is well-characterized from decades of RV observation (e.g., \citealt{Fischer2008}). We chose an arbitrary four-year window over which to fit these RVs, and verified that our choice did not significantly impact our characterization of the outer planet. We detected the hot Jupiter HD 75732 b, but did not detect the four other non-transiting planets. The outermost of these, a super-Jupiter with a period of nearly 14 years, manifested as a trend in our truncated RV time series. We combined this trend with a marginal detection of HGCA acceleration to constrain the companion's mass and separation. Our analysis indicates that the trend is almost certainly planetary ($p_{pl}\sim1$). We show our orbital fit and partial orbit analysis in Figure \ref{fig:75732_radvel_ethraid}.

\subsection{HD 93963}
\begin{figure}[H]
    \includegraphics[width=.49\textwidth]{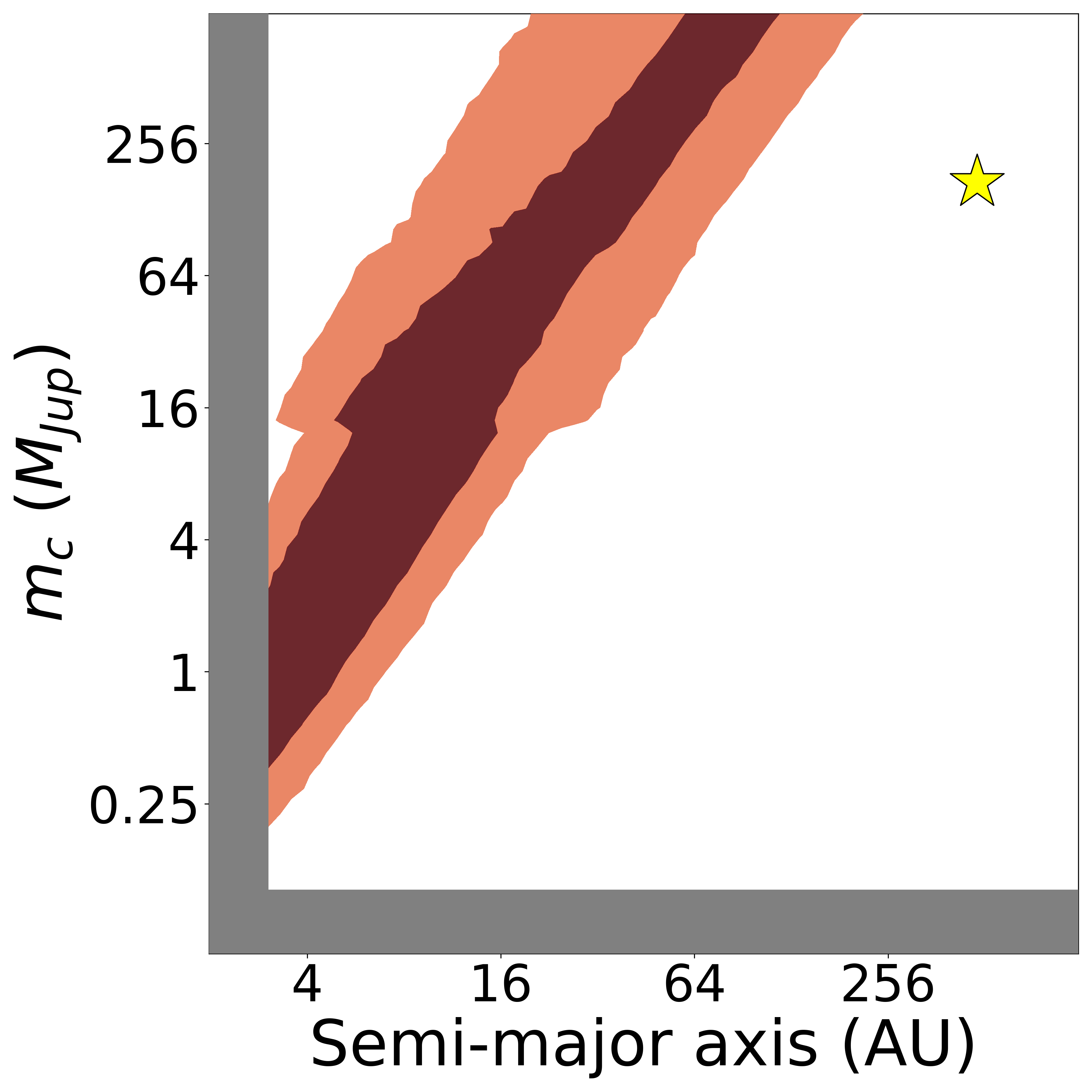}
    \caption{Our second analysis of the trend in the HD 93963 system. We expanded the semi-major axis range over which we tested companion models and therefore did not use the informative mass/separation prior described in Section \ref{subsec:ethraid_prior}, which is defined for separations $\leq 64$ AU. We indicate the position of the stellar companion, HD 93963 B, with a yellow star. Our analysis suggests that an M5 dwarf at a separation of 484 AU is too small/too distant to have caused the observed trend.}
    \label{fig:93963_stellar_companion}
\end{figure}

HD 93963 A is a G2 dwarf at a distance of $\distf$ pc hosting a transiting $\radiusf \, \rearth$ sub-Neptune with a $\perf$-day period \citep{Serrano2022}. Our measured RV trend and curvature of \trendf~ and \curvf, together with 832 nm speckle imaging from 'Alopeke, indicate a 54\% probability of a planetary outer companion. \cite{Serrano2022} estimated that the stellar companion to this star, HD 93963 B, has a separation of $\geq 484$ AU and a spectral type of M5 V ($\approx 170$ \mj; \citealt{PecautMamajek2013}). We show in Figure \ref{fig:93963_stellar_companion} that a companion of that mass and separation is incompatible with the measured RV signature.

\begin{figure}[H]
    \includegraphics[width=.49\textwidth]{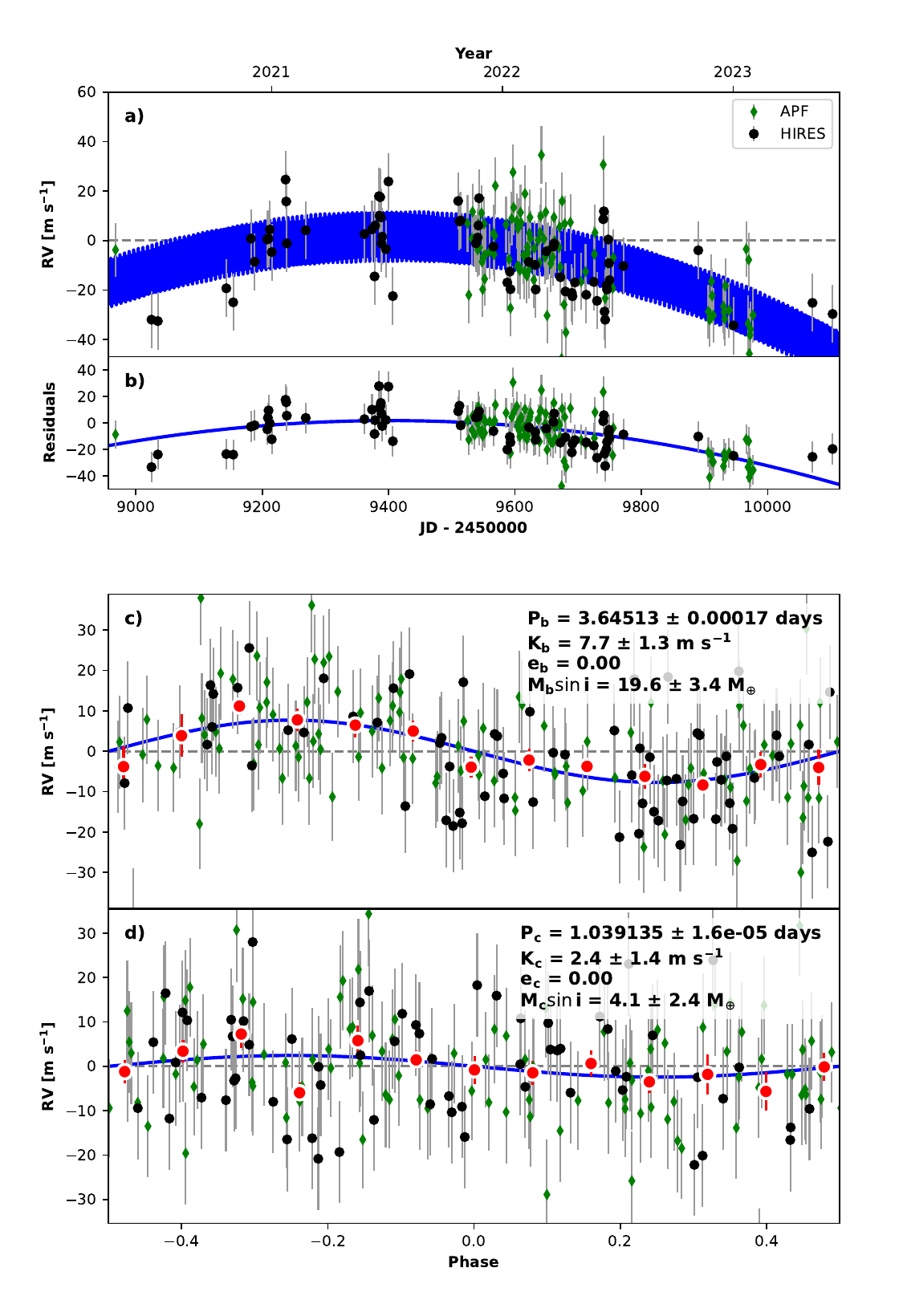}\hfill
    \includegraphics[width=.49\textwidth]{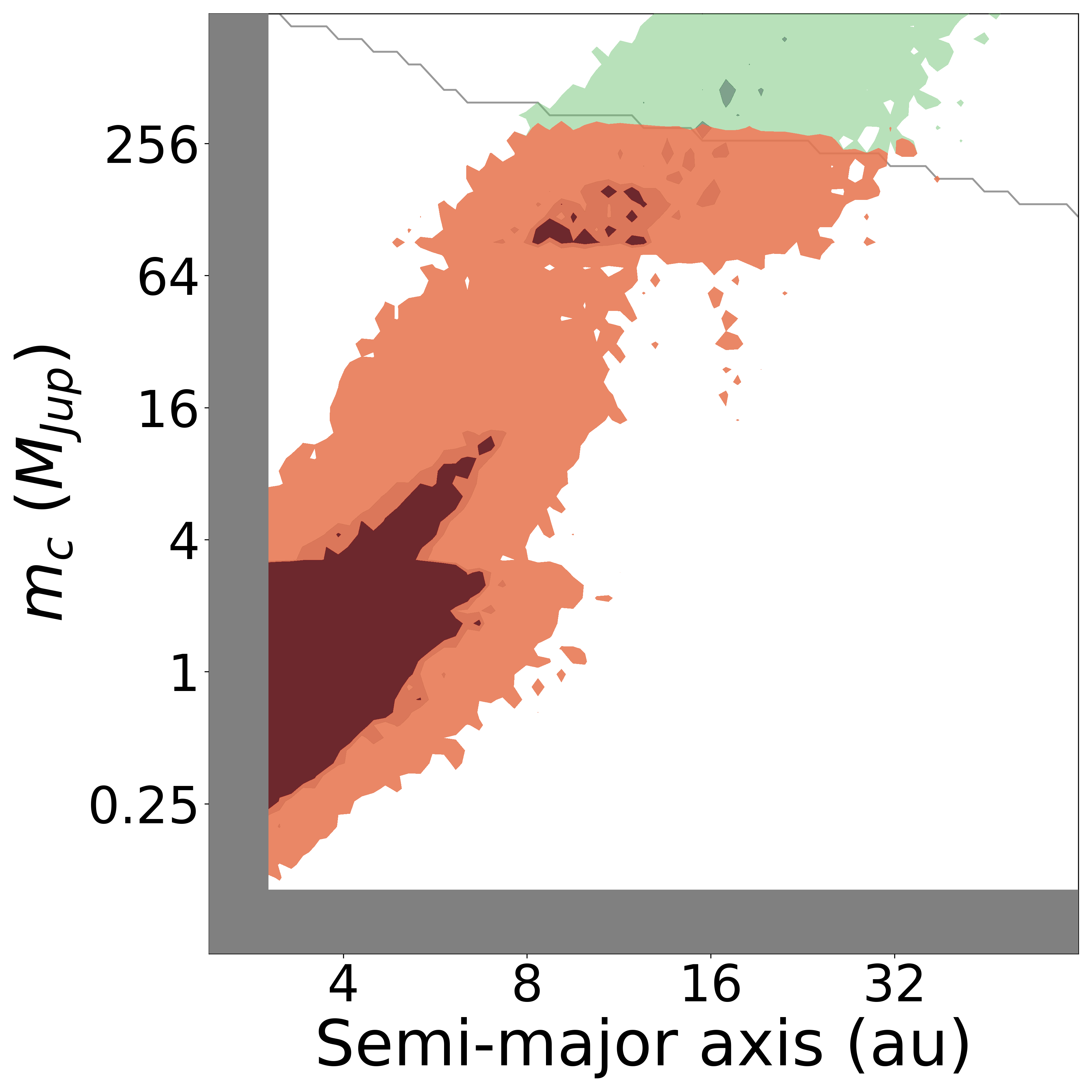}
    \caption{Same as Figure \ref{fig:T001174_radvel_ethraid} for HD 93963. We did not recover either of this system's two transiting planets at high significance. We measured a significant trend and marginal curvature in this system. Our trend analysis showed that the source of this RV trend has a roughly 50\% probability of being a planet.}
    \label{fig:93963_radvel_ethraid}
\end{figure}

\subsection{TIC 142381532}
TIC 142381532 is a K0 dwarf $\distg$ pc away hosting a transiting $\radiusg \, \rearth$ sub-Saturn with a $\perg$-day period \citep{Polanski2024}. We measured RV trend and curvature of \trendg~ and \curvg, and used 832 nm speckle imaging from 'Alopeke to rule out high-mass stellar companions. We calculated a $45\%$ probability that the measured signals originate from a planet. Despite passing our original radius filter of $R_p < 10$ \rearth, the transiting planet in this system does not fit most definitions of a ``small'' planet. We include it for completeness, but exclude it from our conditional occurrence calculations. 

\begin{figure*}
    \includegraphics[width=.49\textwidth]{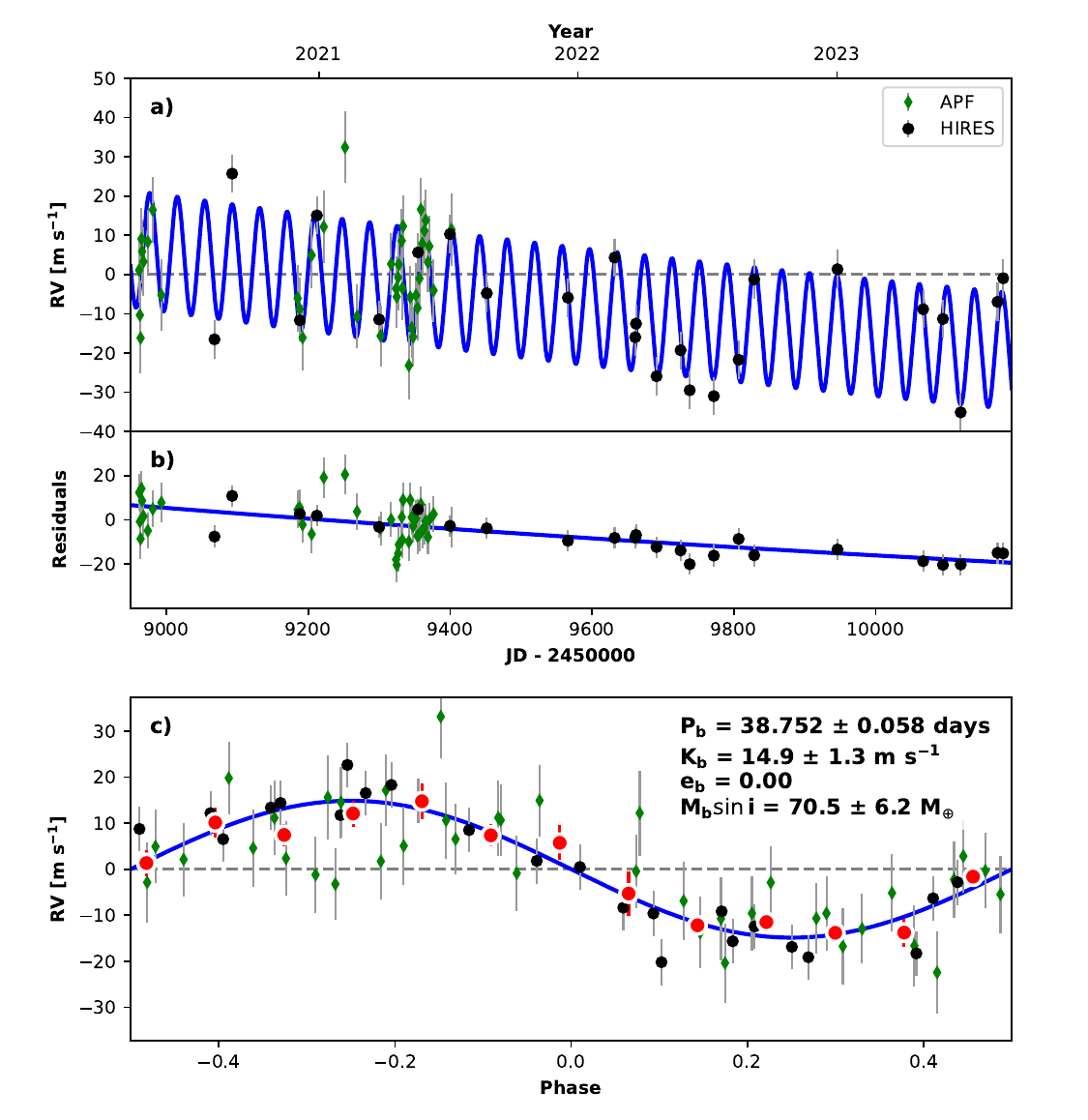}\hfill
    \includegraphics[width=.49\textwidth]{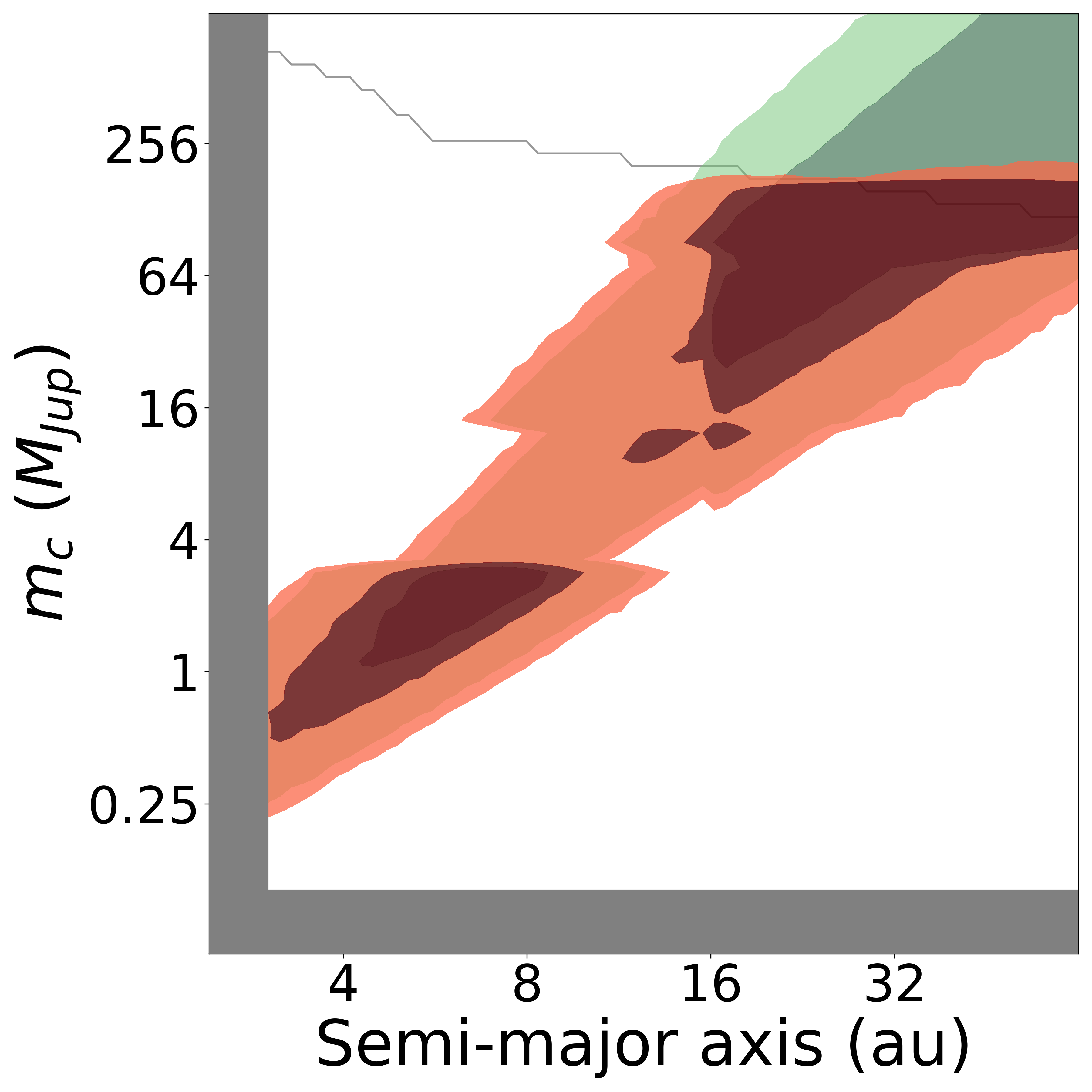}
    \caption{Same as Figure \ref{fig:T001174_radvel_ethraid} for TIC 142381532. The inner transiting planet in this system in a sub-Saturn, which we characterized at high significance. Although this planet passed our $R_p<10 \, \rearth$, it is likely a gas giant, so we exclude this system from our occurrence calculations. Our trend measurement is marginal, and evinces planetary and brown dwarf models with roughly equal probability.}
    \label{fig:TIC142381532_radvel_ethraid}
\end{figure*}

\newpage

\section{Transiting Planet Properties}
\label{appendix:transiting}
\begin{deluxetable*}{ccccccccccc}
\tabletypesize{\scriptsize}
\tablecolumns{8}
\tablewidth{0pt}
\tablecaption{Distant Giants Sample}

\label{tab:dg_transit_table}

\tablehead{
    \colhead{TOI} &
    \colhead{TKS Name} &
    \colhead{RA (deg)} &
    \colhead{Dec (deg)} &
    \colhead{$V$} &
    \colhead{$T_{\text{eff}}$} &
    \colhead{[Fe/H]} &
    \colhead{$R_p \, (\rearth)$} &
    \colhead{P (days)}&
    \colhead{Distant Giant?}&
    \colhead{Trend?}
}
\startdata
 465 &      WASP156 &  32.8 &  2.4 &  11.6 &  5032 &  0.29 & 5.6 &     3.8 &     X &  X \\
 509 &        63935 & 117.9 &  9.4 &   8.6 &  5534 &  0.09 & 3.1 &    18.1 &     X &  X \\
1173 &      T001173 & 197.7 & 70.8 &  11.0 &  5352 &  0.18 & 9.2 &     7.1 &     X &  X \\
1174 &      T001174 & 209.2 & 68.6 &  11.0 &  5124 &  0.00 & 2.3 &     9.0 &     X &   \checkmark \\
1180 &      T001180 & 214.6 & 82.2 &  11.0 &  4790 & -0.01 & 2.8 &     9.7 &     X &  X \\
1194 &      T001194 & 167.8 & 70.0 &  11.3 &  5428 &  0.33 & 8.9 &     2.3 &     X &  X \\
1244 &      T001244 & 256.3 & 69.5 &  11.9 &  4675 & -0.04 & 2.4 &     6.4 &     X &  X \\
1246 &      T001246 & 251.1 & 70.4 &  11.6 &  5158 &  0.17 & 3.3 &    18.7 &     X &  X \\
1247 &       135694 & 227.9 & 71.8 &   9.1 &  5648 & -0.13 & 2.8 &    15.9 &     X &  X \\
1248 &      T001248 & 259.0 & 63.1 &  11.8 &  5272 &  0.22 & 6.6 &     4.4 &     X &  X \\
1249 &      T001249 & 200.6 & 66.3 &  11.1 &  5514 &  0.29 & 3.2 &    13.1 &     X &  X \\
1255 &     HIP97166 & 296.2 & 74.1 &   9.9 &  5214 &  0.28 & 2.7 &    10.3 &     X &  X \\
1269 &      T001269 & 249.7 & 64.6 &  11.6 &  5466 & -0.06 & 2.4 &     4.3 &     X &  X \\
1272 &      T001272 & 199.2 & 49.9 &  11.8 &  5091 &  0.21 & 4.3 &     3.3 &     X &  X \\
1279 &      T001279 & 185.1 & 56.2 &  10.7 &  5414 & -0.10 & 2.6 &     9.6 &     X &  X \\
1288 &      T001288 & 313.2 & 65.6 &  10.4 &  5357 &  0.26 & 4.7 &     2.7 &      \checkmark &  X \\
1339 &       191939 & 302.0 & 66.9 &   9.0 &  5355 & -0.15 & 3.2 &     8.9 &      \checkmark &  X \\
1410 &      T001410 & 334.9 & 42.6 &  11.1 &  4666 &  0.16 & 2.9 &     1.2 &     X &  X \\
1411 &      GJ9522A & 232.9 & 47.1 &  10.5 &  4478 & -0.10 & 1.4 &     1.5 &     X &  X \\
1422 &      T001422 & 354.2 & 39.6 &  10.6 &  5852 & -0.03 & 3.1 &    13.0 &     X &  X \\
1437 &       154840 & 256.1 & 56.8 &   9.2 &  6049 & -0.19 & 2.4 &    18.8 &     X &  X \\
1438 &      T001438 & 280.9 & 74.9 &  11.0 &  5234 &  0.08 & 2.8 &     5.1 &     X &   \checkmark \\
1443 &      T001443 & 297.4 & 76.1 &  10.7 &  5160 & -0.30 & 2.1 &    23.5 &     X &  X \\
1444 &      T001444 & 305.5 & 70.9 &  10.9 &  5466 &  0.14 & 1.3 &     0.5 &     X &  X \\
1451 &      T001451 & 186.5 & 61.3 &   9.6 &  5735 & -0.01 & 2.5 &    16.5 &     X &  X \\
1469 &       219134 & 348.3 & 57.2 &   5.6 &  4839 &  0.11 & 1.2 &     3.1 &      \checkmark &  X \\
1471 &        12572 &  30.9 & 21.3 &   9.2 &  5599 & -0.03 & 4.3 &    20.8 &     X &   \checkmark \\
1472 &      T001472 &  14.1 & 48.6 &  11.3 &  5186 &  0.28 & 4.3 &     6.4 &     X &  X \\
1611 &       207897 & 325.2 & 84.3 &   8.4 &  5091 & -0.04 & 2.7 &    16.2 &     X &  X \\
1669 &      T001669 &  46.0 & 83.6 &  10.2 &  5551 &  0.26 & 2.2 &     2.7 &      \checkmark &  X \\
1691 &      T001691 & 272.4 & 86.9 &  10.1 &  5689 &  0.03 & 3.8 &    16.7 &     X &  X \\
1694 &      T001694 &  97.7 & 66.4 &  11.4 &  5069 &  0.12 & 5.5 &     3.8 &      \checkmark &  X \\
1710 &      T001710 &  94.3 & 76.2 &   9.5 &  5734 &  0.15 & 5.4 &    24.3 &     X &  X \\
1716 &       237566 & 105.1 & 56.8 &   9.4 &  5861 &  0.06 & 2.7 &     8.1 &     X &  X \\
1723 &      T001723 & 116.8 & 68.5 &   9.7 &  5800 &  0.16 & 3.2 &    13.7 &     X &  X \\
1742 &       156141 & 257.3 & 71.9 &   8.9 &  5733 &  0.18 & 2.2 &    21.3 &     X &   \checkmark \\
1751 &       146757 & 243.5 & 63.5 &   9.3 &  5961 & -0.38 & 2.8 &    37.5 &     X &  X \\
1753 &      T001753 & 252.5 & 61.2 &  11.8 &  5620 &  0.03 & 3.0 &     5.4 &     X &  X \\
1758 &      T001758 & 354.7 & 75.7 &  10.8 &  5142 & -0.03 & 3.8 &    20.7 &     X &  X \\
1759 &      T001759 & 326.9 & 62.8 &  11.9 &  4420 & -0.20 & 3.2 &    37.7 &     X &  X \\
1773 &        75732 & 133.1 & 28.3 &   6.0 &  5363 &  0.42 & 1.8 &     0.7 &      \checkmark &  X \\
1775 &      T001775 & 150.1 & 39.5 &  11.6 &  5349 &  0.19 & 8.1 &    10.2 &     X &  X \\
1794 &      T001794 & 203.4 & 49.1 &  10.3 &  5663 &  0.02 & 3.0 &     8.8 &     X &  X \\
1797 &        93963 & 162.8 & 25.6 &   9.2 &  5948 &  0.10 & 3.2 &     3.6 &     X &   \checkmark \\
1823 & TIC142381532 & 196.2 & 63.8 &  10.7 &  4917 &  0.28 & 8.1 &    38.8 &     X &   \checkmark \\
1824 &      T001824 & 197.7 & 61.7 &   9.7 &  5216 &  0.12 & 2.4 &    22.8 &     X &  X \\
2088 &      T002088 & 261.4 & 75.9 &  11.6 &  4902 &  0.31 & 3.5 &   124.7 &     X &  X \\
\enddata
\tablecomments{Properties of the 47 stars in the Distant Giants sample, plus the periods and radii of their inner companions. For multi-transiting systems, we checked planets in the order that TESS detected them, and show the properties of the first one that passed our filters. We truncated period precisions for readability. Median uncertainties are as follows: [Fe/H]--0.06-0.09 dex; $R_p$---9.6\%; P---60 ppm. We calculated metallicity values using the \texttt{SpecMatch-Synthetic} code \citep{PetiguraThesis} for host stars with $T_{\text{eff}}>4800$ K ($\sigma_{\text{[Fe/H]}}=0.06$ dex). We used \texttt{SpecMatch-Empirical} \citep{Yee2017} for host stars above below this limit ($\sigma_{\text{[Fe/H]}}=0.09$ dex). We retrieved all other values from \cite{Chontos2022}. We also indicate in the two right-most columns which systems exhibit either a fully resolved giant planet signal or a long-term RV trend.}
\end{deluxetable*}

\end{document}